\RequirePackage[british]{babel}
\pdfoutput=1
\documentclass[reqno,a4paper,12pt]{amsart}
\usepackage[a4paper,hmargin=2cm,tmargin=3cm,bmargin=3cm]{geometry}
\usepackage[utf8x]{inputenc}
\usepackage[T1]{fontenc}
\usepackage{mathptmx}
\usepackage{amsmath,amssymb,amstext,amsthm,amscd,mathrsfs,eucal,bm,slashed}
\usepackage{graphicx,color}
\usepackage{array}
\usepackage{cite}
\usepackage{hyperref}
\hypersetup{%
  pdftitle   = {},
  pdfkeywords = {rigid supersymmetry, conformal superalgebra, twistor spinors, quantum field theory},
  pdfauthor  = {Paul de Medeiros},
  pdfcreator = {\LaTeX\ with package \flqq hyperref\frqq}
}
\PrerenderUnicode{éÉ}
\newcommand{\half}{\tfrac12}
\newcommand{\cB}{\mathcal{B}}
\newcommand{\cF}{\mathcal{F}}
\newcommand{\cG}{\mathcal{G}}

\newcommand{\cL}{\mathcal{L}}

\newcommand{\cP}{\mathcal{P}}
\newcommand{\cR}{\mathcal{R}}
\newcommand{\cS}{\mathcal{S}}

\newcommand{\ff}{\mathfrak{f}}
\newcommand{\fg}{\mathfrak{g}}

\newcommand{\fS}{\mathfrak{S}}

\newcommand{\fX}{\mathfrak{X}}

\newcommand{\fh}{\mathfrak{h}}

\newcommand{\fso}{\mathfrak{so}}

\newcommand{\fusp}{\mathfrak{usp}}

\newcommand{\fsp}{\mathfrak{sp}}
\newcommand{\fsu}{\mathfrak{su}}
\newcommand{\fu}{\mathfrak{u}}

\newcommand{\Cl}{\mathrm{C}\ell}

\newcommand{\SO}{\mathrm{SO}}

\newcommand{\Spin}{\mathrm{Spin}}

\newcommand{\RR}{\mathbb{R}}
\newcommand{\CC}{\mathbb{C}}
\newcommand{\HH}{\mathbb{H}}
\newcommand{\KK}{\mathbb{K}}
\newcommand{\ZZ}{\mathbb{Z}}

\newcommand{\eA}{\mathscr{A}}

\newcommand{\eJ}{\mathscr{J}}

\newcommand{\eL}{\mathscr{L}}
\newcommand{\eM}{\mathscr{M}}
\newcommand{\eN}{\mathscr{N}}

\newcommand{\eQ}{\mathscr{Q}}
\newcommand{\eR}{\mathscr{R}}

\newcommand{\eG}{\mathscr{G}}

\newcommand{\dS}{\mathrm{dS}}
\newcommand{\AdS}{\mathrm{AdS}}
\newcommand{\Sm}{\mathbb{S}}

\DeclareMathOperator{\End}{End}

\DeclareMathOperator{\Mat}{Mat}

\DeclareMathOperator{\im}{im}
\DeclareMathOperator{\re}{Re}
\DeclareMathOperator{\tr}{tr}

\newcommand{\rf}[1]{[\![#1]\!]}

\theoremstyle{plain}
\newtheorem{lemma}{Lemma}
\newtheorem{proposition}[lemma]{Proposition}
\newtheorem{theorem}[lemma]{Theorem}
\newtheorem{corollary}[lemma]{Corollary}
\theoremstyle{definition}
\newtheorem{definition}[lemma]{Definition}
\newtheorem{remark}[lemma]{Remark}
\newtheorem{example}[lemma]{Example}
\newcommand{\MUNCH}[1]{\relax}

\allowdisplaybreaks
\def\ben{\begin{equation}}
\def\een{\end{equation}}
\def\bena{\begin{eqnarray}}
\def\eena{\end{eqnarray}}
\def\non{\nonumber}

\newcommand{\dv}{{\rm dV}}
\renewcommand{\half}{\tfrac{1}{2}}

\newcommand{\q}{{\rm q}}
\newcommand{\Q}{{\rm Q}}
\renewcommand{\dim}{{\rm dim}}
\newcommand{\Tr}{{\rm Tr}}
\newcommand{\D}{{\rm D}}
\newcommand{\A}{{\rm A}}
\newcommand{\ad}{{\rm ad}}
\renewcommand{\H}{\mathscr{H}}
\newcommand{\M}{\mathscr{M}}

\renewcommand{\d}{\mbox{d}}
\newcommand{\N}{\mathscr{N}}

\newcommand{\mr}{\mathbb{R}}
\newcommand{\mc}{\mathbb{C}}
\renewcommand{\O}{\mathscr{O}}

\newcommand{\V}{\mathcal{V}}
\newcommand{\W}{\mathcal{W}}
\newcommand{\hQ}{\hat {\rm Q}}

\newcommand{\T}{{\rm T}}
\newcommand{\e}{{\rm e}}

\renewcommand{\P}{{\bf P}}

\newcommand{\R}{{\mathscr{R}}}
\newcommand{\C}{{\mathcal{C}}}
\begin{document}
\title[Superconformal quantum field theory in curved spacetime]{Superconformal quantum field theory in curved spacetime}
\author[de Medeiros]{Paul de Medeiros}
\author[Hollands]{Stefan Hollands}
\address{School of Mathematics, Cardiff University, Senghennydd Road, Cardiff CF24 4AG, Wales, UK}
\email{paul.demedeiros@gmail.com}
\address{Universit\"{a}t Leipzig, Institut f\"{u}r Theoretische Physik, Br\"{u}derstrasse 16, D-04103 Leipzig, FRG}
\email{stefan.hollands@uni-leipzig.de}
\date{\today}

\begin{abstract}
By conformally coupling vector and hyper multiplets in Minkowski space, we obtain a class of field theories with extended rigid conformal supersymmetry on any Lorentzian four-manifold admitting twistor spinors. We construct the conformal symmetry superalgebras which describe classical symmetries of these theories and derive an appropriate BRST operator in curved spacetime. In the process, we elucidate the general framework of cohomological algebra which underpins the construction. We then consider the corresponding perturbative quantum field theories. In particular, we examine the conditions necessary for conformal supersymmetries to be preserved at the quantum level, i.e. when the BRST operator commutes with the perturbatively defined S-matrix, which ensures superconformal invariance of amplitudes. To this end, we prescribe a renormalization scheme for time-ordered products that enter the perturbative S-matrix and show that such products obey certain Ward identities in curved spacetime. These identities allow us to recast the problem in terms of the cohomology of the BRST operator. Through a careful analysis of this cohomology, and of the renormalization group in curved spacetime, we establish precise criteria which ensure that all conformal supersymmetries are preserved at the quantum level. As a by-product, we provide a rigorous proof that the beta-function for such theories is one-loop exact. We also briefly discuss the construction of chiral rings and the role of non-perturbative effects in curved spacetime.
\end{abstract}

\maketitle

\clearpage
\vspace*{1cm}
\tableofcontents

\clearpage


\section{Introduction}
\label{sec:introduction}

Supersymmetric quantum field theories are potentially important as extensions of the standard model of elementary particles. They are also apt to study general questions about quantum field theories, often being more amenable to analytic treatments. Such theories are usually considered in Minkowski space. A dynamical curved spacetime can be described within the framework of supergravity, where the metric plays the r\^{o}le of a dynamical field in the theory, rather than a background. As a halfway house between supergravity and supersymmetric field theory in flat space, one can try to construct field theories with rigid supersymmetry on non-dynamical curved backgrounds. When this is possible, the background is typically constrained such that it admits a spinor field, subject to certain differential and/or algebraic constraints, to play the r\^{o}le of a parameter in the rigid supersymmetry variations. Such theories provide an interesting laboratory to study more general properties of quantum field theory in curved spacetime within a somewhat simpler framework, thereby avoiding some of the heavy conceptual and technical complications in quantized supergravity theories, such as non-renormalizability.

The characterisation of rigid supersymmetry for field theories in curved space has attracted much interest in the recent literature \cite{Festuccia:2011ws,Jia:2011hw,Samtleben:2012gy,Klare:2012gn,Dumitrescu:2012he,Cassani:2012ri,Liu:2012bi,deMedeiros:2012sb,Dumitrescu:2012at,Kehagias:2012fh,Closset:2012ru,Martelli:2012sz,Samtleben:2012ua,Kuzenko:2012vd,Hristov:2013spa}. For theories with rigid conformal supersymmetry, the conceptual framework of conformal symmetry superalgebras was developed in \cite{deMedeiros:2013jja}. In this setup, rigid conformal supersymmetry parameters are described by twistor spinors \cite{Penrose:1967wn}, which obey a particular conformally invariant first order differential equation. There have been several previous attempts \cite{Hab:1990,Klinker:2005,Rajaniemi:2006} to define a Lie superalgebra structure for manifolds admitting twistor spinors (see also \cite{Duval:1993hs} for the construction of Schr\"{o}dinger superalgebras which do not involve twistor spinors). The construction in \cite{deMedeiros:2013jja} is distinguished by the inclusion of a non-trivial R-symmetry, which turns out to be crucial in order to solve the graded Jacobi identities for the Lie superalgebra. In a conformal symmetry superalgebra, the even part contains conformal Killing vectors and constant R-symmetries while the odd part contains twistor spinors which are valued in a certain R-symmetry representation.

Much of the recent progress has concerned only the realisation of some amount of rigid supersymmetry at the classical level. A systematic treatment of the quantum structure of rigid supersymmetry in curved space is currently lacking. Indeed, it remains to be seen whether most of the new theories with rigid supersymmetry in curved space exist as well-defined quantum field theories in their own right. It is the modest aim of this paper to address some of these important issues within the context of $\eN =2$ superconformal field theories, reformulated on a Lorentzian four-manifold which admits a twistor spinor. Such backgrounds were classified some time ago by Lewandowski \cite{Lewandowski:1991bx} (see also \cite{Baum:2002,BL:2003,Baum:2012,Leitner:2005,Baum:2008} for the classification of higher-dimensional Lorentzian manifolds which admit a twistor spinor). Up to local conformal equivalence, any such background must be either Minkowski, a pp-wave or a Fefferman space. At the classical level, construction of the theories of interest will be achieved by the conformal coupling of vector and hyper multiplets in four-dimensional Minkowski space. We shall identify what criteria ensure that rigid conformal supersymmetries for such classical theories are preserved at the quantum level. More precisely, we identify when these classical symmetries give rise to conserved quantum charge operators $\eQ_0$ on the Fock space of free particle states which commute with the scattering matrix $\Sm$, i.e.
\ben\label{Scommute}
[\eQ_0, {\Sm} ]=0~.
\een
This will be accomplished within the context of renormalized perturbation theory.

To get an idea about the nature of the problem, let us recall the story in Minkowski space. In that case, the conformal symmetry superalgebra $\cS$ generated by twistor spinors is isomorphic to the simple Lie superalgebra $\fsu(2,2|2)$, i.e. the standard $\eN=2$ conformal superalgebra on $\RR^{3,1}$. The $\fsu(2,2) \cong \fso(4,2)$ component in the even part describes the Lie algebra of conformal isometries of $\RR^{3,1}$. Dilatations are generated by a Lie subalgebra $\fso(1,1) < \fso(4,2)$. However, this classical dilatation symmetry is not realised at the quantum level unless the $\beta$-function of the theory vanishes identically. The exact form of this $\beta$-function \cite{Novikov:1983uc} for a theory with classical $\eN=2$ conformal supersymmetry and gauge symmetry described by a simple Lie algebra $\fg$ is given by
\ben\label{beta}
\beta = - \frac{\hbar}{2\pi^2} \, \left( h^\vee(\fg )- c(U) \right)~,
\een
where $h^\vee(\fg )$ denotes the dual Coxeter number of $\fg$. The complex $\fg$-module $U$ provides the data for hyper multiplet couplings in the theory and $c(U)$ denotes its Dynkin index. As is often claimed but seldom argued, the vanishing of $\beta$ is in fact necessary and sufficient to ensure that the full $\eN=2$ conformal superalgebra is realised at the quantum level. The classification of pairs $(\fg,U)$ with $h^\vee(\fg ) = c(U)$ is provided in our Appendix~\ref{app:A}, completing a previous analysis in \cite{Koh:1983ir}. If $\beta$ does not vanish identically then only the symmetries generated by the $\eN=2$ Poincar\'{e} superalgebra
\footnote{We refer here to the $\eN$-extended Poincar\'{e} superalgebra for $\RR^{3,1}$, without central charges, canonically embedded in the $\fsu(2,2|\eN)$ conformal superalgebra. The even part of this $\eN$-extended Poincar\'{e} superalgebra is the direct sum of the Poincar\'{e} algebra $\fso(3,1) \ltimes \RR^{3,1}$ of $\RR^{3,1}$ isometries and $\fu(\eN)$ R-symmetries.}
can survive.

Based on this example, one might be tempted to infer that the conformal symmetry superalgebra $\cS$ for a given Lorentzian four-manifold can be realised at the quantum level provided all the conformal Killing vectors in its even part are Killing vectors (a feature which may or may not be the case, depending on the background in question). That is, if $\cS$ contains no \lq proper' conformal isometries that are akin to dilatations or special conformal transformations in Minkowski space. However, this is not the case.  For example, as will be seen in Section~\ref{sec:spacetimes}, on a Fefferman space, although the only conformal isometry is generated by a null Killing vector, $\cS$ is still not realised at the quantum level. The reason being that, in curved spacetime, there is typically no relation between the renormalization group (i.e. the $\beta$-function) and geometrical symmetries of the spacetime. Indeed, one can define a constant rescaling of the metric (a homothety) which need not be generated by a conformal Killing vector.

The correct criteria for the preservation of $\cS$ at the quantum level can be summarised as follows:
\begin{itemize}
\item If $\beta=0$, then $\cS$ is always preserved.
\item If $\beta \neq 0$, then $\cS$ is preserved only if all twistor spinors in the odd part are parallel and all conformal Killing vectors in the even part are Killing vectors.
\end{itemize}
The $\beta \neq 0$ case can occur for pp-waves but not for Fefferman spaces. In general, if $\beta \neq 0$, then only a Lie superalgebra $\cS^\prime < \cS$ of symmetries is preserved at the quantum level. By definition, this symmetry superalgebra $\cS^\prime$ contains only those parallel spinors in the odd part of $\cS$ and only those Killing vectors and R-symmetries in even part of $\cS$.

For example, consider the above criteria for a classical $\eN =2$ superconformal field theory in four-dimensional de Sitter space $\dS_4$. Since $\dS_4$ is locally conformally flat, the classical conformal symmetry superalgebra $\cS \cong \fsu(2,2|2)$. The Killing vectors of $\dS_4$ span a Lie subalgebra $\fso(4,1) < \fsu(2,2)$. However, none of the twistor spinors on $\dS_4$ are parallel. Whence, the theory is supersymmetric at the quantum level only if $\beta=0$. This differs markedly from the case of four-dimensional Minkowski space where precisely half of the twistor spinors are parallel (i.e. constant) and $\cS^\prime$ is isomorphic to the $\eN =2$ Poincar\'{e} superalgebra.

We will establish the above criteria by a careful analysis of the renormalization process in curved spacetime, and its compatibility with supersymmetry. Our analysis follows the general procedure developed in~\cite{hw1,hw2,hw3,h1,bf} (see also~\cite{reijzner}). As was previously observed for field theories with $\eN \geq 2$ supersymmetry in four-dimensional Minkowski space~\cite{Maggiore:1994dw,Baulieu:2007dk,White:1992wu,Baulieu:2006tu}, it is helpful to recast the symmetries in the form of \lq BRST-type' transformations. We shall develop and utilise this framework in curved spacetime. To conceptualise the formalism, we work out the general mathematical framework of cohomological algebra that underpins the construction of an appropriate \lq BRST-coboundary' operator, that we will call $\Q$. In our framework, the potential violation of a classical symmetry at the quantum level corresponds to a non-trivial $\Q$-cohomology class, called an anomaly. We carefully derive the consistency conditions (generalising the well-known \lq Wess-Zumino condition') which lead to this conclusion. The general form of the anomaly is presented, whose numerical prefactor is related to the $\beta$-function~\eqref{beta} in Minkowski space. This is what leads to the criteria described above. Our analysis also yields a rigorous mathematical proof of the well-known fact that the $\beta$-function of a field theory with $\eN =2$ supersymmetry in Minkowski space is one-loop exact.

It is worth noting that, since general covariance is manifest in our method of renormalization, if $\beta=0$, it also follows that there is no running of the coupling parameters in the Lagrangian on an {\emph{arbitrary}} curved spacetime with spin structure, which need have no symmetries at all. Due to the subtleties of renormalization in curved spacetime, we have decided to present our arguments in a mathematically rigorous framework~\cite{hw1,hw2,hw3,h1}, which we shall review in some detail. Let us begin though with a short review of the conformal symmetry superalgebras constructed in \cite{deMedeiros:2013jja}. 


\section{Conformal symmetry superalgebras}


\subsection{Conformal Killing vectors and twistor spinors}
\label{sec:twistorspinors}

Let $\eM$ be a $d$-dimensional manifold equipped with a pseudo-Riemannian metric $g$. In this paper, we will always take $d\, {>}\, 2$. The Levi-Civit\`{a} connection of $g$ will be denoted by $\nabla$. It will be assumed that $\eM$ has vanishing second Stiefel-Whitney class so the bundle $\SO(\eM)$ of oriented pseudo-orthonormal frames lifts to $\Spin (\eM)$ by the assignment of a spin structure.

Let $\fX ( \eM )$ denote the space of vector fields on $\eM$. The Lie derivative $\cL_X$ along any $X \in \fX(\eM)$ defines an endomorphism of the space of tensor fields on $\eM$. The Lie bracket of vector fields is defined by $[X,Y] =\cL_X Y = \nabla_X Y - \nabla_Y X \in \fX(\eM)$, for all $X,Y \in \fX(\eM)$. This equips $\fX(\eM)$ with the structure of a Lie algebra. Furthermore
\begin{equation}\label{eq:LieDerHom}
[ \cL_X , \cL_Y ] = \cL_{[X,Y]}~,
\end{equation}
for all $X,Y \in \fX ( \eM )$. Whence, the Lie derivative defines on the space of tensor fields a representation of the Lie algebra of vector fields.

The subspace of {\bf conformal Killing vectors} in $\fX ( \eM )$ is defined by
\begin{equation}\label{eq:CKV}
\fX^c ( \eM ) = \{ X \in \fX ( \eM ) \; |\; \cL_X g = -2 \sigma_X  g\}~,
\end{equation}
for some function $\sigma_X$ on $\eM$. Relative to a coordinate basis $\{ \partial_\mu \}$ on $\fX ( \eM )$, $\sigma_X = -\tfrac{1}{d} \nabla_\mu X^\mu$, for all $X \in \fX ( \eM )$. If $X,Y \in \fX^c ( \eM )$ then $[X,Y] \in \fX^c ( \eM )$. Whence, the restriction of the Lie bracket to $\fX^c ( \eM )$ defines a Lie subalgebra of conformal Killing vectors on $\eM$. Furthermore, $\fX^c ( \eM )$ depends only on the conformal class of $g$. If $\eM$ is conformally isometric to $\RR^{s,t}$ then $\fX^c ( \RR^{s,t} ) \cong \fso(s+1,t+1)$. It is useful to note that any $X \in \fX^c ( \eM )$ obeys
\ben\label{CKVint}
\nabla_\mu \nabla_\nu X_\rho = R_{\rho\nu\mu\sigma} X^\sigma -
g_{\rho\mu} \nabla_\nu \sigma_X -  g_{\rho\nu} \nabla_{\mu} \sigma_X
+ g_{\mu\nu} \nabla_{\rho} \sigma_X~,
\een
where $R_{\mu\nu\rho\sigma}$ is the Riemann tensor of $g$. Indices are lowered and raised using $g$ and its inverse. Any $X \in \fX^c ( \eM )$ with $\sigma_X$ a non-zero constant is called {\bf homothetic}. Any $X \in \fX^c ( \eM )$ with $\sigma_X =0$ is called a {\bf Killing vector} and restricting the Lie bracket on $\fX^c ( \eM )$ to the subspace of Killing vectors on $\eM$ defines a Lie subalgebra.

Let $\Cl (T\eM)$ denote the {\bf Clifford bundle} over $\eM$. To each basis element $\partial_\mu \in \fX(\eM)$, there is an associated section $\Gamma_\mu$ of $\Cl (T\eM)$. These sections obey the product rule
\begin{equation}\label{eq:CliffordAlgebra}
\Gamma_\mu \Gamma_\nu + \Gamma_\nu \Gamma_\mu = 2 g_{\mu\nu} {\bf 1}~,
\end{equation}
where ${\bf 1}$ denotes the identity element. A convenient basis for sections of $\Cl (T\eM)$ is defined in terms of the skewsymmetric products:
\begin{equation}\label{eq:ClBasis}
\Gamma_{\mu_1 ... \mu_k} = \Gamma_{[ \mu_1} ... \Gamma_{\mu_k ]} \equiv \frac{1}{k!} \sum_{\sigma \in S_k} (-1)^{|\sigma|} \Gamma_{\mu_{\sigma(1)}} ... \Gamma_{\mu_{\sigma(k)}}~,
\end{equation}
for all $k=1,...,d$, together with ${\bf 1}$ for $k=0$. To any rank $k$ polyvector field $P$ on $\eM$, we associate a section ${\slashed P} = \tfrac{1}{k!} P^{\mu_1 ... \mu_k} \Gamma_{\mu_1 ... \mu_k}$ of $\Cl (T\eM)$. The same applies to any $k$-form on $\eM$, after identifying it with the dual $k$-polyvector field with respect to the metric $g$. The element $\Gamma_{\mu_1 ... \mu_d}$ of maximal degree $d$ is proportional to an idempotent element $\Gamma$ times the positively oriented volume form $\epsilon_{\mu_1 ... \mu_d}$ on $\eM$. For $d$ odd, $\Gamma$ is central. For $d$ even, $\Gamma \Gamma_\mu = -\Gamma_\mu \Gamma$.

The principle bundle $\Spin (\eM)$ has an associated vector bundle $\$ (\eM)$, called the {\bf spinor bundle} over $\eM$, that is defined by the spinor module for sections of $\Cl (T\eM)$. Let $\fS(\eM)$ denote the space of spinor fields on $\eM$ (i.e. sections of $\$ (\eM)$). For $d$ even, $\fS(\eM) = \fS_+ (\eM) \oplus  \fS_- (\eM)$, where $\fS_\pm (\eM)$ denote the subspaces of {\bf chiral} spinor fields, on which $\Gamma =\pm 1$. One embeds $\fS_\pm (\eM) \subset \fS (\eM)$ via projection operators ${\sf P}_\pm = \half ( {\bf 1} \pm \Gamma )$.

The action of $\nabla$ induced on $\fS ( \eM )$ is compatible with the Clifford action, i.e.
\begin{equation}\label{eq:LCCliffComp}
\nabla_X ( {\slashed Y} \psi ) = ( \nabla_X {\slashed Y} )  \psi + {\slashed Y} \nabla_X \psi~,
\end{equation}
for all $X,Y \in \fX ( \eM )$ and $\psi \in \fS ( \eM )$. Furthermore
\begin{equation}\label{eq:LCBracket}
[ \nabla_\mu , \nabla_\nu ] \psi = \tfrac{1}{4} R_{\mu\nu\rho\sigma} \Gamma^{\rho\sigma} \psi~,
\end{equation}
for all $\psi \in \fS ( \eM )$. The convention for the Ricci tensor is $R_{\mu\nu} = R_{\mu\rho\nu\sigma} g^{\rho\sigma}$, and the scalar curvature is $R = R_{\mu\nu} g^{\mu\nu}$.
The decomposition
\ben\label{schouten}
\begin{split}
& R_{\mu\nu\rho\sigma} = C_{\mu\nu\rho\sigma} - g_{\mu\rho} K_{\nu\sigma} + g_{\nu\rho} K_{\mu\sigma} + g_{\mu\sigma} K_{\nu\rho} - g_{\nu\sigma} K_{\mu\rho} \\
& K_{\mu\nu} = \tfrac{1}{d-2} \left( - R_{\mu\nu} + \tfrac{1}{2(d-1)} \, g_{\mu\nu} \, R \right)~,
\end{split}
\een
of the Riemann tensor defines the Weyl tensor $C_{\mu\nu\rho\sigma}$ and the Schouten tensor $K_{\mu\nu}$.

There always exists on $\fS(\eM)$ a nondegenerate bilinear form $\langle -,- \rangle$ with the properties
\begin{align}
\label{eq:SpinorInnerProd}
\langle \psi , \chi \rangle &= \sigma \langle \chi , \psi \rangle \nonumber \\
\langle {\slashed X} \psi , \chi \rangle &= \tau \langle \psi ,  {\slashed X} \chi \rangle \\
X \langle \psi , \chi \rangle &= \langle \nabla_X \psi , \chi \rangle + \langle \psi , \nabla_X \chi \rangle~, \nonumber
\end{align}
for all $\psi , \chi \in \fS (\eM)$ and $X \in \fX(\eM)$, with respect to a pair of fixed signs $\sigma$ and $\tau$ (see \cite{deMedeiros:2013jja,AlekCort1995math,Alekseevsky:2003vw} for more details). Let $\kappa = \sigma \tau$ denote the sign for which $\langle \psi ,  \Gamma_\mu \chi \rangle = \kappa \langle \chi ,  \Gamma_\mu \psi \rangle$, for all $\psi , \chi \in \fS (\eM)$. The possible choices for $\sigma$ and $\tau$ depend critically on both $d$ and the signature of $g$. The sign $\sigma = \pm 1$ indicates whether $\langle -,- \rangle$ is symmetric or skewsymmetric. The third line in \eqref{eq:SpinorInnerProd} says that $\langle -,- \rangle$ is spin-invariant. For $d$ even, this implies $\langle \Gamma \psi , \chi \rangle = (-1)^{d/2} \langle \psi , \Gamma \chi \rangle$. Whence,
\begin{align}\label{eq:ChiralSpinorProd}
\langle \psi_\pm , \chi_\mp \rangle &= 0 \quad {\mbox{if $d=0$ mod $4$}} \nonumber \\
\langle \psi_\pm , \chi_\pm \rangle &= 0 \quad {\mbox{if $d=2$ mod $4$}}~,
\end{align}
for all $\psi_\pm , \chi_\pm \in \fS_\pm ( \eM )$.

Now let the dual ${\overline \psi}$ of any $\psi \in \fS (\eM)$ with respect to $\langle -,- \rangle$ be defined such that ${\overline \psi} \chi = \langle \psi , \chi \rangle$, for all $\chi \in \fS (\eM)$. From any $\psi , \chi \in \fS (\eM)$, one can define $\psi {\overline \chi}$ as an endomorphism of $\fS (\eM)$. Whence, it can be expressed relative to the basis defined in \eqref{eq:ClBasis}, with coefficients proportional to $k$-polyvectors of the form ${\overline \chi} \Gamma^{\mu_1 ... \mu_k} \psi$. Such expressions are known as {\bf Fierz identities}, full details of which can be found in section 4 of~\cite{deMedeiros:2013jja}. For example, in $d=4$, it is useful to note that
\begin{align}\label{Fierzgeneral}
\psi_\pm {\overline \chi}_\pm &= \half \! \left(  ( {\overline \chi}_\pm \psi_\pm ) {\bf 1} - \tfrac{1}{4} ( {\overline \chi}_\pm \Gamma^{\mu\nu} \psi_\pm ) \Gamma_{\mu\nu} \right) \! {\sf P}_\pm \nonumber \\
\psi_\pm {\overline \chi}_\mp &= \half  ( {\overline \chi}_\mp \Gamma^{\mu} \psi_\pm ) \Gamma_{\mu} {\sf P}_\mp~,
\end{align}
for all $\psi_\pm , \chi_\pm \in \fS_\pm ( \M )$.

The {\bf spinorial Lie derivative} \cite{Lic:1963,Kosmann:1972,BG:1992,Habermann:1996} along any $X \in \fX^c ( \eM )$ is defined by
\begin{equation}\label{eq:SpinLieDer}
\cL_X \psi = \nabla_X \psi + \tfrac{1}{4} ( \nabla_\mu X_\nu ) \Gamma^{\mu\nu} \psi~,
\end{equation}
for all $\psi \in \fS ( \eM )$. It obeys
\begin{align}\label{eq:LieDerBracketSpinors}
[ \cL_X , \nabla_Y ] \psi &= \nabla_{[X,Y]} \psi + \tfrac{1}{2} ( \nabla_\mu \sigma_X ) Y_\nu \, \Gamma^{\mu\nu} \psi \nonumber \\
[ \cL_X , {\slashed \nabla} ] \psi & = \sigma_X {\slashed \nabla} \psi - \left( \tfrac{d-1}{2} \right)  {\slashed{\nabla}} \sigma_X \psi~,
\end{align}
for all $X \in \fX^c (\M)$, $Y \in \fX ( \M )$ and $\psi \in \fS ( \M )$.

For any $X,Y \in \fX^c ( \M )$ and $w \in \RR$, it follows that
\begin{equation}\label{eq:LieDerSpinHom}
[ \cL_X + w\, \sigma_X {\bf 1}, \cL_Y + w\, \sigma_Y {\bf 1}] = \cL_{[X,Y]} + w\, \sigma_{[X,Y]} {\bf 1},
\end{equation}
using the identity $\nabla_X \sigma_Y - \nabla_Y \sigma_X = \sigma_{[X,Y]}$. Whence, $\cL_X + w\, \sigma_X {\bf 1}$ defines on $\fS ( \M )$ a representation of the Lie algebra of conformal Killing vector fields.

The subspace of {\bf twistor spinors} in $\fS(\M)$ is defined by
\begin{equation}
\label{twistorspinor}
\fS^c(\M) = \{ \psi \in \fS(\M) \; |\; \nabla_\mu \psi = \tfrac{1}{d}  \Gamma_\mu {\slashed \nabla} \psi  \}~.
\end{equation}
An important property of twistor spinors is that they \lq square to' conformal Killing vectors. More precisely, from any $\psi , \chi \in \fS^c(\eM)$, one can construct $\xi_{\psi , \chi} \in \fX^c (\eM)$ with components $\xi_{\psi , \chi}^\mu = {\overline \psi} \,\Gamma^\mu \chi$. The defining equation for twistor spinors was first introduced in $d=4$ by Penrose in~\cite{Penrose:1967wn} (see also \cite{Penrose:1986ca,vanNieuwenhuizen:1983wu}). The {\bf Penrose operator}
\begin{equation}
\label{PenroseOperator}
\cP_\mu = \nabla_\mu - \tfrac{1}{d}  \Gamma_\mu {\slashed \nabla}~,
\end{equation}
is an endomorphism of $\fS(\eM)$ whose kernel is precisely $\fS^c(\eM)$. The Penrose operator obeys ${\slashed \cP} =0$ identically. Moreover, for a given spin structure on $\eM$, its kernel $\fS^c(\M)$ depends only on the conformal class of $g$. If $\eM = \RR^{s,t}$ then, at a point $x \in \RR^{s,t}$, any $\psi \in \fS^c ( \RR^{s,t} )$ can be written $\psi = \psi_0 + {\slashed x} \psi_1$, in terms of some constant $\psi_0 , \psi_1 \in \fS ( \RR^{s,t} )$. Any $\psi \in \fS^c ( \eM )$ with $\tfrac{1}{d} {\slashed \nabla} \psi = \lambda \psi$, for some constant $\lambda$, is called a {\bf Killing spinor} if $\lambda \neq 0$ or a {\bf parallel spinor} if $\lambda =0$. The  constant $\lambda \neq 0$ for a Killing spinor is called its {\bf Killing constant}.

Taking $w =\half$ in \eqref{eq:LieDerSpinHom} defines the {\bf Kosmann-Schwarzbach Lie derivative}
\begin{equation}\label{eq:KSLieDer}
{\hat \cL}_X = \cL_X + \half  \sigma_X {\bf 1}~,
\end{equation}
along any $X \in \fX^c (\eM)$. It is noteworthy that only for $w =\half$ does $\cL_X + w\, \sigma_X {\bf 1}$ define an endomorphism of the space of twistor spinors $\fS^c(\eM)$. Indeed, compatibility of the conformal and spin structure on $\eM$ fixes $w=\half$ as the Weyl weight of a spinor field (with $g_{\mu\nu}$ assigned Weyl weight $2$).


\subsection{Lie superalgebras for manifolds with a twistor spinor}
\label{sec:LSA}

Let $\cS = \cB \oplus \cF$ denote the (real) graded vector space on which we shall define a {\bf Lie superalgebra} structure. The even part $\cB = \fX^c ( \M ) \oplus \cR$, where $\cR$ is a real Lie algebra whose elements are constant on $\eM$. The complexification $\cF_\CC = \cF \otimes_\RR \CC$ of the odd part $\cF$ is
\ben
\label{eq:OddComplexification}
\cF_\CC =
\begin{cases}
\fS^c (\M) \otimes_\CC W & \text{if $d$ is odd} \\
\fS^c_+ (\M) \otimes_\CC V \oplus \fS^c_- (\M) \otimes_\CC V^* &\text{if $d=0$ mod $4$} \\
\fS^c_+ (\M) \otimes_\CC W &\text{if $d=2$ mod $4$}~,
\end{cases}
\een
where $V$ and $W$ are certain complex $\cR_\CC$-modules. $V^*$ is the dual module of $V$. $W$ admits a (skew)symmetric $\cR_\CC$-invariant nondegenerate bilinear form $b$, which provides an isomorphism $W^* \cong W$.

The graded Lie bracket on $\cS$ is a bilinear map $[-,-] : \cS \times \cS \rightarrow \cS$, defined such that
\begin{equation}\label{eq:SuperBracket}
[\cB,\cB] \subset \cB \;\; , \quad\quad [\cB,\cF] \subset \cF \;\; , \quad\quad [\cF,\cF] \subset \cB~.
\end{equation}
The $[\cB,\cB]$ and $[\cB,\cF]$ brackets are skewsymmetric and the $[\cF,\cF]$ bracket is symmetric. For all $X , X^\prime \in \fX^c (\eM)$, $\rho , \rho^\prime \in \cR$ and $\epsilon , \epsilon^\prime \in \cF$, these brackets are defined by
\begin{align}
\label{eq:SuperBracket2}
[ X + \rho , X^\prime + \rho^\prime ] &= [ X , X^\prime ] + [ \rho , \rho^\prime ] \nonumber \\
[ X + \rho , \epsilon ] &= {\hat \cL}_X \epsilon + \rho \cdot \epsilon \\
[ \epsilon , \epsilon^\prime ] &= \xi_{\epsilon , \epsilon^\prime} + \rho_{\epsilon , \epsilon^\prime}~. \nonumber
\end{align}
On the right hand side of the $[\cB,\cB]$ bracket, $[-,-]$ is used to denote both the Lie bracket of conformal Killing vector fields on $\fX^c (\eM)$ and the Lie bracket for $\cR$. On the right hand side of the $[\cB,\cF]$ bracket, $\cdot$ denotes the $\cR$-action of $\rho$ on $\epsilon$.  On the right hand side of the $[\cF,\cF]$ bracket, $\xi_{\epsilon , \epsilon^\prime} = \half ( \xi_{\epsilon + \epsilon^\prime} - \xi_{\epsilon} - \xi_{\epsilon^\prime})$ and $\rho_{\epsilon , \epsilon^\prime} = \half ( \rho_{\epsilon + \epsilon^\prime} - \rho_{\epsilon} - \rho_{\epsilon^\prime})$ denote polarisations of particular elements $\xi_\epsilon \in \fX^c (\M)$ and $\rho_\epsilon \in \cR$, defined for any $\epsilon \in \cF$, which we will now specify.

The element $\xi_\epsilon$ is defined as the real part of $\Xi_\epsilon \in \fX^c_\CC (\M)$, where
\begin{equation}\label{eq:XiComplex}
\Xi_\epsilon^\mu  = \begin{cases}
 b_{AB} \, {\overline \epsilon}^A \Gamma^\mu \epsilon^B  & {\text{if $d$ is odd}} \\
2 \, {\overline \epsilon}_+^A \Gamma^\mu \epsilon_{-\, A} & {\text{if $d=0$ mod $4$}} \\
 b_{AB} \, {\overline \epsilon}_+^A \Gamma^\mu \epsilon_+^B  & {\text{if $d=2$ mod $4$}}~,
\end{cases}
\end{equation}
for any $\epsilon \in \cF$, relative to a basis $\{ {\bm e}_A \}$ for either $V$ or $W$.

The element $\rho_\epsilon$ is defined as the real part of $\Pi_\epsilon \in \cR_\CC$, where
\begin{equation}\label{eq:PiComplex}
( \Pi_\epsilon \cdot \psi )^A = \begin{cases}
 {\tt a}  ( {\overline \epsilon}^A {\slashed \nabla} \epsilon_B - \kappa \, {\overline \epsilon}_B{\slashed \nabla} \epsilon^A ) \psi^B & {\text{if $d$ is odd}} \\
{\tt a} ( {\overline \epsilon}_+^A {\slashed \nabla} \epsilon_{-\, B} - \kappa\, {\overline \epsilon}_{-\, B} {\slashed \nabla} \epsilon_+^A ) \psi^B + {\tt b} ( {\overline \epsilon}_+^B {\slashed \nabla} \epsilon_{-\, B} - \kappa \, {\overline \epsilon}_{-\, B} {\slashed \nabla} \epsilon_+^B ) \psi^A & {\text{if $d=0$ mod $4$}} \\
{\tt a} ( {\overline \epsilon}_+^A {\slashed \nabla} \epsilon_{+\, B} - \kappa \, {\overline \epsilon}_{+\, B} {\slashed \nabla} \epsilon_+^A ) \psi^B & {\text{if $d=2$ mod $4$}}~,
\end{cases}
\end{equation}
and $\kappa$ is the sign for which ${\overline \psi} \Gamma_\mu \chi = \kappa \, {\overline \chi} \Gamma_\mu \psi$, for any $\psi , \chi \in \cF$. Permitted values of the constants ${\tt a}$ and ${\tt b}$ will be specified in a moment.

The brackets in \eqref{eq:SuperBracket2} furnish $\cS$ with the structure of a real Lie superalgebra only if they obey the graded Jacobi identity. There are four distinct graded components, of type $[\cB\cB\cB]$, $[\cB\cB\cF]$, $[\cB\cF\cF]$ and $[\cF\cF\cF]$, each of which must vanish identically. As shown in~\cite{deMedeiros:2013jja}, this is indeed automatically the case, but for the $[\cF\cF\cF]$ component,
whose vanishing is equivalent to the condition
\begin{equation}\label{eq:oddoddoddjacobi}
{\hat \cL}_{\xi_\epsilon} \epsilon + \rho_\epsilon \cdot \epsilon = 0~,
\end{equation}
for all $\epsilon \in \cF$. As was shown in~\cite{deMedeiros:2013jja}, generic solutions of \eqref{eq:oddoddoddjacobi} fix the values of ${\tt a}$ and ${\tt b}$ and restrict the dimension of $\eM$ to $d < 7$.

Data for the solutions with $\cR$ compact is summarised in Table~\ref{tab:tableCKS}.
\footnote{Solutions with $\cR$ non-compact were also obtained in~\cite{deMedeiros:2013jja} though, at least if $( \eM , g)$ is Lorentzian, they cannot describe symmetries of a unitary quantum field theory.}
Entries in the \lq type' column denote the ground field $\KK$ over which the representation of $\cR$ is defined. The dimension over $\KK$ of this representation is denoted by $\eN$. If $( \eM , g)$ is conformally flat, each conformal symmetry superalgebra $\cS \cong \cS_\circ$ describes one of the conformal superalgebras classified by Nahm in~\cite{Nahm:1977tg} (and is identified by its \lq Nahm label' in the right-most column of Table~\ref{tab:tableCKS}). Of course, for a geometry $( \eM , g)$ that not conformally flat, the associated conformal symmetry superalgebra $\cS$ is necessarily smaller than $\cS_\circ$. It is interesting to note that conformal symmetry superalgebras with compact $\cR$ exist only when $( \eM , g)$ is Lorentzian or Riemannian.

\begin{table}
\begin{center}
\begin{tabular}{|c|c|c|c|c|c||c|c|}
  \hline
  $(\eM , g)$ & $d$ & ${\tt a}$ & ${\tt b}$ & type & $\cR$ & $\cS_\circ$ & Nahm label \\
  \hline \hline
   &  &  &  &  & & \\ [-.5cm]
  Lorentzian & $6$ & $\tfrac{2}{3}$ & $*$ & $\HH$ & $\fsp (\eN)$ &  $\frak{osp}(6,2| \eN)$ & $\tt{X}$  \\ [.1cm]
  Lorentzian & $5$ & $\tfrac{3}{5}$ & $*$ & $\HH$ & $\fsp(1)$ &  ${\frak f}(4)''$ & $\tt{IX_2}$  \\ [.1cm]
  Riemannian & $5$ & $\tfrac{3}{5}$ & $*$ & $\HH$ & $\fsp(1)$ &  ${\frak f}(4)'$& $\tt{IX_1}$ \\ [.1cm]
  Lorentzian & $4$ & $1$ & $-\tfrac{1}{4}$ & $\CC$ & $\fu (\eN \neq 4)$ & $\fsu(2,2| \eN)$ & $\tt{VIII}$ \\ [.1cm]
  Lorentzian & $4$ & $1$ & $-\tfrac{1}{4}$ & $\CC$ & $\fsu(4)$ &  $\mathfrak{psu}(2,2|4)$ & $\tt{VIII_1}$ \\ [.1cm]
  Lorentzian & $3$ & $\tfrac{2}{3}$ & $*$ & $\RR$ & $\fso(\eN \neq 1)$ & $\mathfrak{osp} ( \eN | 2 )$ & $\tt{VII}$ \\ [.1cm]
  Riemannian & $3$ & $\tfrac{2}{3}$ & $*$ &  $\HH$ & $\fu(1)$ & $\mathfrak{ osp}(2|1,1)$ & $\tt{VII_1}$ \\ [.1cm]
  \hline
\end{tabular} \vspace*{.2cm}
\caption{Data for conformal symmetry superalgebras with compact R-symmetry.}
\label{tab:tableCKS}
\end{center}
\end{table}


\subsubsection{The $d=4$, $\eN =2$ case}
\label{sec:N2d4}

Since we shall be primarily interested in theories with rigid $\eN =2$ superconformal symmetry on Lorentzian four-manifolds with a twistor spinor, let us elucidate a few more details concerning the associated conformal symmetry superalgebras in this case.

At each point $x \in \eM$, the Clifford algebra $\Cl (T_x \eM) \cong \Mat_4 ( \RR )$ has a unique irreducible {\bf Majorana spinor} representation that is isomorphic to $\RR^4$. On the other hand, its complexification (the {\bf Dirac spinor} representation) decomposes into a pair of inequivalent irreducible chiral spinor representations, each isomorphic to $\CC^2$, associated with the two eigenspaces of $\Gamma = i\, \Gamma_{0123}$ on which $\Gamma = \pm 1$. The action of a subalgebra $\Mat_2 (\CC) < \Mat_4 ( \RR )$ on $\CC^2$ which commutes with the complex structure $\Gamma_{0123}$ defines the action of $\Cl (T_x \eM)$ on each chiral projection (the two chiral projections transform in complex conjugate representations).

It is convenient to represent the complexified Clifford algebra such that each basis element $\Gamma_\mu$ is a unitary matrix. It then follows that there exist a pair of unitary matrices ${\sf C}$ and ${\sf B}$ for which
\begin{equation}\label{eq:BCconj}
\Gamma_\mu^{\, t} = - {\sf C} \, \Gamma_\mu {\sf C}^{-1} \; , \quad\quad \Gamma_\mu^* =  {\sf B}\, \Gamma_\mu {\sf B}^{-1}~,
\end{equation}
where $t$ denotes transposition and $*$ denotes complex conjugation. In addition,
\begin{equation}\label{eq:BCsym}
{\sf C}^t = - {\sf C} \; , \quad\quad {\sf B}^* {\sf B} = {\bf 1} \; , \quad\quad  {\sf B}^t {\sf C}^* {\sf B} = {\sf C}~.
\end{equation}
The matrix ${\sf C}$ will be used to represent the bilinear form $\langle -,- \rangle$ in \eqref{eq:SpinorInnerProd}, with $\sigma = \tau = -1$ and $\kappa =1$. Whence, the dual, or {\bf Majorana conjugate}, of a Dirac spinor $\psi$ is ${\overline \psi} = \psi^{\, t} {\sf C}$. The matrix ${\sf B}$ defines a real structure on the Dirac spinor representation and a Dirac spinor $\psi$ is Majorana only if $\psi^* = {\sf B} \psi$.

Table~\ref{tab:tableCKS} shows that $\cR \cong \fu(2)$ with $V \cong \CC^2$. Let $\{ {\bm e}_A \}$ denote a basis on $\CC^2$. An element $\epsilon \in \cF$ corresponds to a pair of bosonic Majorana twistor spinors, written $( \epsilon_+^A , \epsilon_{-\, A} )$ in terms of their complex chiral projections. The Majorana reality condition relates chiral projections via complex conjugation such that $\epsilon_{-\, A} = ( {\sf B} \, \epsilon_+^A )^*$. The parameters \eqref{eq:XiComplex} and \eqref{eq:PiComplex} for the $[\cF,\cF]$ bracket are defined by
\ben\label{xirhoNequals2}
 \begin{split}
\xi_\epsilon^\mu &= 2 \, {\overline \epsilon}_+^A \Gamma^\mu \epsilon_{-\, A} \\
( \rho_\epsilon )^A{}_B &= ( {\overline \epsilon}_+^A {\slashed \nabla} \epsilon_{-\, B} - {\overline \epsilon}_{-\, B} {\slashed \nabla} \epsilon_+^A )  -\tfrac{1}{4} ( {\overline \epsilon}_+^C {\slashed \nabla} \epsilon_{-\, C} - {\overline \epsilon}_{-\, C} {\slashed \nabla} \epsilon_+^C ) \delta^A_B~.
 \end{split}
\een
The Majorana condition for $\epsilon \in \cF$ implies that the vector $\xi_\epsilon$ is real and that the 2$\times$2 complex matrix $( \rho_\epsilon )^A{}_B$ is skewhermitian. Let ${\hat \rho}_\epsilon \in \fsu(2) < \fu(2)$ denote the trace-free part of $\rho_\epsilon$ and define $\rho_\epsilon^\prime = \tfrac{1}{4} ( {\overline \epsilon}_+^A {\slashed \nabla} \epsilon_{-\, A} - {\overline \epsilon}_{-\, A} {\slashed \nabla} \epsilon_+^A ) \in \fu(1)$, which acts on $( \epsilon_+^A , \epsilon_{-\, A} )$ with charges $(+1 , -1)$.


\subsection{Lorentzian manifolds with a twistor spinor}
\label{sec:spacetimes}

The existence of one or more twistor spinors puts strong restrictions on the geometry of $(\eM,g)$. For instance, it must admit certain conformal Killing vectors obtained by squaring twistor spinors in the manner described in section~\ref{sec:twistorspinors}. Moreover, by taking higher derivatives of the twistor
spinor equation, one obtains certain integrability conditions which must also be satisfied. In particular, for any $\epsilon \in \fS^c(\eM)$, it follows using \eqref{eq:LCBracket} that
\begin{align}
\label{consistency}
C_{\mu\nu\rho\sigma} \Gamma^{\rho\sigma} \epsilon &= 0 \nonumber \\
\nabla_\mu {\slashed \nabla} \epsilon &= \tfrac{d}{2} K_{\mu\nu} \Gamma^\nu \epsilon \\
{\slashed \nabla}^2 \epsilon &= -\tfrac{d}{4(d-1)} \, R\epsilon~, \nonumber
\end{align}
involving the Weyl and Schouten tensors defined below \eqref{schouten}. Further conditions can be obtained by taking two derivatives of the twistor spinor equation and using the Bianchi identities.

There are a number of classification results concerning the existence of twistor spinors (with and without zeros) on $(\eM,g)$ in different dimensions and signatures. The number of linearly independent twistor spinors is bounded above by $2^{\lfloor \tfrac{d}{2} \rfloor +1}$ and this bound is saturated only if $(\eM,g)$ is locally conformally flat (i.e. $C_{\mu\nu\rho\sigma} =0$). Let us now focus on the most physically interesting case; where $(\eM,g)$ is a spacetime in Lorentzian signature.

The classification in $d\geq 3$ of all local conformal equivalence classes of Lorentzian spin manifolds admitting generic twistor spinors without zeros was established by Baum and Leitner~\cite{Baum:2002,BL:2003,Baum:2012,Leitner:2005,Baum:2008}. Their results generalise the classification in $d=4$ obtained earlier by Lewandowski in~\cite{Lewandowski:1991bx}. Since it is precisely this $d=4$ case that will be relevant in our forthcoming analysis, let us now recall the classification in more detail.

Let $\epsilon \in \fS^c (\eM)$, with chiral projections $\epsilon_\pm \in \fS_\pm^c (\eM)$. Without loss of generality, we assume that $\epsilon$ is Majorana, whence $\epsilon_- = ( {\sf B} \epsilon_+ )^*$. Squaring $\epsilon$ defines a conformal Killing vector $\xi^\mu = {\overline \epsilon} \Gamma^\mu \epsilon = 2\, {\overline \epsilon_-} \Gamma^\mu \epsilon_+$. Since $\epsilon$ is Majorana, it implies that $\xi$ is real and null. From the first equation in~\eqref{consistency}, it follows that $\epsilon_+$ is a four times repeated principal spinor, i.e. the Weyl tensor is of algebraic type N or O in the Petrov classification~\cite{Wald} (the type O case is when $C_{\mu\nu\rho\sigma} =0$). It follows that $\xi$ is the corresponding principal null vector. Conversely, by a theorem of Lewandowski~\cite{Lewandowski:1991bx}, given a four times repeated null vector $\xi$ which is also a conformal Killing vector, there always exists a twistor spinor $\epsilon_+$ such that $\xi^\mu = 2\, {\overline \epsilon_-} \Gamma^\mu \epsilon_+$, where $\epsilon_- = ( {\sf B} \epsilon_+ )^*$. Up to a complex multiple, $\epsilon_+$ is unique. Furthermore, unless the Weyl tensor vanishes identically, any other twistor spinor must be a complex multiple of $\epsilon_+$, since there can obviously be only one four times repeated principal null direction. Thus, we have two possible cases:

{\bf (1)} Type O. The Weyl tensor vanishes, whence $(\eM,g)$ is locally conformally flat. Locally, $(\eM,g)$ has the same number of linearly independent twistor spinors and conformal Killing vectors as Minkowski space. Up to a Weyl transformation, every twistor spinor can be written $\epsilon = \epsilon_0 + {\slashed x} \epsilon_1$, in terms of a pair of constant spinors $\epsilon_0 , \epsilon_1$. The same is true for locally conformally flat solutions in higher dimensions.

{\bf (2)} Type N. The Weyl tensor does not vanish. The {\bf twist} of $\xi$ is defined as the three-form $\xi^\flat \wedge \d\xi^\flat$, where $\xi^\flat = \xi_\mu \d x^\mu$ is the one-form dual to $\xi$ with respect to $g$. Then either \\ [.2cm]
{\bf (2a)} The twist of $\xi$ is zero. In this case, $g$ is locally conformally equivalent to a {\bf pp-wave} metric. By definition, a pp-wave is a Lorentzian manifold which admits a parallel null vector. This vector is identified with $\xi$. In higher dimensions, pp-waves are defined in the same way and correspond to a subclass of a broader class of Lorentzian geometries with special holonomy, called Brinkmann waves. Having special holonomy implies that they admit parallel spinors and it can be shown that every twistor spinor on a pp-wave is parallel.

In {\bf Brinkmann coordinates} $( u,v, {\bm x} )$, the pp-wave metric is
\ben\label{ppwave}
g = 2 \d u \d v + h(u,{\bm x}) \d u^2  + \d {\bm x}^2~,
\een
where $h$ is an arbitrary smooth function of $u$ and ${\bm x}$. In $d$ dimensions, $\d {\bm x}^2$ denotes the flat Euclidean metric on $\RR^{d-2}$. The parallel null vector is $\xi = \partial / \partial v$ in these coordinates. For generic $h$, $\xi$ is the only conformal Killing vector on a pp-wave. All scalar curvature invariants of \eqref{ppwave} vanish identically (in particular, the scalar curvature $R=0$). The metric is Ricci-flat only if $h$ is a harmonic function of $\bm{x}$.

Let ${\bm e}^+ = \d u$, ${\bm e}^- = \d v + \half h \d u$ and ${\bm e}^a = \d x^a$ define a local null frame on the pp-wave, where $a=1,...,d-2$. With respect to this choice of frame, it is straightforward to show that any twistor spinor $\epsilon$ is parallel and obeys ${\slashed \xi} \epsilon =0$. Since ${\slashed \xi}^2 =0$, the maximum number of linearly independent twistor spinors on a pp-wave is precisely half the rank of the spinor bundle.

An interesting subclass of pp-waves, called {\bf plane waves}, is defined by taking $h(u,{\bm x}) = h_{ab}(u) x^a x^b$ (see \cite{Blau:Notes} for a nice review). In fact, plane waves describe the geometry of a spacetime in an infinitesimal neighbourhood of a null-geodesic, via the \lq Penrose limit'. Whence, they have a somewhat universal character. In addition to $\xi = \partial_v$, plane waves have several other conformal Killing vectors that are defined as follows. Let ${\bm h} = ( h_{ab} )$ and define \lq propagators' ${\bm A} = ( A_{ab} )$ and ${\bm B} = ( B_{ab} )$ such that
\ben\label{abprop}
\partial_u^2 {\bm A}(u,u^\prime) = {\bm h}(u) {\bm A}(u,u^\prime) \; , \quad\quad \partial_u^2 {\bm B}(u,u^\prime) = {\bm h}(u) {\bm B}(u,u^\prime)~,
\een
with \lq boundary conditions' ${\bm A}(u,u) = \partial_u {\bm B}(u,u) = 1$ and $\partial_u {\bm A}(u,u) = {\bm B}(u,u) = 0$. There are $2(d-2)$ Killing vectors defined by
\ben
\label{XYKV}
X_a = A_{ab} \partial_b - ( \partial_u A_{ab} ) \, x^b \partial_v \; , \quad\quad Y_a = B_{ab} \partial_b - ( \partial_u B_{ab} ) x^b \partial_v~,
\een
taking $u^\prime =0$, and one additional (homothetic) conformal Killing vector
\ben
\label{kKV}
k = 2 v \partial_v + x^a \partial_a~.
\een
The non-vanishing Lie brackets of \eqref{XYKV}, \eqref{kKV} and $\xi$ are
\ben
[X_a, Y_b] = \delta_{ab} \xi \; , \quad [k,X_a] = -X_a \; , \quad [k,Y_a] = -Y_a \; , \quad [k,\xi] = -2\xi~.
\een
The isometry algebra spanned by \eqref{XYKV} and $\xi$ is therefore isomorphic to the $(2d-3)$-dimensional {\bf Heisenberg algebra} ${\mathfrak{heis}}_{d-2}$. Thus, $\fX^c(\M) \cong {\mathfrak{heis}}_{d-2} \ltimes \RR$ when $\M$ is a plane wave, with the extension by $\RR$ generated by $k$. For general pp-waves, the classification of conformal isometries is rather complicated (see \cite{Maartens:1991mj,KT:2004}). \\ [.2cm]
{\bf (2b)} The twist of $\xi$ is not zero. In this case, $g$ is locally conformally equivalent to the metric of a Fefferman space \cite{Lewandowski:1991bx,Nurowski}. Fefferman spaces exist in any even spacetime dimension $d=2n+2$, with $n>0$, and always admit a twistor spinor. The basic class of examples for such a spacetime can be described as follows~\cite{Feff:1976}.

Let $\Omega \subset \CC^{n+1}$ be a convex open domain with smooth boundary $\partial \Omega = \Sigma$. $\Sigma$ is viewed as a real $(2n+1)$-dimensional manifold. Now consider a real-valued solution $F$ to the complex Monge-Amp\`{e}re equations
    \ben
    \det \left(
    \begin{matrix}
    F_{,i{\bar j}} & F_{,{\bar j}}\\
    F_{,i} & F
    \end{matrix}
    \right) =1
    \quad \text{on} \;\; \Omega \; , \quad\quad F = 1 \quad \text{on} \;\; \partial \Omega~,
    \een
    where the subscript \lq$,i$' denotes $\partial / \partial z_i$, with respect to coordinates $(z_1, \dots, z_{n+1})$ on $\CC^{n+1}$, and bars denote complex conjugation. It can be shown that a smooth solution to this equation always exists~\cite{chenyau}. An example of a Fefferman space is defined by the manifold $\M=\RR \times \Sigma$, equipped with Lorentzian metric
    \ben
     g= \frac{i}{n+2} \, \d r \, (F_{,i} \, \d z_i - F_{,{\bar i}} \, \d {\bar z}_i) + F_{,i{\bar j}} \, \d z_i \d {\overline z}_j~,
    \een
    where $r$ is the coordinate of the $\RR$ factor in $\M$. The normalisation of $r$ is chosen such that a biholomorphism of $\Omega$ corresponds to a conformal transformation of $g$. This class of metrics is closely related to the concept of a CR structure,
\footnote{Whether one takes CR to be an acronym for \lq Cauchy-Riemann' or \lq Complex-Real' is a matter of taste.}
which is the correct framework in which to define the general Fefferman metric.

An {\bf almost CR structure} consists of the following data:
\begin{itemize}
\item
 A $(2n+1)$-dimensional real manifold $\Sigma$ with a $2n$-dimensional subbundle $L \subset T\Sigma$.
\item
An almost complex structure $J$ on $L$ (i.e. a bundle isomorphism $J : L \rightarrow L$ with $J^2 =-1$).
\end{itemize}
Let $\Gamma (L)$ denote the space of smooth sections of $L$. The almost CR structure above is integrable provided
\begin{equation} [JX,Y] + [X, JY] \in \Gamma(L) \quad\quad {\mathrm{and}} \quad\quad [X,Y] + J[JX,Y] + J[X,JY] - [JX,JY] =0~,
\end{equation}
for all $X,Y \in \Gamma (L)$. An integrable almost CR structure is called a {\bf CR structure} and a manifold that is equipped with a CR structure is called a {\bf CR-manifold}. Any smooth real hypersurface $\Sigma \subset \mc^{n+1}$ of dimension $2n+1$
is a CR-manifold. If $\tilde J$ is the standard complex structure on the ambient
$\mc^{n+1}$, then $L=T\Sigma \cap \tilde J(T\Sigma)$ and $J = \tilde J|_{T\Sigma}$. More generally, any $(2n+1)$-dimensional real submanifold of a complex manifold with complex dimension $n+1$ is a CR-manifold.

For a given CR structure, one can fix a pseudo-Hermitian one-form $\theta$ such that $\theta|_L = 0$. If the associated {\bf Levi-form} $L_\theta$, defined by
\ben
L_\theta(X,Y) = \d \theta(X,JY)~,
\een
for all $X,Y \in \Gamma(L)$, is positive-definite, then the CR structure is called {\bf strictly pseudo-convex}. In this case, the tensor $g_\theta = L_\theta + \theta \otimes \theta$ defines a Riemannian metric on $\Sigma$. There is a distinguished connection $\nabla^W$ on $\Sigma$ that is compatible with both $g_\theta$ and $J$, called the {\bf Tanaka-Webster connection}. The Tanaka-Webster connection $\nabla^W$ has non-trivial torsion ${\rm Tor}^W$, given by
\ben
{\rm Tor}^W(X,Y) = L_\theta(X,Y) \zeta  \; , \quad\quad {\rm Tor}^W(\zeta,X) = -\tfrac{1}{2}([\zeta,X]+J[\zeta,JX])~,
\een
for all $X,Y \in \Gamma(L)$, where $\zeta$ is a vector field on $\Sigma$ with $\theta(\zeta)=1$ and $g_\theta (\zeta , X) =0$ for all $X \in \Gamma(L)$.

A spin structure on any such CR-manifold defines a canonical line bundle $\eM$ over $\Sigma$. Let $\pi : \eM \rightarrow \Sigma$ denote the projection for this bundle. There is a unique connection $A^W$ on $\eM$ for which the Tanaka-Webster connection $\nabla^W$ is induced on $\Sigma$. In terms of this data, one may define the {\bf Fefferman metric}
\begin{equation}
\label{eq:FeffermanMetric}
g =  - \tfrac{8}{n+2} \left( \tfrac{1}{4(n+1)} \, R^W \pi^* \theta +i \, A^W \right) \pi^* \theta + \pi^* L_\theta~,
\end{equation}
on $\eM$, in terms of the scalar curvature $R^W$ of $\nabla^W$. It follows that the conformal class of $g$ does not depend on $\theta$, but only on the CR structure. The Lorentzian manifold $(\eM ,g)$, with even dimension $d=2n+2$, is called the {\bf Fefferman space} of the strictly pseudo-convex CR-manifold $(\Sigma,J,\theta)$, equipped with its canonical spin structure. More details of this construction, including the explicit form of twistor spinors, can be found in \cite{Baum:1999} (see also \cite{Lewandowski:1991bx} in $d=4$, in terms of a somewhat different formalism). In general, up to a complex multiple, Fefferman spaces admit just one linearly independent twistor spinor of a given chirality.

In $d>4$, there are a few more distinct classes of Lorentzian manifolds $(\eM ,g)$ which admit a non-vanishing twistor spinor (see \cite{deMedeiros:2013jja} for more details). Any such $(\eM ,g)$ is locally conformally equivalent to either a Lorentzian Einstein-Sasaki manifold (if $d$ is odd) or the direct product of a Lorentzian Einstein-Sasaki manifold with a Riemannian manifold admitting Killing spinors. Since our focus in this paper is on $d=4$, we shall not concern ourselves further with these other Lorentzian geometries in higher dimensions.

The type O solutions in class {\bf (1)} include Minkowski $\RR^{d-1,1}$, de Sitter $\d S_d$ and anti-de Sitter $\AdS_d$ spacetimes. We will give the explicit form below of twistor spinors on any even-dimensional de Sitter spacetime, in Example~\ref{deSspinors}. The spacetimes in class {\bf (2a)} are of some physical interest because they can describe the region of a gravitational wave far from the source. The spacetimes in class {\bf (2b)} have limited physical interest, since a Fefferman metric is never conformally Einstein in $d=4$ \cite{Lewandowski:1991bx} (the conformally flat case is in class {\bf (1)}). Unfortunately, perhaps the most physically interesting spacetimes in $d=4$, namely the Kerr-Newman black holes, do not admit twistor spinors because they are of Petrov type D (a.k.a. type II-II in \cite{Wald}).

Recall from below \eqref{PenroseOperator} that a Killing spinor $\epsilon$ corresponds to a special type of twistor spinor, obeying
\ben\label{kspinor1}
\nabla_\mu \epsilon = \lambda \, \Gamma_\mu \epsilon \ ,
\een
with Killing constant $\lambda \neq 0$. This implies $\tfrac{1}{d} {\slashed \nabla} \epsilon = \lambda \epsilon$. Whence, if $\epsilon$ is nowhere-vanishing, the third condition in~\eqref{consistency} implies $R = -4d(d-1)\lambda^2$. Consequently, the scalar curvature $R$ must be constant and $\lambda$ must be either real (if $R<0$) or imaginary (if $R>0$). If $\epsilon$ is parallel, with $\lambda =0$, then $R=0$. In $d=4$, any Majorana Killing spinor $\epsilon$ must have $\lambda$ real (whence $R<0$). For example, both $\AdS_4$ and $\dS_4$ admit Dirac Killing spinors but only $\AdS_4$ admits Majorana Killing spinors.

Conversely, if $R$ is a non-zero constant, then each twistor spinor $\epsilon$ defines a pair of spinors
\ben
\epsilon^\pm =
\epsilon \pm {\sqrt{\frac{4}{R} \frac{(1-d)}{d}}} {\slashed \nabla} \epsilon
\een
that are eigenvectors of ${\slashed \nabla}$, with eigenvalues $\mp {\sqrt{\frac{R}{4} \frac{d}{(1-d)}}}$. Furthermore, using the second condition in~\eqref{consistency}, one can fix a conformal class such that $\epsilon^\pm$ are Killing spinors. In this sense, on a manifold with constant non-zero scalar curvature, any twistor spinor can be expressed as a linear combination of Killing spinors. We conclude this section with a concrete example.

\begin{example}\label{deSspinors} Let $\M = \dS_d$ with metric
\ben
\label{eq:dSmetric}
g = \ell^2 (-\d t^2 + \cosh^2 t \; \d \sigma^2_{d-1} )~,
\een
where $\ell$ is a constant scale and $\d \sigma^2_{d-1}$ is the round metric on the unit sphere $S^{d-1}$. The de Sitter metric is locally conformally flat and Einstein, with constant scalar curvature $R= \frac{d(d-1)}{\ell^2}$. Let us now assume that $d$ is even.

Since the twistor spinor equation is conformally invariant, the local form of any twistor spinor on $\dS_d$ can be written as a twistor spinor in Minkoswki space multipled by the appropriate conformal factor. Let us instead obtain the global form of twistor spinors on $\dS_d$, in the coordinates used in \eqref{eq:dSmetric}, via the observations noted above. Since $R$ is constant, any twistor spinor on $\dS_d$ must be a linear combination of Killing spinors. Moreover, since $R= \frac{d(d-1)}{\ell^2}$, the Killing constants are $\lambda = \pm \tfrac{i}{2\ell}$. It is straightforward to obtain explicitly the general solution of \eqref{kspinor1} on $\dS_d$, e.g. by an analytic continuation of the Killing spinors on $S^d$ given in \cite{Lu:1998nu}. Let
\ben
{\bm e}^{0} = \ell \ \d t \; , \quad
{\bm e}^{1} = \ell \cosh t \, \d \vartheta_1 \; ,
... \; , \quad
{\bm e}^{d-1} = \ell \cosh t \sin \vartheta_1 \cdots \sin \vartheta_{d-2} \, \d \vartheta_{d-1} \ ,
\een
define a local frame on $\dS_d$, in terms of polar angles $\vartheta_i$ on $S^{d-1}$. Relative to this local frame, the general solution of~\eqref{kspinor1}, with $\lambda = \pm \tfrac{i}{2\ell}$, is
\ben\label{psidef}
\epsilon^\pm = \Big(
\cosh\tfrac{t}{2}\; {\bf 1} \pm  \sinh\tfrac{t}{2}\; \Gamma_{0}
\Big)
\prod_{j=1}^{d-1} \Big(
\cos\tfrac{\vartheta_j}{2} \; {\bf 1} + i\, \sin \tfrac{\vartheta_j}{2} \; \Gamma_{j} \Gamma_{j+1}
\Big) \epsilon_0 \ ,
\een
where $\epsilon_0$ is an arbitrary constant Dirac spinor. The gamma matrices are written with respect to frame indices (i.e. they are just as in Minkowski space). Whence, we conclude that any twistor spinor on $\dS_d$ can be expressed as a linear combination of Killing spinors of the form~\eqref{psidef}.
\end{example}

Since $\dS_4$ is locally conformally flat, we see from Table~\ref{tab:tableCKS} that it admits a conformal symmetry superalgebra $\cS_\circ \cong \fsu(2,2| \eN)$ for both $\eN=1$ and $\eN=2$ while $\cS_\circ \cong {\mathfrak{psu}}(2,2| 4)$ for the $\eN =4$ case. In the $\eN =1$ case, given a Dirac Killing spinor $\epsilon$ on $\dS_4$, one can define a pair of Majorana twistor spinors $\epsilon + ( {\sf B} \epsilon )^*$, $i ( \epsilon - ( {\sf B} \epsilon )^* )$ in the odd part of $\fsu(2,2| 1)$. The $\eN =2$ case gives rise to a conformal symmetry superalgebra of the type discussed in section~\ref{sec:N2d4}. Given a Dirac Killing spinor $\epsilon$ on $\dS_4$ (with imaginary Killing constant $\lambda$), one can define a pair of Majorana twistor spinors $( \epsilon_+^A , \epsilon_{-\, A} )$ in the odd part of $\fsu(2,2| 2)$ by identifying $\epsilon_+^1 = \epsilon_+$, $\epsilon_{-\, 2} = \epsilon_-$ and $\epsilon_{-\, 1} = ( {\sf B} \epsilon_+ )^*$, $\epsilon_+^2 = ( {\sf B} \epsilon_- )^*$. (There is no relation between the chiral projections $\epsilon_\pm$ of the Dirac spinor $\epsilon$.) This identification implies $\nabla_\mu \epsilon_+^A = \lambda \varepsilon^{AB} \, \Gamma_\mu \epsilon_{-\, B}$, where $\varepsilon^{AB} = - \varepsilon^{BA}$ and $\varepsilon^{12} =1$. Whence, $( \epsilon_+^A , \epsilon_{-\, A} )$ are indeed Majorana twistor spinors on $\dS_4$.


\section{Cohomology constructions related to $\cS$}

In a field theory which has conformal symmetry superalgebra $\cS$, the action of $\cS$ is described by transformations of the constituent fields, and functionals thereof such as the Lagrangian. Mathematically, this corresponds to a particular action of $\cS$ on a configuration space $\C$ of fields (including gauge and matter fields for the theory in question). The transformations relevant in this paper will be written down concretely below, in the context of field theory with rigid $\eN=2$ conformal supersymmetry in curved spacetime. However, before we do this, we would like to explain in some generality what the general structure of the action of $\cS$ is. In particular, we shall explain in full generality the nature of an associated cohomological construction. This will help in understanding the quantum structure of the theory in our investigation of anomalies. This setup is a generalised version of the familiar BRST construction in physics.
\footnote{There is a related construction, involving a generalised 
\lq master equation' \`{a} la Batalin-Vilkovisky, described in \cite{Brandt:1997cz}. This is based on the existence of certain conserved currents and it would be interesting to understand more precisely how this relates to our construction.}
%


\subsection{Standard BRST structure}
\label{sec:brststandard}
 As a warm up, consider an (over-)simplified prototype model of the actual situation below. We take a finite-dimensional manifold $\C$, with a smooth group action $G \times \C \rightarrow \C$ given by $(g,\phi) \mapsto g \cdot \phi$, where $G$ is a Lie group with corresponding Lie algebra $\cG$ and Lie bracket $[-,-]$. For each $X \in \cG$, we can define a vector field $\delta_X \in \fX(\C)$ by its action on a function $F \in C^\infty(\C)$ through
\ben
\delta_X F(-) = \tfrac{d}{d\tau} \, F({\rm exp}(\tau X) \cdot \, - \, ) \bigg|_{\tau=0}~.
\een
By construction,
\ben\label{alg}
[ \delta_X , \delta_Y ] \equiv \delta_X \delta_Y - \delta_Y \delta_X = \delta_{[X,Y]}~,
\een
so the map $X \mapsto \delta_X$, for all $X \in \cG$, defines a representation of $\cG$ on $C^\infty(\C)$. A functional $F$ is invariant under $G$ if $\delta_X F=0$, for all $X \in \cG$. The \lq BRST-type' construction can now be explained as follows. Let
\ben
\V^n = C^\infty(\C) \otimes \wedge^n \cG^*,
\een
i.e. $\V^n$ consists of smooth maps from $\C$ into the skewsymmetric $n$-fold multi-linear functionals on $\cG$. Now define the differential $\d_{\cG}: \V^n \rightarrow \V^{n+1}$ by
\ben
\label{eq:dG}
\begin{split}
(\d_{\cG} \alpha)_{n+1}(X_1, \dots, X_{n+1}) &= \sum_{i=1}^{n+1}
\delta_{X_i} \alpha_n(X_1, \dots, \hat X_i, \dots, X_{n+1}) \\
&\quad -\sum_{i<j} (-1)^{i+j} \, \alpha_n([X_i, X_j], \dots, \hat X_i, \dots, \hat X_j, \dots)~,
\end{split}
\een
for all $\alpha_n \in \V^n$ and $X_i \in \cG$, where a hat denotes omission. Using~\eqref{alg} and the Jacobi identities for $\cG$ and $\fX(\C)$, it is straightforward to check that $\d_{\cG}^2 = 0$. The differential $\d_{\cG}$ is known to physicists as a {\bf BRST operator} and the cohomology groups
\ben
H^n(\d_{\cG}) =\frac{\{{\rm ker} \ \d_{\cG}: \V^n \rightarrow \V^{n+1} \}}{\{ {\rm im} \
\d_{\cG}: \V^{n-1} \rightarrow \V^n \}}~,
\een
especially $H^1(\d_{\cG})$, are important in the general discussion of anomalies. The space $H^0(\d_{\cG})$ is just the space of $G$-invariant functions on $\C$. The above construction is closely related to the mathematical concept of a Chevalley-Eilenberg complex. To make contact with the notation used in physics,
let us choose a basis $\{ X_a \} \in \cG$, where $a=1,..., \dim \, \cG$, with $[X_a , X_b ] = f_{ab}^c X_c$, in terms of structure constants $f_{ab}^c$, and let $\{ c^a \} \in \cG^*$ denote a dual basis. Let $\phi^i$ be local coordinates on $\C$, where $i=1,..., \dim \, \C$. Relative to this basis, any $\alpha_n = \alpha_{a_1 ... a_n}(\phi) c^{a_1} \, \wedge ...  \wedge c^{a_n} \in \V^n$. In the physics literature, the $c^a$ are called {\bf ghosts}. Since they only appear in wedge products, we think of them as Grassmann-odd variables, i.e. elements in the exterior algebra $\wedge^\bullet \cG^* =\oplus_n \wedge^n \cG^*$, and we shall drop the $\wedge$'s. If $V^i_a(\phi) \partial_i$ are components of the vector field $\delta_{X_a}$ on $\C$, for some $X_a \in \cG$, the action of $\d_{\cG}$ in \eqref{eq:dG} gives
\ben
\d_{\cG} c^a = -\half f^a_{bc} c^b c^c \ , \qquad \d_{\cG} \phi^i
= V^i_a(\phi) c^a \ .
\een
These are the familiar BRST transformation rules described in the physics literature, where $\d_{\cG}$ is often called $s$.
\begin{example}
The standard example of this construction in an infinite-dimensional setting is the following. Let $\C$ be the space of all smooth gauge fields on the trivial principal bundle over $\M$. Gauge connections in this bundle may be viewed as $\fg$-valued 1-forms $A=A_\mu \d x^\mu$. The Lie algebra of gauge transformations is $\cG=C^\infty(\M, \frak{g})$. Of course, in this case,
both $\C$ and $\cG$ are infinite-dimensional. For any $\Lambda \in \cG$, the corresponding vector field $\delta_\Lambda \in \fX (\C)$ is given by
\ben
\delta_\Lambda = \int_\M D_\mu \Lambda(x) \frac{\delta}{\delta A_\mu (x)} \ .
\een
The resulting BRST-type transformations are $s c(x) =  -\half [c(x), c(x)]$
and $s A_\mu (x) = D_\mu c(x)$. In this example, the $\frak{g}$-valued ghost $c$ must be a function on $\M$ because $\cG=C^\infty(\M, \frak{g})$.
\end{example}
\begin{example}
Another example follows by taking $\cG = \cB = \fX^c(\M) \oplus \cR$, corresponding to the bosonic part of a conformal symmetry superalgebra $\cS$. Let $V$ be a representation of $\cR$. Let $\C=C^\infty(\M, V)$ be the configuration space of smooth $V$-valued scalar fields $\varphi$ on $\eM$. A conformal Killing vector $X \in \fX^c(\M)$ acts by $\delta_X \varphi = (\cL_X + w_\varphi \sigma_X) \varphi$, where $w_\varphi \in \RR$ specifies the {\bf Weyl weight} of $\varphi$. An element $\rho \in \cR$ acts by $\delta_\rho \varphi = \rho \cdot \varphi$, where $\cdot$ denotes the action of $\cR$ on $V$. It may be checked that~\eqref{alg} is satisfied (for any $w_\varphi$). Relative to a basis $\{ {\bm e}_A\} \in V$, the corresponding BRST transformations are
\ben
sX = -\half [X,X] \; , \quad\quad
s\alpha_B{}^A = -\alpha_B{}^C \alpha_C{}^A \; , \quad\quad
s\varphi^A = (\cL_X + w_\varphi \sigma_X) \varphi^A + \alpha_B{}^A \varphi^B~,
\een
where $\alpha_B{}^A$ is the ghost of $\rho^A{}_B$, and $X$ denotes the ghost of $X \in \fX^c (\eM)$. More precisely, relative to a basis $\{ \xi_i \} \in \fX^c (\eM)$, any $X = X^i \xi_i \in \fX^c(\eM)$ has ghost $X = \theta^i \xi_i$, with Grassmann-even components $X^i$ replaced by Grassmann-odd components $\theta^i$ for the ghost field.
\end{example}


\subsection{Extended BRST structure}
\label{sec:brstextension}
As we will see, the simple BRST construction just described is not quite adequate to accommodate $\eN=2$ superconformal field theory. The reason is that rather than having a simple representation of the type~\eqref{alg}, we have additional terms on the right hand side corresponding to field-dependent gauge transformations, and also involving terms proportional to the equations of motion. To simplify the situation for the moment, let us ignore the equations of motion, and let us continue to pretend that all symmetries are bosonic. Then the situation is schematically the following: We have a manifold of field configurations, $\C$, together with two Lie algebras, $\cG, \fh$.
 These will later be the local gauge transformations ($\cG$), and conformal symmetry superalgebra ($\fh=\cS$). However, in this section, they are arbitrary (bosonic) Lie algebras.
 These Lie algebras act by vector fields $\fh \oplus \cG \owns X \mapsto \delta_X \in \fX(\C)$. The Lie-bracket structure is assumed to be schematically $[\frak{h}, \frak{h}] \subset \frak{h}, [\cG, \cG] \subset \cG, [\frak{h}, \cG] \subset \cG$. The vector fields $\delta_X$ now assumed to obey instead a relation of the form
\ben\label{vxy}
[\delta_X, \delta_Y] = \delta_{[X,Y]} + \delta_{\Lambda(X,Y)}~,
\een
where
\ben
\Lambda: \frak{h} \times \frak{h} \to C^\infty(\C, \cG) \; , \quad\quad (X,Y) \mapsto  \Lambda(X,Y)~,
\een
is a linear, skewsymmetric map into the $\cG$-valued {\emph{functions}} on $\C$. $X \mapsto \delta_X$ is extended to Lie algebra valued {\em functions} in a the canonical way, i.e. if $F \in C^\infty(\C), X \in \cG$, then
$\delta_{F \otimes X} = F \delta_X$. The fact that $\Lambda(X,Y)$ is a {\em function} on $\C$ corresponds to the fact that the commutator of two symmetries in $\frak{h}$ closes onto a {\emph{field-dependent}} gauge transformation. Consistency of the above relations implies that $\Lambda$ must satisfy the cocycle-type condition ($X,Y,Z \in \frak{h}$)
\ben\label{cocyLa}
\begin{split}
&(\delta_X + \ad_X) \Lambda(Y,Z) + (\delta_Z + \ad_Z) \Lambda(X,Y) + (\delta_Y + \ad_Y) \Lambda(Z,X) \\
&+ \Lambda([X,Y],Z)] + \Lambda([Y,Z],X) + \Lambda([Z,X],Y) \\
&= 0 \ ,
\end{split}
\een
and the condition ($X \in \cG, Y,Z \in \frak{h}$)
\ben\label{vxlambda}
\delta_X \Lambda(Y,Z) + [X,\Lambda(Y,Z)] = 0 \ .
\een
If $\Lambda$ did not depend on $\phi \in \C$, then the second condition states
 that $\Lambda(X,Y) \in Z(\cG)$, implying also $[\cG, \frak{h}]=0$. Then the first and last terms are absent in the cocycle type condition, $\Lambda$ would
correspond to a central charge, and \eqref{vxy} to a central extension of $\frak{h}$
by $\cG$.

We define a differential $\d_{\frak{h} | \cG}$
on the complex $\V^n = C^\infty(\C) \otimes \wedge^n (\cG \oplus \frak{h})^*$
by the formula
\ben
\begin{split}
(\d_{\frak{h} | \cG} \alpha)_{n+1}(X_1, \dots, X_{n+1}) &= \sum_{i=1}^{n+1}
\delta_{X_i} \alpha_n(X_1, \dots, \hat X_i, \dots, X_{n+1})  \\
&\quad -\sum_{i<j} (-1)^{i+j} \, \alpha_n([X_i, X_j], \dots, \hat X_i, \dots, \hat X_j, \dots) \\
&\quad -\sum_{i<j} (-1)^{i+j} \, \alpha_n(\Lambda(X_i, X_j), \dots, \hat X_i, \dots, \hat X_j, \dots)~.
\end{split}
\een
The term involving $\Lambda$ is by definition present only when $X_i, X_j \in \frak{h}$.
Again, one verifies that
\begin{lemma}
$\d_{\frak{h} | \cG}^2=0$.
\end{lemma}
{\em Proof:} \\
\ben
\begin{split}
(\d_{\frak{h} | \cG}^2 \alpha)_{n+2}(X_1, \dots, X_{n+2}) &=
\sum_{i<j} [\delta_{X_i}, \delta_{X_j}] \alpha_n(X_1, \dots, \hat X_i, \dots, \hat X_j, \dots, X_{n+2}) \\
&\quad - \sum_{i<j} \delta_{[X_i,X_j]} \alpha_n(X_1, \dots, \hat X_i, \dots, \hat X_j, \dots, X_{n+2}) \\
&\quad - \sum_{i<j} \delta_{\Lambda(X_i,X_j)} \alpha_n(X_1, \dots, \hat X_i, \dots, \hat X_j, \dots, X_{n+2}) \\
&\quad + \sum_{i<j} \sum_{k \neq i,j} (-1)^{i+j} \, \alpha_n([X_k,[X_i, X_j]], \dots, \hat X_i, \dots, \hat X_j, \dots, \hat X_k, \dots) \\
&\quad + \sum_{i<j} \sum_{k \neq i,j} (-1)^{i+j} \, \alpha_n(\Lambda([X_k,X_i], X_j), \dots, \hat X_i, \dots, \hat X_j, \dots, \hat X_k, \dots)  \\
&\quad + \sum_{i<j} \sum_{k \neq i,j} (-1)^{i+j} \, \alpha_n((\ad_{X_k} + \delta_{X_k}) \Lambda(X_i, X_j), \dots, \hat X_i, \dots, \hat X_j, \dots, \hat X_k, \dots) \\
&=0~,
\end{split}
\een
where the last equality follows from the Jacobi identity and the cocycle identity for $\Lambda$,~\eqref{cocyLa},~\eqref{vxlambda}. \qed

Let us display the differential in \lq physics notation'.
To this end, denote the ghosts relative to the basis $\{X_I\}$ of $\frak{h}$ by $\xi^I$, and those with $\cG$ as before by
$c^a$. Also denote the vector field $\delta_{X_a}$ by $V^i_a \partial_i$, and the vector field $\delta_{X_I}$ by $V_I^i \partial_i$. In that notation, the differential $\d_{\frak{h} | \cG}$ is given by
\ben\label{onshells}
\begin{split}
 \d_{\frak{h} | \cG} c^a &= -\half f^a_{bc} c^b c^c - f^a_{bI} c^b \xi^I - \half \Lambda^a_{IJ}(\phi) \xi^I \xi^J \ , \\
\d_{\frak{h} | \cG} \xi^I &= -\half f^I_{JK} \xi^J \xi^K \ , \\
\d_{\frak{h} | \cG} \phi^i &= V^i_a(\phi) c^a + V^i_I(\phi) \xi^I \ .
\end{split}
\een

The situation described by~\eqref{vxy} is still not quite yet exactly what we have in
our application below. Instead, the situation will be closer to the following one. We have a manifold, $\C$, of (off-shell) field configurations, together with a smooth action functional $S: \C \to \mr$, which is invariant in the sense that $\delta_X S=0$ for
all $X \in \frak{h} \oplus \cG$. As before, we have Lie algebras $\frak{h}, \cG$, and a map
$\Lambda: \frak{h} \times \frak{h} \to C^\infty(\C, \cG)$ with the same properties as before.
Additionally, we have, for each $X,Y \in \frak{h}$ a map $E: \frak{h} \times \frak{h} \to  {\rm Sect}^\infty(
T\C \wedge T\C), (X,Y) \mapsto E_{X,Y}$, and
\ben\label{vxye}
[\delta_X, \delta_Y] = \delta_{[X,Y]} + \delta_{\Lambda(X,Y)} - i_{\d S} E_{X,Y} \ ,
\een
where the last term vanishes by definition
if $X$ or $Y$ are in $\cG$. In our model problem
where all symmetries are bosonic (i.e. $\frak{h}, \cG$ is an ordinary Lie algebra), we have
\ben\label{esym}
E_{X,Y}=-E_{Y,X}~.
\een
$i$ is the operator of interior multiplication, i.e. contraction of a vector with the first index of a tensor. If we introduce local coordinates $\phi^i$ on $\C$, then $E = E^{ij}(\phi)\partial_i \wedge \partial_j$, and the last term in~\eqref{vxye} reads $E^{ij}(\phi) \partial_i S(\phi) \partial_j$. Whence, it vanishes on-shell, i.e. on $\C_0 = \{\phi \in \C \mid \partial_j S(\phi)=0\}$. We are dealing with a situation where the symmetry algebra closes only \lq on-shell'.

The algebraic relation~\eqref{vxye} implies consistency conditions on $E$. To write these down, it is convenient to define the space ${\rm Poly}^k(\C)$ of rank $k$ {\bf polyvector fields}, where
\ben
v = v^{i_1 \dots i_k}(\phi) \partial_{i_1} \wedge \cdots \wedge \partial_{i_k}
\in {\rm Poly}^k(\C) \ ,
\een
equipped with {\bf Schouten-Nijenhuis bracket}
\ben
(-,-) : {\rm Poly}^k(\C) \times {\rm Poly}^l(\C) \rightarrow {\rm Poly}^{k+l-1}(\C)~,
\een
defined such that
\ben
(v,w)^{i_1 \dots i_{k+l-1}} = k \, v^{j[i_1...i_{k-1}} \partial_j w^{i_k...i_{k+l-1}]}
+l \, (-1)^{kl} \, w^{j[i_1...i_{l-1}} \partial_j v^{i_k...i_{k+l-1}]}~,
\een
for all $v \in {\rm Poly}^k(\C)$ and $w \in {\rm Poly}^l(\C)$. This bracket is graded symmetric and satisfies the graded Jacobi identity,
\ben
(-1)^{|u||v|}(u,(v,w)) + (-1)^{|v||w|}(v,(w,u)) + (-1)^{|w||u|}(w,(u,v)) = 0 \ ,
\een
where the degree $|u|=k$ for any $u \in {\rm Poly}^{k}(\C)$.

Besides the cyclic identity for $\Lambda$, consistency of \eqref{vxye} requires
\ben\label{cyclE0}
(\delta_X, E_{Y,Z}) + (\delta_Y, E_{Z,X}) + (\delta_Z, E_{X,Y}) +
E_{X,[Y,Z]} + E_{Y,[Z,X]} + E_{Z,[X,Y]} = 0 \ .
\een
We would now like to define a nilpotent BRST-type differential extending $\d_{\frak{h}| \cG}$ to this more complicated setting. The previously defined differential $\d_{\frak{h} | \cG}$ is now no longer nilpotent, due to the presence of
the additional term involving $E$ in the algebraic relation~\eqref{vxye}. This problem can be
remedied if we {\em assume} that $E$ satisfies the \lq Jacobi identity'
\ben\label{cyclE}
(E_{X_1, X_2},E_{X_3,X_4}) + (E_{X_3, X_1},E_{X_2, X_4}) + (E_{X_2, X_3},E_{X_1,X_4})= 0 \ ,
\een
and if we {\em assume} $(E_{X_1,X_2},\Lambda_{X_3,X_4}) \pm$ permutations $=0$.
Both conditions will be satisfied in our application. Then we define the set of \lq chains' as
\ben
\V^n = \bigoplus_{k-l=n} \Bigg( {\rm Poly}^l(\C) \otimes \wedge^k (\frak{h}^* \oplus \cG^*) \Bigg) \otimes {\sf S}(\cG) \ .
\een
Here, ${\sf S}(\cG)=\bigoplus_n \odot^n \cG$ is the symmetric algebra over $\cG$,
which consists of the totally symmetric tensors of arbitrary rank over $\cG$.
Our new nilpotent differential is defined as the map $\Q:
\V^n \to \V^{n+1}$, where
\ben\label{Qdef}
\begin{split}
[\Q (p \otimes \alpha)]_{k,l}(X_1, \dots, X_{k}) &=
\langle \delta, p \rangle \wedge  \alpha_{k,l-1}(X_1, \dots, X_{k}) \\
&\quad + \sum_{i=1}^{k}
[\pi(X_i)p] \otimes \alpha_{k-1,l}(X_1, \dots, \hat X_i, \dots, X_{k})  \\
&\quad + \sum_{i=1}^{k}
p \otimes (\delta_{X_i}, \alpha_{k-1,l})(X_1, \dots, \hat X_i, \dots, X_{k})  \\
&\quad - \sum_{i<j} (-1)^{i+j} \, p \otimes \alpha_{k-1,l}([X_i, X_j], \dots, \hat X_i, \dots, \hat X_j, \dots) \\
&\quad - \sum_{i<j} (-1)^{i+j} \, p \otimes \alpha_{k-1,l}(\Lambda(X_i, X_j), \dots, \hat X_i, \dots, \hat X_j, \dots) \\
&\quad - \sum_{i<j} (-1)^{i+j} \, p \otimes (E_{X_i, X_j}, \alpha_{k-2,l-1})(\dots, \hat X_i, \dots, \hat X_j, \dots) \\
&\quad - \sum_{i<j} (-1)^{i+j} \, (\Lambda(X_i, X_j) \odot p,\alpha_{k-2,l+1})(\dots, \hat X_i, \dots, \hat X_j, \dots) \\
&\quad + p \otimes (S, \alpha_{k,l+1})(X_1, \dots, X_{k}) \ ,
\end{split}
\een
where $X_j \in \frak{h} \oplus \cG$, $\alpha_{k,l} \in {\rm Poly}^l(\C) \otimes
\wedge^k (\frak{h}^* \oplus \cG^*)$, and $p \in {\sf S}(\cG)$. If $p=Y_1 \odot \cdots \odot Y_n$, then the \lq contraction' appearing in the first line is defined by
\ben
\langle p, \delta \rangle = \sum_j Y_1 \odot \dots \hat Y_j \dots \odot Y_n \otimes \delta_{Y_j} \in
{\sf S}(\cG) \otimes \fX(\C) \ ,
\een
which is a vector field on $\C$ that gets multiplied with the rank $l-1$
polyvector field $\alpha_{k,l-1}$ in the first line to give a rank $l$ polyvector field.
If $X \in \frak{h} \oplus \cG$, then $\pi_X p$ denotes the natural action of $X$
on ${\sf S}(\cG)$.
Note that, unlike $\d_{\frak{h} | \cG}$,
$\Q$ depends on the action functional $S \in C^\infty(\C)$.
We then have:
\begin{lemma}\label{lemma6}
The differential $\Q:
\V^n \to \V^{n+1}$ is nilpotent.
\end{lemma}
{\em Proof:} We apply $\Q$ to \eqref{Qdef},
using the cyclic identities~\eqref{cyclE},~\eqref{cyclE0} for $E$, and~\eqref{vxlambda},~\eqref{cocyLa}
for $\Lambda$, the Jacobi identity for the Schouten-Nijenhuis bracket and the Lie bracket, the fact $\delta_X S=0$, and the algebra~\eqref{vxye}. We omit the
somewhat lengthy calculation. \qed

Let us end this subsection with the presentation of $\Q$ in local coordinates. In physics, the coordinate vector fields $\partial/\partial \phi^i$ are referred to as
\lq anti-fields', and are written $\hat \phi_i$. The generators of ${\sf S}(\cG)$ relative to a basis $\{X_a\}$ of $\cG$ are called $\hat c_a$ and referred to as \lq anti-ghosts'. Then on the ghosts and fields:
\ben\label{bhatdef}
\begin{split}
 \Q c^a &= -\half f^a_{bc} c^b c^c - \pi_{I}{}^a{}_b c^b \xi^I - \half \Lambda^a_{IJ}(\phi) \xi^I \xi^J \ , \\
\Q \xi^I &= -\half f^I_{JK} \xi^J \xi^K \ , \\
\Q \phi^i &= V^i_a(\phi) c^a + V^i_I(\phi) \xi^I - \half E^{ij}_{IJ}(\phi)\xi^I \xi^J \hat \phi_j \ , \end{split}
\een
where $\pi_{I}{}^b{}_a = f_{Ia}^b$, whereas on the anti-fields and anti-ghosts:
\ben\label{antibhat}
\begin{split}
\Q \hat \phi_i &= \partial_i S(\phi) - \tfrac{1}{4} \partial_i E^{kl}_{IJ}(\phi) \xi^I \xi^J \hat \phi_k \hat \phi_l - \partial_i V^j_a(\phi) c^a \hat \phi_j - \partial_i V^j_I(\phi) \xi^I \hat \phi_j + \half \hat c_a \partial_i \Lambda^a_{IJ}(\phi) \xi^I \xi^J \\
 \Q \hat c_a &=  V_a^i(\phi) \hat \phi_i - f^b_{ac} c^c \hat c_b - \pi_{I}{}^b{}_a \hat c_b \xi^I \ .
\end{split}
\een
The original action functional $S \in C^\infty(\C)$ is not invariant under $\Q$, but it turns out that one can write down, in a canonical fashion,
a modified action, which is. Unlike the original action, the modified action $\hat S=S+\frac{1}{4}E$ is
not simply a function on $\C$, but rather a more general element $\hat S  \in \V^0$, i.e. it depends on anti-fields, ghosts, etc. In components:
\ben\label{hatsmodel}
\hat S =
S(\phi) + \tfrac{1}{4} E^{ij}_{IJ}(\phi)
\xi^I \xi^J \hat \phi_i \hat \phi_j \ .
\een
That $\Q \hat S=0$ follows from a straightforward but rather lengthy calculation using~\eqref{cocyLa},~\eqref{vxlambda},~\eqref{cyclE},~\eqref{cyclE0},~\eqref{vxye}.


\subsection{Supersymmetric generalisation and gauge-fixing}

Let us finally describe a supersymmetric extension of the previous extended BRST structure. In this case, the space of field configurations $\C$ is not an ordinary manifold, but a supermanifold. Likewise, the Lie algebra $\frak{h}$ is replaced by a Lie superalgebra $\cS$ (concretely the conformal symmetry superalgebra in our application). There exist different mathematical definitions of the concept of a supermanifold. The one which is suitable for our purposes is that of de Witt-Rogers, see e.g.~\cite{rogers}. In that approach, a supermanifold $\C$ of bosonic dimension $n$ and fermionic dimension $m$ is locally modelled upon $\RR^{n|m}$, which is defined, roughly speaking, as $n$ Cartesian powers of the even part of an infinite-dimensional (real) Grassmann algebra, times $m$ powers of the odd part. A suitable inductive limit is understood here to deal with the infinite-dimensional nature of the Grassmann algebra, the somewhat subtle details of which are explained in \cite{rogers}.  A generic element of $\RR^{n|m}$ is written as $(\phi^i, \theta^A)$, where $\phi^i$ are even coordinates and $\theta^A$ odd coordinates. The bosonic coordinates are thus not real numbers, but just even elements of the Grassmann algebra. This seemingly too large space for the bosonic coordinates is compensated in the deWitt-Rogers~setting by a correspondingly restrictive notion of smooth function on $\RR^{n|m}$, the space of which is called $GH^\infty(\RR^{n|m})$. On functions of class
$GH^\infty(\RR^{n|m})$, one can unambiguously
define the derivative operators $\partial/\partial \phi^i$ and $\partial/\partial \theta^A$ with respect to even and odd coordinates in a consistent fashion. Vector fields are then
defined
\footnote{The symbol $\partial/\partial \theta^A$ denotes the \lq left' derivative.} %
as first order derivative operators with coefficient functions in $GH^\infty(\RR^{n|m})$. Any vector field can naturally be decomposed into even and odd part. The commutator of two odd vector fields is always zero, so one replaces this notion by its anti-commutator. With this convention understood, the vector field (super) commutator satisfies the graded Jacobi identity. The notion of a smooth (i.e. $GH^\infty$) supermanifold $\C$ modelled on $\RR^{n|m}$ is then defined by declaring what one means by open sets, an atlas etc. (see \cite{rogers} for more details).
The notion of a polyvector field is generalised in such a way that the indices for the odd coordinates are symmetric rather than skewsymmetric. The Schouten-Nijenhuis bracket is defined accordingly.

In the supersymmetric setting, the Lie algebra $\frak{h}$ is replaced with the Lie superalgebra $\cS$, which has a graded decomposition $\cS = \cB \oplus \cF$ into an even (bosonic) part $\cB$ and an odd (fermionic) part $\cF$. $\cS \oplus \cG \owns X \mapsto \delta_X \in \fX(\C)$ is now a linear map respecting the even/odd grading. However, since
 $\cG$ is to remain a Lie algebra, we can no longer combine it with $\cS$ into a single Lie superalgebra. Therefore, we are in a somewhat more general situation than above, in the sense that the bracket $[\cG,\frak{h}] \subset \cG$ above must be replaced by a an action $\pi: \cS \to {\rm End}(\cG)$. $\pi$ {\em cannot} be a representation, so it has a \lq curvature',
\ben\label{curvatureend}
R_\pi (X,Y) = [\pi(X),\pi(Y)]-\pi([X,Y]) \neq 0 \ , \qquad X,Y \in \cS \ .
\een
Here, and from now on, all brackets are understood in the graded sense.
The vector fields are to satisfy the following analog of relation~\eqref{vxye}:
\ben\label{vxye1}
\begin{split}
[\delta_X, \delta_Y] &= \delta_{\pi(X)Y} \ , \quad \text{if $X \in \cS, Y \in \cG$}, \\
[\delta_X, \delta_Y] &= \delta_{[X,Y]} + \delta_{\Lambda(X,Y)} + (E_{X,Y},S) \ , \quad \text{if $X,Y \in \cS$}, \\
[\delta_X, \delta_Y] &= \delta_{[X,Y]} \ , \quad \text{if $X,Y \in \cG$} \ .
\end{split}
\een
$E:\cS \times \cS \to {\rm Poly}^2(\C)$
in \eqref{vxye} is now a map which, instead of the skewsymmetry property in \eqref{esym}, has a graded symmetry (i.e. symmetric if both entries are in $\cF$ and skewsymmetric otherwise). Likewise, $\Lambda: \cS \times \cS \to GH^\infty(\C, \cG)$ now also has a graded symmetry, and is in the class $GH^\infty$ of functions.  The consistency relation satisfied by $\Lambda$ is now (replacing \eqref{vxlambda})
\ben\label{vxlambda1}
R_\pi(Y,Z) X + \delta_X \Lambda(Y,Z) + [X,\Lambda(Y,Z)] = 0 \ .
\een
In the graded setting, we define the set of \lq chains' $\V^n$ similarly as before,
with the difference that the factor $\wedge^k (\frak{h}^* \oplus \cG^*)$ corresponding to the
anti-commuting ghosts must now be replaced by a tensor factor whose symmetrisation properties
reflect the fact that, for $\frak{h} = \cS$, the ghosts in $\cB$ are still anti-commuting,
while those in $\cF$ are commuting.
The differential $\Q$ is finally defined by a formula identical
to that given before, but where the various brackets are now to be understood in a graded sense. A local formula for $\Q$ can again be given. Compared to the bosonic case, there are
now also fermionic coordinates $\theta^A$, and corresponding anti-fields $\hat \theta_A
= \partial/\partial \theta^A$.

The final complication is that $\C$ is infinite-dimensional in our case, since it is a space of field configurations. Here, one has to be very careful, in principle, what topology one wishes to take for $\C$, what precise notion of smoothness, etc. In general, it seems that infinite-dimensional manifolds modelled over locally convex spaces are best (see~\cite{reijzner} for a discussion of these issues). However, we will find that such subtleties have little impact on our specific application below since all functionals are only polynomial, and hence smooth under any reasonable notion of smoothness.

{\bf Gauge-fixing.} To perform the gauge-fixing procedure below, it is necessary to further enlarge the setting by introducing the \lq $(B,\overline{c})$-system'. Abstractly, this system is constructed as follows from the data $\cS, \cG$ and the map $\pi: \cS \to
{\rm End}(\cG)$.
Relative to a basis $X_a$, $a=1, ..., \dim\, \cG$, and a basis $X_I$, $I=1,..., \dim\, \cS$ we introduce new generators by $\overline{c}^a, B^a$ and we denote by
$\xi^I$ the ghosts which were defined previously. Then we extend $\Q$ by
\ben\label{bcbar}
\begin{split}
\Q \overline{c}^a &= B^a + \pi_{Ib}{}^a \xi^I \overline{c}^b \\
\Q B^a &= \pi_{Ib}{}^a \xi^I B^b - \half R_{IJb}{}^a \xi^I \xi^J \overline{c}^b \ ,
\end{split}
\een
where $R_{IJb}{}^a = \pi_I{}_b{}^c \pi_J{}_c{}^a \pm \pi_J{}_b{}^c \pi_I{}_c{}^a -
f_{IJ}^K \pi_K{}_b{}^a$ are the components of the curvature endomorphism~\eqref{curvatureend}, and the plus sign is chosen only if both $\xi^I$ and $\xi^J$ are commuting. The definition is made so that $\Q$ is still nilpotent\footnote{Here it is helpful to note the \lq Bianchi identity'
$\pi(X) R_\pi(Y,Z) - R_\pi(X,Y) \pi(Z) + R_\pi([Y,Z],X) +$ cyclic permutations $=0$.}. The difference from the
more commonly used gauge-fixing system is the presence of this curvature.
The gauge-fixed action is defined as $\hat S \to \hat S + \Q \eG$ for a conveniently chosen element of $\eG \in \V^{-1}$, sometimes called the \lq gauge fermion'. An associated modified differential is given by
\ben
\label{qhatdef}
\hat \Q = \Q - (\Q {\mathscr G}, -) = \e^{(\eG, -)} \circ \Q \circ \e^{-(\eG,-)} \ .
\een
It differs from $\Q$ only by a term that is \lq pure gauge', hence the cohomology of $\hat \Q$ is isomorphic to that of $\Q$ under conjugation by $\e^{(\eG,-)}$.


\section{BRST structure of rigid supermultiplets in curved spacetime}
\label{sec:BRSTNequals2}


\subsection{Supermultiplets and Lagrangians}

Let us now describe the formulation of rigid $\eN =2$ conformal supermultiplets on a Lorentzian four-manifold $\eM$ that admits a twistor spinor. The structure of the associated conformal symmetry superalgebra $\cS$ was described explicitly in section~\ref{sec:N2d4}. The rigid $\eN =2$ conformal supermultiplets are defined by an action of $\cS$ and, for each supermultiplet, we shall obtain a gauge-invariant Lagrangian that is invariant under $\cS$, up to boundary terms. If $\eM$ is conformally flat, then $\cS \cong \fsu(2,2|2)$ and we recover the well-known description of $\eN =2$ vector and hyper multiplets on $\RR^{3,1}$. Rather remarkably, the description of these supermultiplets on $\eM$ follows by applying a straightforward procedure of conformal coupling to the supersymmetry variations and Lagrangians in Minkowski space.


\subsubsection{Symmetry transformations}
\label{sec:symmetrytransform}
The action $\delta_X$ of any conformal Killing vector $X \in \cB$ on a field $\Phi$ with Weyl weight $w_\Phi$ is given by
\begin{equation}
\label{eq:deltaX}
\delta_X \Phi = \cL_X \Phi + w_\Phi \sigma_X \Phi~,
\end{equation}
in terms of the Lie derivative $\cL_X$ along $X$ and the real function $\sigma_X = -\tfrac{1}{4} \nabla_\mu X^\mu$. For example, the background metric $g$ has $w_g =2$ and $\delta_X g =0$. In general, $w_\Phi$ is defined as the tensorial rank of $\Phi$ minus its canonical dimension (a spinor has tensorial rank zero).

The R-symmetry generated by $\cS$ is $\cR \cong \fu(2)$, with parameters that are constant on $\eM$. It is convenient to write $\fu(2) \cong \fusp(2) \oplus \fu(1)$, where $\fusp(2) = \fu(2) \cap \fsp_2 (\CC) \cong \fsu(2)$. Relative to a basis $\{ {\bm{e}}_A \}$ on $\CC^2$, let $u^A$ denote the components of a complex vector $u$ which transforms in the fundamental representation of $\fusp(2)$ acting on $\CC^2$. Identifying the complex conjugate with the dual of any such vector, the components of $u^*$ are written $u^*_A$ (whereby $u^A u^*_A$ is $\fu(2)$-invariant). With respect to this basis, let $\varepsilon_{AB}$ denote the components of the $\fsp_2 (\CC)$-invariant symplectic form ${\bm e}^1 \wedge {\bm e}^2$ on $\CC^2$. Indices can be raised and lowered using this tensor such that $u_A  = \varepsilon_{AB} \, u^B$ (and $u^A = u_B \, \varepsilon^{BA}$ via the identity $\varepsilon_{AC} \varepsilon^{BC} = \delta_A^B$). A second rank symmetric tensor $w$ on $\CC^2$ obeying the reality condition $( w^{AB} )^* = \varepsilon_{AC} \varepsilon_{BD} w^{CD}$ corresponds to the adjoint representation of $\fusp(2)$. We shall denote the elements in $\cR$ by $\rho = ( {\hat \rho} , \rho^\prime )$, where ${\hat \rho} \in \fusp(2)$ and $\rho^\prime \in \fu(1)$.

The action $\delta_\rho$ of any constant $\rho \in \cR < \cB$ on a field $\Phi$ in some representation of $\fusp(2)$ and with $\fu(1)$ R-charge $r_\Phi$ is given by
\begin{equation}
\label{eq:deltarho}
\delta_\rho \Phi = {\hat \rho} \cdot \Phi + r_\Phi \rho^\prime \Phi~,
\end{equation}
where $\cdot$ denotes the $\fusp(2)$-action defined by the representation $\Phi$ is in. For example, if $\Phi$ is in the fundamental representation of $\fusp(2)$, then $( {\hat \rho} \cdot \Phi )^A = {\hat \rho}^A{}_B \Phi^B$.

Let $\fg$ be a real Lie algebra and let $\cG = C^\infty(\M, \fg)$ denote the space of all $\fg$-valued functions, which generate infinitesimal gauge transformations on $\eM$. For any $\Lambda \in \cG$, the action of an infinitesimal gauge transformation $\delta_\Lambda$ on a $\fg$-valued gauge field $A_\mu$ or on a scalar or spinor field $\Phi$ valued in some $\fg$-module $U$ is given by
\ben
\label{eq:gaugetransform}
\delta_\Lambda A_\mu = D_\mu \Lambda \ , \quad\quad
\delta_\Lambda \Phi = - \Lambda \cdot \Phi~,
\een
where $\cdot$ in \eqref{eq:gaugetransform} denotes the $\fg$-action for $U$. If $U = \fg$ is the adjoint representation then the $\fg$-action is defined by the Lie bracket on $\fg$, whence $\delta_\Lambda \Phi = [ \Phi , \Lambda]$.

Having defined the field-theoretic action of bosonic symmetries generated by $\cB = \fX^c (M) \oplus \cR$ and $\cG$, let us now turn to the fermionic symmetries generated by $\cF$. The supersymmetry parameter $\epsilon \in \cF$ corresponds to a pair of bosonic Majorana twistor spinors $( \epsilon_+^A , \epsilon_{-\; A} )$, both with Weyl weight $\half$. The components $\epsilon_+^A$ have positive chirality and transform in the fundamental representation of $\fusp(2)$, with $\fu(1)$ R-charge $+1$. The components $\epsilon_{-\; A}$ have negative chirality and transform in the anti-fundamental representation of $\fusp(2)$, with $\fu(1)$ R-charge $-1$. The Majorana condition relates chiral projections via complex conjugation such that $\epsilon_{-\; A} = ({\sf B} \epsilon^A_+)^*$. For any $\epsilon \in \cF$, the action of the supersymmetry variation $\delta_\epsilon$ on fields is fermionic (i.e. it acts as a Grassmann-odd derivation on fields in the supermultiplet). We will now provide a detailed account of the relevant (on-shell) supermultiplets.


\subsubsection{Vector multiplet}
\label{sec:vector}
All fields in the vector multiplet are $\fg$-valued. The bosonic sector contains a gauge field $A_\mu$ and a complex scalar $\varphi$. The fermionic sector contains a pair of Majorana spinors $( \lambda_+^A , \lambda_{-\; A} )$, with $\lambda_{-\; A} = ({\sf B} \lambda^A_+)^*$.

The fields $( A_\mu , \lambda_+^A , \varphi   )$ have Weyl weights $(0,-\tfrac{3}{2},-1)$ and $\fu(1)$ R-charges $(0,1,2)$. The bosonic fields $A_\mu$ and $\varphi$ are $\fusp(2)$-invariant while $\lambda_+^A$ transforms in the fundamental representation of $\fusp(2)$. Plugging this data into \eqref{eq:deltaX} and \eqref{eq:deltarho} specifies the action of $\cB$ on the vector multiplet.

The action of $\cF$ on the vector multiplet is defined by the supersymmetry variations
\begin{align}
\label{eq:4dsusyvectorcurved}
\delta_\epsilon A_\mu &= {\overline \epsilon}_{-\; A} \Gamma_\mu \lambda_+^A + {\overline \epsilon}_+^A \Gamma_\mu \lambda_{-\; A} \nonumber \\
\delta_\epsilon \varphi &= {\overline \epsilon}_+^A  \lambda_{+\; A} \\
\delta_\epsilon \lambda_+^A &= -{\slashed F} \epsilon_+^A -2\, [\varphi,\varphi^*] \epsilon_+^A + 2 \, {\slashed D} \varphi \,\epsilon_-^A
+ \varphi {\slashed \nabla} \epsilon_-^A~, \nonumber
\end{align}
for all $\epsilon \in \cF$. The variations in \eqref{eq:4dsusyvectorcurved} are Weyl-covariant and can be obtained by simply conformally coupling their counterparts in Minkowski space.

For convenience, let us assume now that $\fg$ is simple with Killing form $(-,-)$. The conformally coupled Lagrangian for the on-shell vector multiplet on $\eM$ is given by
\begin{align}\label{eq:4dSYMlagNequals2curved}
\eL_{\rm V}=& -\tfrac{1}{4} ( F_{\mu\nu} , F^{\mu\nu} ) -2 ( D_\mu \varphi , D^\mu \varphi^* ) - \tfrac{1}{3} R ( \varphi , \varphi^* ) + 2 ( [\varphi,\varphi^*] ,  [\varphi,\varphi^*] )
\nonumber \\
&-\half ( {\overline \lambda}_{-\; A} , {\slashed D} \lambda_+^A ) -\half ( {\overline \lambda}_+^A , {\slashed D} \lambda_{-\; A} ) + (\varphi , [ {\overline \lambda}_-^A , \lambda_{-\; A} ] ) + (\varphi^* , [ {\overline \lambda}_+^A , \lambda_{+\; A} ] )~.
\end{align}
This Lagrangian is manifestly invariant under both $\cG$ and $\cR$. Being conformally coupled, the integral of $\eL_{\rm V}$ on $\eM$ is also Weyl-invariant. Whence, it is $\cB$-invariant.

Furthermore, up to boundary terms, $\eL_{\rm V}$ is also $\cF$-invariant under the supersymmetry variations \eqref{eq:4dsusyvectorcurved}. The proof of this statement follows using the Fierz identities
\begin{align}\label{eq:oddoddoddjacobiNequals2}
\epsilon_+^A {\overline \epsilon}_-^B &= \half ( {\overline \epsilon}_-^B \Gamma^\mu \epsilon_+^A ) \Gamma_\mu {\sf P}_- \; , \quad\quad \Gamma^{\mu\nu} \epsilon_+^A {\overline \epsilon}_-^B \Gamma_{\mu\nu} = 0 \; , \quad\quad \epsilon_{+\; A} {\overline \epsilon}_-^A = -\tfrac{1}{4} {\slashed \xi}_\epsilon {\sf P}_- \nonumber \\
\epsilon_+^A {\overline \epsilon}_+^B &= -\tfrac{1}{4} \left( \varepsilon^{AB} \, ( {\overline \epsilon}_+^C  \epsilon_{+\; C} ) + \half (  {\overline \epsilon}_+^A \Gamma^{\mu\nu}  \epsilon_+^B ) \Gamma_{\mu\nu}  \right) {\sf P}_+ \; , \quad\quad \epsilon_{+\; A} {\overline \epsilon}_+^A = \tfrac{1}{2} ( {\overline \epsilon}_+^A  \epsilon_{+\; A} ) {\sf P}_+  \nonumber \\
\Gamma^{\mu\nu} \epsilon_+^A {\overline \epsilon}_+^B \Gamma_{\mu\nu} &= \left( 3 \varepsilon^{AB} \, ( {\overline \epsilon}_+^C  \epsilon_{+\; C} ) - \half (  {\overline \epsilon}_+^A \Gamma^{\mu\nu}  \epsilon_+^B ) \Gamma_{\mu\nu}  \right) {\sf P}_+~,
\end{align}
derived from~\eqref{Fierzgeneral}, together with the fact that $\epsilon$ is a twistor spinor, which implies $\nabla_\mu {\slashed \nabla} \epsilon^A_+ = -(R_{\mu\nu} - \tfrac{1}{6} g_{\mu\nu} R ) \Gamma^\nu \, \epsilon^A_+$ using \eqref{consistency}. One also requires the Lichnerowicz-Weitzenb\"{o}ck identity ${\slashed D}^2 \psi = D^2 \psi + \tfrac{1}{4} R \psi + {\slashed F} \, \psi$, which is valid for any $\fg$-valued spinor $\psi$ on $\eM$.


\subsubsection{Hyper multiplet}
\label{sec:hyper}
Let $U$ be a complex representation of $\fg$ and let $U^*$ denote its complex conjugate representation. The fermionic sector of the hyper multiplet contains a $U\oplus U^*$-valued chiral spinor $\psi_+^A$ which transforms in the fundamental representation of $\fusp(2)$ (i.e. $\psi_+^1$ is $U$-valued and $\psi_+^2$ is $U^*$-valued). It is convenient to define the $U^* \oplus U$-valued anti-chiral spinor $\psi_{-\; A} = ({\sf B} \psi^A_+)^*$. The bosonic sector contains scalars $\phi^{AB}$, subject to the reality condition $(\phi^{AB} )^* = \varepsilon_{AC} \varepsilon_{BD} \phi^{CD}$, which transform in the bifundamental representation of $\fusp(2)$. The component $\phi^{11} = ( \phi^{22} )^*$ is $U$-valued and the component $\phi^{12} = - ( \phi^{21} )^*$ is $U^*$-valued.

The fields $( \phi^{AB} , \psi_+^A  )$ have Weyl weights $(-1,-\tfrac{3}{2})$ and $\fu(1)$ R-charges $(0,-1)$. This data specifies the action of $\cB$ on the hyper multiplet, using \eqref{eq:deltaX} and \eqref{eq:deltarho}.

The action of $\cF$ on the coupled on-shell vector and hyper multiplet is defined by the supersymmetry variations
\begin{align}
\label{eq:4dsusyhyper}
\delta_\epsilon A_\mu &= {\overline \epsilon}_{-\; A} \Gamma_\mu \lambda_+^A + {\overline \epsilon}_+^A \Gamma_\mu \lambda_{-\; A} \nonumber \\
\delta_\epsilon \varphi &= {\overline \epsilon}_+^A  \lambda_{+\; A} \nonumber \\
\delta_\epsilon \lambda_+^A &= -{\slashed F} \epsilon_+^A -2\, [\varphi,\varphi^*] \epsilon_+^A + 2 \, {\slashed D} \varphi \,\epsilon_-^A + \varphi {\slashed \nabla} \epsilon_-^A - \phi^{AB}  \phi_{CB} \, \epsilon_{+}^C \\
\delta_\epsilon \phi^{AB} &= {\overline \epsilon}_+^A \psi_+^B +
{\overline \epsilon}_-^A \psi_-^B \nonumber \\
\delta_\epsilon \psi_+^A &= 2\, {\slashed D} \phi^{BA} \, \epsilon_{-\; B} +
 \phi^{BA} {\slashed \nabla} \epsilon_{-\; B}
+4 \varphi^* \cdot \phi^{BA} \, \epsilon_{+ \; B}~, \nonumber
\end{align}
for all $\epsilon \in \cF$. The on-shell variations in \eqref{eq:4dsusyhyper} are Weyl-covariant and can be obtained by simply conformally coupling their counterparts in Minkowski space.

The conformally coupled Lagrangian for the hyper multiplet on $\eM$ is given by
\begin{align}
\label{eq:4dMlagNequals2}
\eL_{\rm H} =& - \half D_\mu \phi^{AB} D^\mu \phi_{AB}  -
\tfrac{1}{12} R \, \phi^{AB} \phi_{AB}
-\half   {\overline \psi}_{-\; A}  {\slashed D} \psi_+^A \nonumber \\
&+ 2 ( \varphi^* \cdot \phi^{AB} )( \varphi \cdot \phi_{AB} ) + \half {\overline \psi}_+^A  ( \varphi \cdot \psi_{+ \; A} ) -  \half {\overline \psi}_-^A  ( \varphi^* \cdot \psi_{- \; A} ) \\
&
+ \tfrac{1}{4} \phi_{CD} \phi^{CB}  \phi_{AB} \phi^{AD}
+ ( {\overline \lambda}_+^A \cdot \psi_+^B ) \, \phi_{AB} - ( {\overline \lambda}_{-\; A} \cdot \psi_{-\; B} ) \, \phi^{AB}~. \nonumber
\end{align}
To avoid clutter, we have omitted in $\eL_{\rm H}$ the obvious inner products with respect to the canonical dual pairing between $U$ and $U^*$. The Lagrangian \eqref{eq:4dMlagNequals2} is manifestly invariant under both $\cG$ and $\cR$. Being conformally coupled, the integral of $\eL_{\rm H}$ on $\eM$ is also Weyl-invariant. Whence, it is $\cB$-invariant.

Moreover, up to boundary terms, the full on-shell Lagrangian $\eL_{\rm V} + \eL_{\rm H}$ is $\cF$-invariant under the supersymmetry variations \eqref{eq:4dsusyhyper}. The proof follows by direct calculation.


\subsection{BRST operator}
\label{sec:QSYM}

The action of the conformal symmetry superalgebra $\cS$ on the super-manifold $\C$ of smooth field configurations $\Phi = ( A_\mu , \lambda^A_+, \varphi, \phi^{AB}, \psi^A_+)$ defined in the previous subsection is not a representation. Instead, it turns out that we have a structure identical to that
described in \eqref{vxye1}. Indeed, by evaluating the
graded commutators of superconformal symmetry transformations $\delta_X, \delta_\rho, \delta_\epsilon$, and gauge transformations $\delta_\Lambda$, with parameter $\Lambda \in \cG=C^\infty(\M, \fg)$, one finds that
the map $E: \cS \times \cS \to {\rm Poly}^2(\C)$ is given by
\ben\label{Edef}
\begin{split}
E_{\epsilon,\epsilon} &= 2
\int_\M \dv \left( \varepsilon^{C(A} \varepsilon^{B)D} \Big(
\epsilon_{-\; C} \frac{\delta }{\delta \lambda_-^D} +
\epsilon_{+\; C} \frac{\delta }{\delta \lambda_+^D}
 \Big)
\Big(
\epsilon_{+\; A}\frac{\delta }{\delta \lambda^B_+} +
\epsilon_{-\; A}\frac{\delta }{\delta \lambda^B_-}
\Big) \right. \\
&\quad\quad\quad\quad\quad + \left.
2(\overline \epsilon_+^B \Gamma_\mu \epsilon_{-\; B})
\frac{\delta}{\delta \psi_{-}^A} \Gamma^\mu \frac{\delta }{\delta \overline \psi_{+\; A}}
- (\overline \epsilon_-^B \epsilon_{-\; B}) \frac{\delta}{\delta \psi_+^A} \frac{\delta}{
\delta \overline \psi_{+\; A}}
- (\overline \epsilon_+^B \epsilon_{+\; B}) \frac{\delta}{\delta \psi_-^A} \frac{\delta}{
\delta \overline \psi_{-\; A}} \right)~,
\end{split}
\een
where $\dv = \sqrt{|g|} \, \d^4 x$ denotes the canonical volume form on $(\eM , g)$. For different entries, $E$ is defined by polarisation $E_{\epsilon,\epsilon'} = \half(E_{\epsilon+\epsilon',\epsilon+\epsilon'}-E_{\epsilon,\epsilon}-
E_{\epsilon',\epsilon'})$. All other components vanish identically (e.g. $E_{X,Y} =0$, for all $X,Y \in \fX^c(\M)$). Similarly, one finds the map $\Lambda :
\cS \times \cS \to GH^\infty(\C, \cG)$ is given by
 \ben
 \Lambda_{\epsilon,\epsilon} =  -2(\overline \epsilon_{-\; A} \Gamma^\mu
 \epsilon_+^A) A_\mu -4\, {\mathrm{Re}} \left( ( {\overline \epsilon}_-^A  \epsilon_{-\; A} ) \, \varphi \right) \ .
 \een
All other components are again determined by polarisation or vanish identically.
The explicit form of the map $\pi: \cS \to {\rm End}(\cG)$ is given by $\pi_X \Lambda = \cL_X \Lambda$, $\pi_\epsilon \Lambda = \pi_\rho \Lambda = 0$.

The general formalism described in section~\ref{sec:brstextension} provides a nilpotent differential $\Q$, which we now present explicitly. Gauge parameters in $\cG$ are assigned a fermionic ghost $c$. Conformal Killing vectors, twistor spinors and $\fu(2)$ R-symmetries in $\cS$ are assigned ghosts $X$, $\epsilon^A_+$ and $\alpha_B{}^A$. The ghosts $X$ and $\alpha_B{}^A$ are fermionic while $\epsilon_+^A$ is bosonic. We also define $\alpha^{AB}= \alpha_C{}^B \varepsilon^{CA}$, etc.  The components $\alpha= \alpha_A{}^A$ and $\alpha^{(AB)}$ describe ghosts for the $\fu(1)$ and $\fusp(2)$ R-symmetry factors. Each field $\Phi$ has anti-field ${\hat \Phi}$.

For convenience, we summarise the relevant data in Table~\ref{table2} for fields and ghosts and in Table~\ref{table3} for the $(B, \overline c)$ system.
\begin{table}
\begin{center}
\begin{tabular}{|c||ccccc|c||cccc|}
\hline
Field & $A_\mu$ & $\lambda^A_+$ & $\varphi$ & $\psi^A_+$ & $\phi^{AB}$ & Ghost &
$\alpha^{AB}$ & $\epsilon^A_+$ & $X$ & $c$ \\
\hline
Dimension & 1 & $\tfrac{3}{2}$ & 1 &  $\tfrac{3}{2}$ & 1 &  & 0 & $-\half$ & $-1$ & 0\\
Ghost number & 0 & 0 & 0 & 0 & 0 & & 1 & 1 & 1 & 1 \\
Spin & 1 & $\tfrac{1}{2}$ & 0 &  $\tfrac{1}{2}$ & 0 & & 0 & $\half$ & $1$ & 0\\
Grassmann parity & 0 & 1 & 0 & 1 & 0 & & 1 & 0 & 1 & 1\\
Weyl weight & 0 & $-\tfrac{3}{2}$ & $-1$ & $-\tfrac{3}{2}$ & $-1$ &  & 0 & $\half$ & 0 & 0\\
$\frak{u}(1)$ R-charge & 0 & $1$ & $2$ & $-1$ & 0 & & 0 & $1$ & 0 & 0 \\
Dynamical & yes & yes & yes & yes & yes & & no & no & no & yes\\
\hline
\end{tabular}
\vspace{.5cm}
\caption{Data for fields and ghosts.}
\label{table2}
\end{center}
\end{table}
\begin{table}
\begin{center}
\begin{tabular}{|c||cc|}
\hline
Field & $B$ & ${\bar c}$ \\
\hline
Dimension & 2 & 2 \\
Ghost number & $0$ & $-1$ \\
Spin & 0 & 0 \\
Grassmann parity & 0 & 1 \\
Weyl weight & $-2$ & $-2$ \\
$\frak{u}(1)$ R-charge & 0 & 0 \\
Dynamical & yes & yes \\
\hline
\end{tabular}
\vspace{.5cm}
\caption{Data for the $(B, \overline c)$ system.}
\label{table3}
\end{center}
\end{table}
This also fixes the data for the corresponding anti-fields. By construction, a field $\Phi$ and its anti-field $\hat \Phi$ have opposite Grassmann parity.
\footnote{The Grassmann parity $p$ of a field is defined as $p=0$ if it is bosonic (commuting) or $p=1$ if it is fermionic (anti-commuting).}
If $\Phi$ has dimension $\Delta_\Phi$, ghost number ${\rm g}_\Phi$, Weyl weight $w_\Phi$ and $\fu(1)$ R-charge $r_\Phi$ then, by definition, $\hat \Phi$ has $\Delta_{\hat \Phi} = 4 - \Delta_\Phi$, ${\rm g}_{\hat \Phi} = -1- {\rm g}_\Phi$, $w_{\hat \Phi} = -4 - w_\Phi$ and $r_{\hat \Phi} = - r_\Phi$. 

Using the data in Table~\ref{table2}, the action of $\Q$ derived from the first equation in~\eqref{bhatdef} on the vector and hyper multiplet fields is given by
\ben
\begin{split}\label{b1}
\Q A_\mu &= D_\mu c + \cL_X A_\mu + {\overline \epsilon}_{-\; A} \Gamma_\mu \lambda_+^A + {\overline \epsilon}_+^A \Gamma_\mu \lambda_{-\; A}\ , \\
\Q \lambda^A_+ &= [\lambda^A_+,c]  + \alpha \lambda^A_+ - \alpha^{(BA)} \lambda_{+B}
+ (\cL_X - \tfrac{3}{2} \sigma_X) \, \lambda^A_+ \\
&\quad -{\slashed F} \epsilon_+^A -2\, [\varphi,\varphi^*] \epsilon_+^A + 2 \, {\slashed D} \varphi \,\epsilon_-^A
+ \varphi {\slashed \nabla} \epsilon_-^A +
2 (\epsilon^{(A}_+\hat \lambda_+^{B)}+\epsilon^{(A}_- \hat
\lambda_-^{B)})  \epsilon_{B+} \ , \\
\Q \varphi &= [\varphi,c] + 2\alpha \varphi + (\cL_X - \sigma_X) \varphi +
\varepsilon_{AB} \, {\overline \epsilon}_+^A  \lambda_+^B
 \ , \\
\Q \psi_+^A &= [\psi^A_+, c] - \alpha \psi_+^A - \alpha^{(BA)} \psi_{B+} +
(\cL_X - \tfrac{3}{2}\sigma_X) \psi^A_+
 +\\
&\quad +
2\, {\slashed D} \phi^{BA} \, \epsilon_{-\; B} +
 \phi^{BA} {\slashed \nabla} \epsilon_{-B}
+4\, \varphi^* \cdot \phi^{BA} \, \epsilon_{+ \; B}
- {\slashed \xi} \hat \psi_-^A - 2 (\overline \epsilon_{-B} \epsilon^B_-) \hat \psi^A_+ \ , \\
\Q \phi^{AB} &= [\phi^{AB}, c] - \alpha^{(AC)} \phi_C{}^B
 -\alpha^{(BC)} \phi^{A}{}_C + (\cL_X - \sigma_X) \phi^{AB}   + {\overline \epsilon}_+^A \psi_+^B +
{\overline \epsilon}_-^A \psi_-^B~,
\end{split}
\een
where $\xi^\mu = 2\, {\overline \epsilon}_{-\; A} \Gamma^\mu \epsilon_+^A$. Using the brackets \eqref{eq:SuperBracket2} for $\cS$, one also obtains from \eqref{bhatdef} the action of $\Q$ on ghosts:
\ben
\begin{split}\label{b2}
 \Q c &= -\half [c,c]- \cL_X c + \xi^\mu A_\mu +4\, {\mathrm{Re}} \left( ( {\overline \epsilon}_-^A  \epsilon_{-\; A} ) \, \varphi \right) \ , \\
\Q  X &= -\half [X,X] - \xi \ , \\
\Q \epsilon^A_+ &= -(\cL_X + \tfrac{1}{2} \sigma_X) \, \epsilon^A_+ +
\alpha \epsilon_+^A + \alpha^{(AB)} \epsilon_{+B} \ , \\
\Q \alpha^{AB} &= -\alpha^{AC} \alpha^{DB} \varepsilon_{CD} -
({\overline \epsilon}_+^{(A} {\slashed \nabla} \epsilon_-^{B)} - {\overline \epsilon}_-^{(A} {\slashed \nabla} \epsilon_+^{B)}) \\
&\quad + \tfrac{1}{2} \varepsilon^{AB} ( {\overline \epsilon}_+^{C} {\slashed \nabla} \epsilon_{-\; C} - {\overline \epsilon}_{-\; C} {\slashed \nabla} \epsilon_+^C ) \ .
\end{split}
\een
Using \eqref{antibhat}, we can derive the action of $\Q$ on anti-fields. The action on bosonic anti-fields is given by
\ben
\begin{split}\label{b3}
\Q \hat \lambda_{+\; A} &= -{\slashed D} \overline \lambda_{-\; A} - [\varphi^*, \overline \lambda_{+\; A}] + \overline \psi^B_+ \phi_{AB}\\
&\quad +(\cL_X - \tfrac{5}{2} \sigma_X) \hat \lambda_{+\; A} + [\hat \lambda_{+\; A}, c] - \alpha \hat \lambda_{+\; A} - \alpha_{(AB)} \hat \lambda_{+}^B - \hat \varphi \overline \epsilon_{+\; A} + \hat A^\mu \overline \epsilon_{-\; A} \Gamma_\mu
 \ , \\
\Q \hat \psi_{+\; A} &= -\half {\slashed D} \overline \psi_{-\; A}+
\varphi \cdot \overline \psi_{+\; A} + \overline \lambda^B_+ \cdot \phi_{BA} \\
&\quad +(\cL_X - \tfrac{5}{2} \sigma_X) \hat \psi_{+\; A} - [\hat \psi_{+\; A}, c] + \alpha
\psi_{+\; A} - \alpha_{(AB)} \psi^B_+ + \hat \phi_{BA} \overline \epsilon^B_+~.
\end{split}
\een
Similar expressions follow for the action of $\Q$ on the fermionic anti-fields $( {\hat A}_\mu , \hat \varphi, \hat \phi^{AB} )$ and anti-ghost $\hat c$, which we omit since their explicit form is not required.

For the $(B, \overline c)$ system, using the data in Table~\ref{table3}, the action of $\Q$ follows from \eqref{bcbar} and is given by
\ben\label{b4}
\begin{split}
\Q \overline c &= (\cL_X - 2\sigma_X) \overline c + B~, \\
\Q B &= -(\cL_\xi - 2 \sigma_\xi) \overline c + (\cL_X - 2 \sigma_X) B \ .
\end{split}
\een

These transformations reduce to those of \cite{Maggiore:1994dw} in Minkowski space, after restricting to the $\eN=2$ Poincar\'{e} superalgebra in $\cS\cong \fsu(2,2|2)$. For $U = \fg_\CC$, in Minkowski space, they also recover the transformations of \cite{White:1992wu} with respect to the conformal superalgebra $\fsu(2,2|2) < \fsu(2,2|4)$.

 It follows that,
\begin{corollary}
$\;$ \\ [.1cm]
$\bullet$ $\Q$ is nilpotent. \\
$\bullet$ If $F$ is a Weyl-invariant functional of the fields, then so is $\Q F$. \\
$\bullet$ $\Q$ leaves invariant the dimension and R-charge, and adds one to the ghost number.
\end{corollary}
\noindent
{\em Proof:} $\Q^2=0$ follows by the construction described in
section~\ref{sec:brstextension}. Weyl-invariance of $F$ follows because the action of $\Q$ is Weyl-equivariant. The remaining properties follow using the data in Tables~\ref{table2} and~\ref{table3}.
\qed


\subsection{BRST cohomology}

The nilpotent differential $\Q$ has an associated cohomology ring, defined by the vector space of functionals of fields in the kernel of $\Q$, modulo those in the image of $\Q$. More precisely, on $(\eM,g)$, we shall consider functionals of fields taking values in the exterior algebra $\wedge^\bullet ( T\M )$. The form-valued functionals we are concerned with -- such as the Lagrangian density, counterterms in renormalized perturbation theory (see section~\ref{sec:pert}) or composite operators -- have a specific \lq local covariant' dependence on the fields. This characteristic is very important, and is therefore built into the next definition.
\begin{definition}
For any $p,q \in \ZZ_{\geq 0}$, $\P^p_q (\M)$ is defined as the space of all $\wedge^p (T\eM )$-valued, polynomial, local covariant functionals $\O$ with ghost number $q$ on the space of fields $\Phi$, which comprise the metric $g$, the vector and hyper multiplet fields $( A_\mu , \varphi, \lambda^A_+, \phi^{AB}, \psi_+^A )$, their anti-fields, and the ghosts $(c, \alpha^{AB},\epsilon_+^A, X)$. A functional $\O$
is called {\bf local and covariant} if it is defined on a spacetime $(\M, g)$ with analytic dependence on $g$ and such that
\ben
f^* \O[\Phi] = \O[f^* \Phi] \ ,
\een
for any isometric embedding $f: \M^\prime \to \M$ which preserves both the spin and causal structure of the spacetime.
\end{definition}
Concerning the space $\P^p_q  (\M)$, there are two basic theorems that we shall make use of. The first one, sometimes called the {\bf Thomas replacement theorem}~\cite{iyerwald}, states that any $\O \in \P^p_q  (\M)$ may only depend on derivatives of the metric in the form of the Riemann tensor and its (symmetrised) covariant derivatives $\nabla_{(\mu_1} ... \nabla_{\mu_k)} R_{\alpha\beta\gamma\delta}$. Furthermore, at any
point $x \in \M$, $\O$ may only depend on fields $\Phi$ and their (symmetrised) covariant derivatives $\nabla_{(\mu_1} ... \nabla_{\mu_k)} \Phi$ evaluated {\em at the point} $x$. This result allows us to assign a dimension to terms in $\P^p_q  (\M)$, by assigning $\nabla_\mu$ to have dimension $1$.
The second important structural theorem about $\P^p_q  (\M)$ is sometimes called the {\bf algebraic Poincar\'{e} lemma}, see e.g.~\cite{wald1}. It states that, if $\d \O = 0$, for some $\O \in \P^p_q  (\M)$, then $\O = \d \O^\prime$, for some $\O^\prime \in P^{p-1}_q  (\M)$. This is true, even if the de Rham cohomology of $\M$ is non-trivial. The point is that closedness of $\O[\Phi]$ as a {\emph{functional}} of $\Phi$, for {\em{all}} field configurations $\Phi$, is a much stronger restriction than merely being a closed form.

The cohomology of $\Q$ is defined by
\ben
H^p_q(\Q, \M) = \frac{\{{\rm ker} \ \Q: \P^p_q  (\M) \rightarrow \P^{p}_{q+1}  (\M) \}}{\{ {\rm im} \ \Q: \P^{p}_{q-1}  (\M) \rightarrow \P^{p}_q  (\M) \}} \ .
\een
The structure of these cohomology groups depends critically on the nature of the conformal symmetry superalgebra $\cS$ associated with $( \M ,g)$, particularly on its odd part $\cF$. It is convenient to define the action $\Q : \P^p_q(\M) \rightarrow \P^p_{q+1}(\M)$ with an additional sign $(-1)^p$ for the form degree. In addition to $\Q^2 = 0$ and $\d^2 =0$, it then follows that $\Q \d + \d \Q = 0$. Thus, the pair of differentials $(\Q, \d)$ equip $\bigoplus_{p,q \geq 0} \P^p_q  (\M)$ with the structure of a {\bf bi-complex}, on which the operator $\Q + \d$ is nilpotent. Thus, each element $\Gamma=
(\O^p_q, \O^{p-1}_{q+1}, \dots)$  in the kernel of $\Q+\d$ obeys the {\bf descent equations}
\ben\label{ladder}
\begin{split}
\Q \O^p_q &= \d \O^{p-1}_{q+1} \\
\Q \O^{p-1}_{q+1} &= \d \O^{p-2}_{q+2} \\
& \vdots \\
\Q \O^0_{p+q} &= 0 \ .
\end{split}
\een
Conversely, using the algebraic Poincar\'{e} lemma, it is straightforward to show that from each element $\O^p_q \in \P^p_q  (\M)$ which is $\Q$-closed modulo a $\d$-exact expression, one can construct a series of elements $\O^{p-k}_{q+k} \in \P^{p-k}_{q+k}  (\M)$, for all $k = 0,1,... p$, obeying \eqref{ladder}. If $\O^p_q$ is Weyl-invariant, then so are all the other $\O^{p-k}_{q+k}$. This follows because, whilst $\Q$ is Weyl-equivariant, $\d$ is only trivially Weyl-equivariant on Weyl-invariant functionals.

We refer to the set of equations in \eqref{ladder} as a {\bf ladder}. Notice that each rung of the ladder in \eqref{ladder} has the same total degree (i.e. form degree plus ghost number) $p+q$. It is convenient to
\begin{definition}
Let $H^n(\Q | \d, \M)$ denote the space of all ladders with total degree $n$.
\end{definition}

We now summarise some important facts about the structure of these spaces for the theories of interest on $\eM$.

\begin{theorem}
\label{cohothm}
$\;$\\ [.1cm]
1) The following elements represent a non-trivial class in
$H^4(\Q| {\rm d} , \M)$: \\ [.1cm]
a) $\O^4_0 = \hat \eL \dv$, where
\ben
\label{Lhat}
\hat \eL = \eL_{\rm V} + \eL_{\rm H} +
\tfrac{1}{4} Y^{AB} Y_{AB}
- \hat \psi_-^A {\slashed \xi} \hat{\overline{\psi}}_{+\; A}
 + (\overline \epsilon_-^B \epsilon_{-\; B}) \hat \psi_{+}^A\hat{\overline{\psi}}_{+\; A}
 + (\overline \epsilon_+^B \epsilon_{+\; B}) \hat \psi_{-}^A \hat{\overline{\psi}}_{-\; A} \ ,
\een
in terms of $Y^{AB} = 2(\epsilon_-^{(A} \hat \lambda^{B)}_- + \epsilon^{(A}_+ \hat \lambda^{B)}_+)$ and $\xi^\mu = 2\overline \epsilon_{-\; A} \Gamma^\mu \epsilon_+^A$. \\ [.1cm]
b) Any local curvature term with dimension $4$. \\ [.2cm]
2) The following elements represent a non-trivial class in $H^5(\Q|{\rm d}, \M)$: \\ [.1cm]
a) $\O^4_1 = {\mathscr A}$, where
    \ben
\begin{split}
\eA &=  {\rm Re} \Big({\overline{{\slashed \nabla} \epsilon}}_{-\; A} \Big[ -{\slashed F} \lambda_+^A - 2\, [ \varphi , \varphi^* ] \lambda_+^A + 2\, {\slashed D} \varphi \lambda_-^A - Y^{AB} \lambda_{+\; B} + \phi^{AB}  (\lambda_{+}^C \cdot \phi_{CB})  \Big] \Big) \dv \\
&\quad + {\rm Re} \Big({\overline{{\slashed \nabla} \epsilon}}_{-\; A} \Big[
\half {\slashed D} \phi^{BA} \, \psi_{-\; B}
+ \varphi^* \cdot \phi^{BA} \, \psi_{+ \; B}~
+ 2\,  \hat \psi_{+\; B} \psi^B_+ \, \epsilon_+^A
\Big] \Big) \dv \\
&\quad -2 \sigma_X \hat \eL \dv + 2i\alpha \, {\rm Tr}(F \wedge F)~,
\end{split}
\label{Adef}
\een
and $\Tr$ denotes the Killing form on $\fg$ (omitted in the other terms to avoid clutter). \\ [.1cm]
b) $\O^4_1$ given by any linear combination of $\alpha \, \tr ( {\bm C} \wedge {\bm C} )$ and $\alpha \tr (\bm{C} \wedge{}^* \bm{C} )$, where ${\bm C}$ denotes the Weyl tensor, written as an $\End (T\M)$-valued $2$-form on $\M$.
\footnote{The term $\tr ( {\bm C} \wedge {\bm C} )$ is proportional to the Pontryagin class $\tr(\bm{R} \wedge \bm{R})$, where ${\bm R}$ is the Riemann tensor, written as an $\End (T\M)$-valued $2$-form on $\M$.}
 \\ [.2cm]
3) If $(\M, g)$ is conformally flat, then at operator dimension $4$, and (top) form degree $4$, the cohomology rings are generated by the above elements together with the obvious dimension 4 curvature terms.
\end{theorem}

\begin{remark}\label{rem:12}
$\;$ \\ [-.6cm]
\begin{itemize}
\item If $(\M,g)$ does not have any symmetries, either bosonic or fermionic, then $\cS$ consists only of R-symmetry transformations. In that case, $\Q$ is closely related to the ordinary BRST operator for both gauge- and R-symmetry transformations, adorned with anti-fields. A detailed description of the associated cohomologies can be extracted from~\cite{Barnich:2000zw}. $H^4(\Q| \d, \M)$ is generated by the space of all gauge and R-symmetry invariant operators with form degree $4$, together with certain operators involving anti-fields which are present due to the existence of conserved R-symmetry currents.
\footnote{The existence of such operators follows from general results in \cite{Barnich:2000zw}. For example, 
\ben
\alpha\, \Tr \left( {\hat \varphi} \varphi - {\hat \varphi}^* \varphi^* + \half \hat \lambda_{+\, A}^{} \lambda^A_+ - \half \hat \lambda_{-\, A}^{} \lambda_-^A - \half \hat \psi_{+\, A}^{} \psi^A_+ + \half \hat \psi_{-\, A}^{} \psi_-^A  \right) \dv \in H^4(\Q|\d, \M)~, \nonumber
\een
corresponds to the $\fu(1)$ component of the R-symmetry. These operators are not in the cohomology if $\cS$ is non-trivial.}
Most of these elements are not in the cohomology when $\cS$ is non-trivial. Moreover, at ghost number $1$, $H^5(\Q| \d, \M)$ is generated by terms of the form $\alpha \O$ (for any gauge and R-symmetry invariant $\O \in \P^4_0 (\M)$), the {\bf gauge anomaly}
\footnote{The gauge anomaly in \eqref{eq:nonabeliananomaly} is written with respect to a basis $\{ {\bm e}_i \}$ for $\fg$. Relative to this basis, the structure constants $f_{ij}^k$ are defined by the Lie bracket $[ {\bm e}_i , {\bm e}_j ] = f_{ij}^k {\bm e}_k$ on $\fg$. The tensor $t_{ijk} = \Tr \left( {\bm e}_i {\bm e}_j {\bm e}_k +  {\bm e}_i {\bm e}_k {\bm e}_j \right)$ is totally symmetric and $\fg$-invariant. Since we have taken $\fg$ to be simple, a non-zero $t_{ijk}$ exists only when $\fg$ is either $\fsu(N>2)$, $F_4$ or $E_7$.}
\ben\label{eq:nonabeliananomaly}
t_{ijk} \, \d c^i \wedge \left( A^j \wedge \d A^k + \tfrac{1}{4} \, f_{lm}^k \, A^j \wedge A^l \wedge A^m \right)~,
\een
and the curvature terms in 2b). In the absence of any non-trivial parameters $(X , \epsilon^A_+ ) \in \cS$, \eqref{Adef} reduces to the topological term $\alpha \, {\rm Tr} (F \wedge F)$. Note in particular that the gauge anomaly is {\em not} in $H^5(\Q| \d, \M)$ if $\cF \subset \cS$ is non-trivial.

\item Notice that the structure of elements in Theorem~\ref{cohothm} for the theory based on vector and hyper multiplets is considerably simpler than for a generic field theory with minimal supersymmetry (whose explicit form in Minkowski space is given in section 7 of \cite{Brandt:2002pa}).

\item It is possible to calculate all ladders~\eqref{ladder} for the class of theories of interest. An explicit example for the vector multiplet is provided in Appendix~\ref{app:B}.

\item If $(\M,g)$ admits a twistor spinor, then there exist relations between the dimension $4$ local curvature terms. In particular, since $d=4$, given a twistor spinor $\epsilon_+ \in \fS^c_+ (\eM)$, the first equation in \eqref{consistency} implies $0 = C^{\mu\nu\alpha\beta} C_{\mu\nu\rho\sigma} \Gamma_{\alpha\beta} \Gamma^{\rho\sigma} \epsilon_+ = 2( i \, C_{\mu\nu\rho\sigma} {}^* C^{\mu\nu\rho\sigma} -  C_{\mu\nu\rho\sigma} C^{\mu\nu\rho\sigma} ) \epsilon_+$. Assuming $\epsilon_+$ has no zeros then, since the Weyl tensor is real, this means both $\tr(\bm{C} \wedge \bm{C})$ and $\tr(\bm{C} \wedge {}^* \bm{C})$ vanish identically on $\eM$. Using also that the Euler class, proportional to $\tr(\bm{R} \wedge {}^* \bm{R})$, is always locally $\d$-exact, it follows that the most general dimension 4 curvature term which is not locally exact is a multiple of $R^2 \dv$. Moreover, if $\eM$ admits a parallel spinor (e.g. if it is a pp-wave), then $R =0$ too.
\end{itemize}
\end{remark}

\noindent
{\em Proof:} Statement 1) in Theorem~\ref{cohothm} follows by applying the $\Q$-action defined by \eqref{b1}.~\eqref{b2},~\eqref{b3} to the relevant operators and showing that the result is $\d$-exact. At ghost number $0$, this follows from the general arguments in section~\eqref{sec:brstextension}, because
$\hat S = \int_\M \hat \eL \dv$ is the extended $\Q$-invariant action \eqref{hatsmodel}.

Statement 2a) is similarly verified via a rather difficult calculation, making extensive use of the Fierz identities \eqref{eq:oddoddoddjacobiNequals2}, the twistor spinor equation for $\epsilon_+^A$ and various subsidiary identities, such as \eqref{consistency}. We omit the details.

Let us instead comment on the curvature terms in $H^5(\Q|\d, \M)$ in statement 2b). If $(\M,g)$ admits a twistor spinor then, as noted in the third point of Remark~\ref{rem:12}, all these curvature terms must vanish. On the other hand, if $(\M,g)$ does not admit a twistor spinor, then it follows from \eqref{b2} that $\Q \alpha = 0$. Whence, in either case, we obtain the desired result. Similarly, if $(\M,g)$ does not admit a conformal Killing vector, then \eqref{b2} implies that $\Q \sigma_X = -\half \sigma_{[X,X]}=-\half \cL_X \sigma_X$. Therefore
\ben
\Q\Big( \sigma_X \tr(\bm{C} \wedge {}^* \bm{C})\Big)
= -\half \cL_X \sigma_X \tr(\bm{C} \wedge {}^* \bm{C})
= -\half \d \Big(\sigma_X i_X \tr(\bm{C} \wedge {}^* \bm{C}) \Big) \ ,
\een
using the conformal transformation property of the Weyl tensor and $\cL_X = \d i_X + i_X \d$. However, this term can actually be ignored since it is $\d$-closed. This follows by a theorem of~\cite{DC:1975} which states that either:
\vspace*{-.2cm}
\begin{itemize}
\item $\fX^c(\M)$ can be realised by Killing vectors with respect to a particular metric $\tilde g = \Omega^2 g$ in the conformal class of $g$. In this case, $\sigma_X = \cL_X \Omega$ which implies that $\sigma_X \tr(\bm{C} \wedge {}^* \bm{C})$ is $\d$-exact, by the same calculation
as above.

\item $( \M , g)$ is conformally flat or a plane wave. (Recall the example in section~\ref{sec:spacetimes} where $\fX^c(\M) \cong {\mathfrak{heis}}_2 \ltimes \RR$, with ${\mathfrak{heis}}_2$ generated by $5$ Killing vectors and $\RR$ generated by a homothetic conformal Killing vector.) In this case, all quadratic invariants of the Weyl tensor vanish identically.
\end{itemize}
Similarly, $\sigma_X \tr ( \bm{R} \wedge {}^* \bm{R} )$ is seen to be $\Q$-closed modulo $\d$,
but again, it is actually $\d$-closed.

If $(\M,g)$ is conformally flat, let us outline an indirect argument why the elements noted in Theorem~\ref{cohothm} are sufficient to generate the relevant cohomologies (i.e. that there cannot exist further, non-exact independent expressions). One might be tempted to first solve the problem in Minkowski space, e.g. by a brute force calculation such as \cite{Maggiore:1994dw}, before reducing to this solution via a Weyl transformation on a general conformally flat background. However, this would be too naive since, a priori, we do not know how the generators of the cohomology transform under Weyl transformations. Thus, we shall employ the following \lq scaling type' argument, which will provide some new insight.

Let $y^\mu$ denote a choice of local {\bf Riemannian normal coordinates} centred around an arbitrary point $p \in \M$. Let $\psi_s: y^\mu \mapsto s\, y^\mu$ with $g^{(s)} = s^{-2} \, \psi_s^* g$, in terms of some positive real number $s$. In Riemannian normal coordinates,
\ben
g^{(s)}_{\mu\nu}(y) = g_{\mu\nu}(sy) =
\eta_{\mu\nu} -  \tfrac{1}{3}s^2  \ R_{\mu\rho\nu\sigma} y^\rho y^\sigma
+O(s^3)~.
\een
All terms involving higher powers of $s$ in the expansion can be written in terms of components of the Riemann tensor and its covariant derivatives at $y = 0$.
\footnote{Since $C_{\mu\nu\rho\sigma} =0$, the Riemann tensor can be expressed in terms of the Ricci tensor and scalar curvature via the identity $R_{\mu\nu\rho\sigma} = \half ( g_{\mu\rho} R_{\nu\sigma} - g_{\nu\rho} R_{\mu\sigma} - g_{\mu\sigma} R_{\nu\rho} + g_{\nu\sigma} R_{\mu\rho} ) - \tfrac{1}{6} ( g_{\mu\rho} g_{\nu\sigma} - g_{\nu\rho} g_{\mu\sigma} ) R$.}
By definition, $g^{(s)}$ is locally diffeomorphic to $g$ multiplied by the conformal factor $s^{-2}$. Consequently, $g^{(s)}$ must admit the same number of conformal Killing vectors and twistor spinors as $g$.

The expansion for a twistor spinor is given by
\ben\label{rnc2}
\begin{split}
\epsilon^{(s)} &= \epsilon + {\slashed y} \, \epsilon^\prime \\
&\quad - \tfrac{1}{8} s^2 \, y^\mu (R_{\mu\nu} - \tfrac{1}{6} \eta_{\mu\nu} R )
\Big(  {\slashed y} \, \Gamma^\nu \epsilon
 -\tfrac{1}{3} ( y^\rho y_\rho \Gamma^\nu - 2 y^\nu \, {\slashed y} ) \epsilon^\prime \Big)
+ O(s^3)
\ ,
\end{split}
\een
where, with respect to our choice of frame, $\epsilon$ and $\epsilon^\prime$ are constant spinors and the gamma matrices are those with respect to the Minkowski metric $\eta$, which is used to raise and lower indices. All terms involving higher powers of $s$ in \eqref{rnc2} involve only $\epsilon$, $\epsilon^\prime$ and covariant derivatives of the Riemann tensor at $y=0$.

Using \eqref{CKVint}, the expansion for a conformal Killing vector is given by
\ben\label{rnc3}
\begin{split}
X^{(s)\; \mu} &= a^\mu +  b \, y^\mu + \tfrac{1}{2} y^\nu y_\nu \, c^\mu - y^\nu c_\nu \, y^\mu + d^{\mu\nu} y_\nu \\
&\quad - \tfrac{1}{3} s^2 \, R^\mu{}_{\nu\rho\sigma} y^\nu y^\rho \, a^\sigma +
\tfrac{1}{36} s^2 \, R^\mu{}_\nu y^\nu y^\rho y_\rho \, b \\
&\quad + \tfrac{1}{144} s^2  \left( R^{\mu\nu} (
\half y^\rho y_\rho \, c_\nu -  y_\nu y^\rho c_\rho  ) - ( y^{\mu} \, c^{\nu} - y^{\nu} \, c^{\mu} ) R_{\nu \rho} y^\rho \right) y^\sigma y_\sigma \\
&\quad + \half s^2 R^{\mu\nu\rho\sigma} y_\nu y_\sigma y^\tau \, d_{\rho\tau}
+ O(s^3)~,
\end{split}
\een
where $a^\mu$, $b$, $c^\mu$ and $d^{\mu\nu} = - d^{\nu\mu}$ are constant.

The leading terms in \eqref{rnc2} and \eqref{rnc3} which do not involve the curvature give precisely the generic form of a twistor spinor and conformal Killing vector in Minkowski space. For example, in \eqref{rnc3}, $a^\mu, d^{\mu\nu}$, $b$ and $c^\mu$ respectively parameterise translations, Lorentz transformations, dilatations and special conformal transformations in Minkowski space. This key fact will now be used to argue that, at least at the level of cohomological perturbation theory, the calculation of $H^4(\Q| \d, \M)$ at dimension four must reduce to that of $H^4(\Q| \d, \RR^{3,1})$, which is known to consist precisely of the ladders described in Theorem~\ref{cohothm}.

First, let $\Q^{(s)}$ be the differential associated with $g^{(s)}$. Since $g^{(s)}$, together with its associated twistor spinors and conformal Killing vectors, depend {\em smoothly} on $s$, $\Q^{(s)}$ must also depend smoothly on $s$. Thus $\Q^{(s)}$ has a Taylor expansion in powers of $s$ at $y=0$. In particular, since $g^{(0)} = \eta$ and twistor spinors and conformal Killing vectors reduce to their counterparts in Minkowski space at $s=0$, $\Q^{(0)}$ must give the differential for the $\N=2$ superconformal field theory in Minkowski space. Moreover, any ladder $\Gamma = \{ ...,\O^p_q ,..., \O^0_{p+q}  \}$ of operators in the theory with respect to differential $\Q$ and metric $g$ must follow at some value of $s$ from a smooth one-parameter family of ladders $\Gamma^{(s)}$ with respect to differential $\Q^{(s)}$ and metric $g^{(s)}$. In particular, the zeroth order Taylor coefficient $\Gamma^{(0)}$ in Minkowski space must be one of the ladders described in Theorem~\ref{cohothm}. Note that this ladder can be defined with respect to any metric. Whence, defining it with respect to $g^{(s)}$ allows us to subtract it off as the leading term in the Taylor expansion of $\Gamma^{(s)}$. This defines a new ladder which, by construction, has only terms involving positive powers of $s$ in its Taylor expansion. Furthermore, if the leading term is not identically zero, it must define a non-trivial ladder with respect to $\Q^{(0)}$. Equations \eqref{rnc2} and \eqref{rnc3} imply that any such term must contain the Riemann tensor or its covariant derivative at the point $p \in \M$. In addition, it must be $y$-independent in order to preserve the translation symmetry with parameter $a^\mu$, whose ghost forms part of $\Q^{(0)}$. By simple dimension counting, one sees that the only option must be of the form $R \varphi \varphi^*$, or involve a local curvature term of dimension 4 (which is trivially in the cohomology). However, it is easily verified that $R \varphi \varphi^*$ does not define a ladder with respect to $\Q^{(0)}$. Whence, we conclude that this term cannot be present in the Taylor expansion. Iterating this argument order by order in the Taylor expansion, one finds that there can in fact be no non-trivial contributions to the ladder from any higher order terms. \qed


\section{Quantum field theory and conformal supersymmetry in curved spacetime}
\label{sec:pert}

We would like to understand when the conformal symmetry superalgebra $\cS$ is realised in the quantum field theory associated with the Lagrangian $\eL$ of the $\eN=2$ theory. The precise meaning of this statement depends on one's framework for quantum field theory. In the framework of scattering theory, one says that the symmetry is realized if the scattering matrix $\Sm$ commutes with the charge operators ${\mathscr Q}_0$ representing the symmetries of $\cS$ in Fock space, $[ {\mathscr Q}_0 , {\Sm} ]=0$. This implies the invariance of the scattering amplitudes. To define these objects concretely, one normally splits the Lagrangian of the theory $\eL=\eL_0+\eL_1$ into a part $\eL_0$ that is quadratic in the fields, and higher order terms $\eL_1$ describing the interactions of the fields. The free field theory associated with $\eL_0$ is quantized in the \lq standard' fashion, while the scattering matrix is defined in perturbation theory by the heuristic formula due to Dyson,
\ben
\label{smatrix}
{\Sm} = \T \exp \frac{i}{\hbar} \int_\M \eL_1 \dv \ .
\een
Both ${\mathscr Q}_0$ and $\eL_1$ (hence $\Sm$) are operators on the Hilbert space of the free field theory defined by $\eL_0$, and the time-ordered exponential is defined by its Taylor expansion (perturbation theory). Amplitudes are defined by sandwiching $\Sm$ between particle states in the Hilbert space with definite value of the \lq momenta' and other quantum numbers. Applying \lq Wick's theorem' gives these amplitudes in terms of contributions
corresponding to Feynman graphs. This is, more or less, the strategy we shall adopt here. However, there are a number of complications:

\begin{enumerate}
\item[a)] On a curved manifold, there is no canonical way
to quantize the free field theory associated with $\eL_0$, i.e. there is, even for the restricted classes of manifolds $(\M,g)$ admitting twistor spinors, no preferred Hilbert space representation, apart from
certain special cases.
\item[b)] On a curved manifold, the integral over all of $\M$ in~\eqref{smatrix} typically gives rise to infra-red divergences.
\item[c)] In any spacetime, the expression~\eqref{smatrix} contains ultra-violet divergences.
\item[d)] We are dealing with a gauge theory, so one has work with an appropriately gauge-fixed theory.
\end{enumerate}

a) The reference to any highly non-canonical representation is avoided by working with a scattering matrix taking values in an abstract algebra, $\W$ associated with $\eL_0$. b) in that abstract
algebra, $[{\mathscr Q}_0, -]$ gets replaced with a (graded) derivation $\hat \Q_0$, $\Sm$ gets replace with an \lq infra-red cutoff version', and the desired relation $[{\mathscr Q}_0, {\Sm} ]=0$ gets replaced by the more complicated relation~\eqref{WA} which formally reduces to the former if we choose a representation of $\W$ on some Hilbert space, and if we remove the infra-red cutoff. c) is dealt with by a suitable renormalization prescription in curved space, and d) is dealt with via a version of the BRST-method described in previous sections in the classical case.


\subsection{Definition of $\W(\M,g)$}

Here we recall the construction of $\W(\M,g)$ for the toy model of a single scalar field $\phi$ with $\eL_0 = - \half g^{\mu\nu} \partial_\mu \phi \partial_\nu \phi - \tfrac{1}{12} R\phi^2$
on a globally hyperbolic Lorentzian spacetime $(\M,g)$. The algebra $\W$ incorporates (i) the standard commutation relations expected from the quantized field $\phi$, appropriately generalised to curved space. (ii) It is large enough to include \lq Wick powers' and their \lq time-ordered products'. (iii) \lq Wick's theorem' is built into the product of $\W$. (iv) The definition of $\W$ is covariant. (v) If an appropriate representation of $\W$ is chosen on a Hilbert space, then the usual formulas of fields in terms of creation- and destruction operators are recovered. However, crucially, $\W$ itself does not depend on any choice of such a non-canonical representation.

The crucial input into $\W$ is the {\bf Hadamard parametrix}, $H$. It is a distribution defined near the diagonal in $\M {\times} \M$. The formula is
\ben
H(x,x') = \frac{1}{4\pi^2} \left( \Delta^{\half} (x,x') \, (s+i0\, t)^{-1} + \sum_{j=0}^\infty u_j(x,x') \, \log(s + i0\, t) \right) \ .
\een
Here, $s$ is the signed squared geodesic distance between $x,x' \in \M$, $\Delta$ is the van Vleck determinant, and $u_j$ are certain recursively defined transport coefficients. $t$ is a sign function which is $\pm 1$ provided $x \in \eJ^\pm (x')$, i.e. $x$ is in the causal future/past of $x'$.  The sum over $j$ converges in a convex normal neighbourhood if $(\M,g)$ is real analytic, as we assume for simplicity. $H$ satisfies
\begin{itemize}
\item $[(\nabla^2 - \tfrac{1}{6}R) \otimes {\bf 1}] H = 0 = [{\bf 1} \otimes (\nabla^2 - \tfrac{1}{6}R)] H$ modulo $C^\infty(\M {\times} \M)$.
\item The {\bf wave front set} \cite{hormander} of $H$ near the diagonal in $\M {\times} \M$ is given by
    \ben
    \label{Hwfs}
    {\rm WF}(H) \subset \{ (x,k; x',-k') \in T^*(\M {\times} \M) \setminus 0 \mid k \ \
    \text{future directed}, \ \ k \sim k' \} \ ,
    \een
        where $k \sim k'$ means that $x$ and $x'$ can be joined by a null geodesic, and that $k$ and $k'$ are tangent to that null geodesic and parallel transported into each other using $\nabla$. The wave front set is an invariant of a distribution characterising its singular set, see e.g.~\cite{hormander}.
    \item Twice the imaginary part of $H$ is equal to the \lq commutator function' $E$, which is defined as the difference between the unique retarded and advanced Green's function for $\nabla^2 - \tfrac{1}{6}R$.
\end{itemize}
Details can e.g. be found in~\cite{rad,junker} and references therein.

\begin{example}\label{planepp}
Consider the plane wave metric $g = 2\, \d u \d v + h_{ab}(u) x^a x^b \d u^2 + \delta_{ab} \d x^a \d x^b$,
where $a,b=1,2$. A pair of $2{\times}2$ matrix \lq propagators' $\bm{A}(u,u')$ and $\bm{B}(u,u')$ can be defined as in \eqref{abprop}. The Hadamard parametrix may be written
\ben
H(x,x') = \frac{1}{4\pi^2}\frac{u-u'}{\sqrt{\det \, \bm{B}(u,u')} \left( s(x,x') - i 0\,  {\rm sgn}(v-v') \right)}~,
\een
in a convex normal neighbourhood, with $x=(u,v,x^a)$ and $x' = (u',v',x'^a)$. The signed squared geodesic distance is given by
\ben
s = \half (u-u') \bigg(
2(v-v') +  \partial_u A_{ab}(u,u') x^a x'^b
+(x^c \partial_u B_c{}^a(u,u') - x'^a ) ( B^{-1} )_{ab}
(x^b - A^b{}_d (u,u') x'^d )
\bigg) \ .
\een
For further discussion of the properties of these quantities, see~\cite{Harte:2012uw}.
The plane wave spacetime is actually not globally hyperbolic, so unique retarded and
advanced propagators -- and therefore their difference $E$ -- do not exist. Hence,
the algebra $\W(\M,g)$ is not defined globally in that case, although it can be defined locally.
\end{example}

Let us now define $\W(\M,g)$, as a $*$-algebra.

\begin{definition}
The $*$-algebra $\W(\M,g)$ is the linear space of the identity ${\bf 1}$ and
functionals of the form
\ben
W(f) = \int_{\M^n} f(x_1, \dots, x_n) \prod_{j=1}^n \phi(x_j) \dv_j \ ,
\een
for any positive integer $n$, where $f$ is a symmetric distribution with compact support on $\M^n$ and ${\rm WF}(f) \cap [V_+^n \cup V_-^n] = \emptyset$, where $V_\pm = \cup_x (V_\pm)_x \subset T\M$ and $(V_\pm)_x$ is the future/past lightcone at $x\in \M$. These elements are subject to the following relations:
\begin{enumerate}
\item If $f,f' \in C^\infty_0(\M)$ are smooth with compact support and have only one argument, then the commutation relation $[W(f),W(f')] = i\, \hbar \, E(f, f') {\bf 1}$.
\item If $f$ and $f'$ are supported in a convex normal neighbourhood, then the product rule (\lq Wick's theorem') is $W(f) W(f') = \sum_k \hbar^k W(f \otimes_k f')$
with $\otimes_k$ meaning the \lq $k$-times contracted' tensor product:
\ben
\label{contr}
\begin{split}
(f \otimes_k f')(x_1, \dots, x_{n+m-2k}) &=  k! {n \choose k} {m \choose k} {\sf S}\int_{\M^{2k}} \prod_{i=1}^k H(y_{i}, y_{k+i}) \, \dv (y_{i}) \dv (y_{k+i}) \\
& \times f(y_{1}, \dots, y_{k}, x_1, \dots, x_{n-k}) f'(y_{k+1}, \dots, y_{k+i}, x_{n-k+1},
\dots, x_{n+m-2k})~.
\end{split}
\een
$\sf S$ denotes the symmetrisation map.
\item The *-operation in the algebra is denoted $\dagger$ and is defined as $W(f)^\dagger = W(f^*)$, and where $f^*$ is the complex conjugate of $f$.
\end{enumerate}
\end{definition}

\begin{remark}
The first property implements the usual covariant commutation relation, written informally $[\phi(x), \phi(x')] = i\, \hbar \, E(x, x')$. The second property implements Wick's theorem. The wave-front conditions on $f$ and $f'$ are imposed so that $\W$ is \lq sufficiently large' and the distributional products implicit in~\eqref{contr} are well-defined. This follows from~\eqref{Hwfs}, together with standard theorems of microlocal analysis (Theorem~8.10.2 of~\cite{hormander}). $\W$ is defined locally and covariantly in terms of the spacetime geometry, because $H$ only depends on the geometric data $(\M,g)$, and the time-orientation. The last relation informally means that $\phi(x) = \phi(x)^\dagger$. For details and further explanation of this construction, see \cite{hw4}.
\end{remark}


\subsection{Time-ordered products}

We would now like to define {\bf Wick products} and {\bf time-ordered products} in the free field theory defined by $\eL_0$. According to our philosophy, these objects are not operators on some (non-canonical)
Hilbert space, but instead elements of $\W(\M,g)$. It is well-known that naive definitions lead to infinities, which have to be dealt with using some form of \lq renormalization'. Rather than presenting an explicit, specific renormalization scheme, it is conceptually clearer to point out, and work with, the general properties of the time-ordered products that should be satisfied in any reasonable scheme, and then prove that there actually exists a concrete construction having those general properties. We will now present a list of the most important properties.

We define a {\bf renormalization scheme} to be a collection
$\T \equiv (\T_1, \T_2, \dots, \T_n, \dots)$ of linear maps
\ben
\T_n: \P(\M)^{\otimes  n}  \to {\mathcal D}'(\M^n;  \W) \, ,
\een
taking values in the distributions over $\M^n$, with target space $\W$, satisfying
properties T0)--T11) below. The time-orderedness is expressed by T8), whereas the other properties correspond to various natural and important features. Whence, $\T_n$ are also called {\bf time-ordered products}
\footnote{This is standard, but somewhat misleading, because they are not products in the usual sense used in algebra, i.e. an $n$-times multi-linear map from a vector space to itself.}
. $\T_n$ takes as argument the tensor product of $n$ local covariant classical forms $\O_1, ..., \O_n$, and it gives an expression $\T_n(\O_1(x_1) \otimes \dots \otimes
\O_n(x_n))$, which is itself a distribution in $n$ spacetime variables
$x_1, \dots, x_n$, with values in the algebra $\W$. That is, $\T_n(\O_1(x_1) \otimes \dots \otimes \O_n(x_n))$ is itself a map
that needs to be smeared with $n$ test functions $f_1(x_1), ...,
f_n(x_n) \in C^\infty_0 (\M)$.

The properties, T0)--T11), to be satisfied by any such renormalization scheme are:

\medskip
\noindent
{\bf T0) One factor.}
For time-ordered products of $n=1$ factors, there is nothing to order. We therefore define $\T_1(\O(x))$ to simply be the Wick-product
\footnote{
As discussed in~\cite{hw1}, in principle, one should allow adding
to the Wick-product terms of lower order times curvature terms of
the right dimension. For the purposes of this paper though the given prescription will suffice.
}
, i.e. $\T_1(\O(x)) = \O(x)$, which is an element of $\W$ after smearing $x$
against some smooth $f$.

\medskip
\noindent
{\bf T1) Locality and covariance.}
The time-ordered products are locally and covariantly
constructed in terms of the metric. This means more precisely the following: let
$\psi:\M \to \M'$ be a causality preserving isometric embedding
between two spacetimes preserving the causal structure, so that $\psi^* g'=g$. Denote by $\alpha(\psi)$ the corresponding canonical homomorphism $\W(\M,g) \to \W(\M',g')$, which maps $W(f) \mapsto W(\psi_* f)$. Then we require
\ben
\alpha(\psi) \circ \T_g = \T_{g'} \circ \bigotimes \psi_*
\een
where $\T_g$ and $\T_{g'}$ denote the time-ordered products on $(\M,g)$ and $(\M',g')$ respectively. The mapping $\psi_*: \P(\M) \to \P(\M')$ is the natural push-forward
map. Thus, the local and covariance condition imposes a relation
between the construction of time-ordered products on locally isometric
spacetimes.

\medskip
\noindent
{\bf T2) Scaling.} We would like the time-ordered products to \lq scale naturally',
up to logarithms, under a rescaling $g \mapsto \mu^2 g$, for any positive real number $\mu$. We first note that $\W(\M, g)$ is $*$-isomorphic to $\W(\M, \mu^2 g)$ under the map $W(f) \mapsto \mu^n W(f)$. This follows from the corresponding
transformation of the Hadamard parametrix $H_{\mu^2 g} = \mu^{-2} H_g$  Let $\alpha_\mu$ be this $*$-isomorphism. Then consider
\ben\label{tmudef}
\T^{(\mu)}_g = \alpha^{-1}_\mu \circ \T_{\mu^2 g} \circ
\bigotimes {\rm exp}(
( \log \mu )\, {\tt N}_{\rm dim})~,
\een
where ${\tt N}_{\rm dim}$ is the dimension counter.
Because we have put the identification map $\alpha_\mu$ on the right
side, $\T^{(\mu)}$ defines a new time-ordered product in the algebra
associated with the unscaled metric, $g$. In the
absence of scaling anomalies, this would be equal to the original
$\T$ for all $\mu \in \mr_+$. It is generally not
possible to achieve this exactly homogeneous scaling behaviour. We postulate the poly-homogeneous scaling behaviour
\begin{equation}
\left( \mu \tfrac{\d}{\d\mu} \right)^{n+1} \T_n^{(\mu)} = 0~.
\end{equation}
which expresses that $\T^{(\mu)}_n$ is a polynomial in $\log \mu$ of order at most $n$.

\medskip
\noindent
{\bf T3) Microlocal spectrum condition.}
Consider a time-ordered product $\T_n(\O_1(x_1) \otimes \cdots
\otimes \O_n(x_n))$ as a $\W$-valued distribution on $\M^n$.
Then we require that
\begin{equation}
\label{microcond}
{\rm WF}(\T_n) \subset C_T(\M, g)~,
\end{equation}
where the set $C_T(\M, g) \subset T^*\M^n \setminus 0$
is described as follows (we
use the graph-theoretic notation introduced in \cite{bf, bf1}):
Let $G(p)$ be a \lq decorated embedded graph'
in $(\M, g)$. By this we mean a graph embedded in $\M$ such that its
vertices are points $x_1,..., x_n \in \M$
and its edges $e$ are oriented null geodesic curves. Each such null
geodesic is equipped with a co-parallel, co-tangent co-vector field $p_e$.
If $e$ is an edge in $G(p)$ connecting the points $x_i$ and $x_j$
with $i < j$, then $s(e) = i$ is its source
and $t(e) = j$ its target. It is required that
$p_e$ is future/past directed if $x_{s(e)} \notin \eJ^\pm(x_{t(e)})$.
With this notation, we define
\begin{align}
\label{gamtdef}
C_T(\M, g) &=
\left\{ (x_1, k_1; \dots; x_n, k_n) \in T^*\M^n \setminus 0 \; \bigg| \;
\exists \; G(p)\; {\mbox{with vertices}} \;\; x_1,..., x_n \in \M \right. \nonumber \\
&\left. \hspace*{6cm} {\mbox{and}} \;\; k_i = \sum_{\{ e | s(e) = i\}} p_e \;\; - \sum_{\{ e | t(e) = i \}} p_e \; , \;\; \forall \; i=1,...,n \right\}~.
\end{align}

\medskip
\noindent
{\bf T4) Smoothness.}
The functional dependence of the time-ordered products
on spacetime metric $g$ is such that, if the metric is varied smoothly, then the time-ordered products vary smoothly.

\medskip
\noindent
{\bf T5) Analyticity.} A corresponding condition in the real analytic setting.

\medskip
\noindent
{\bf T6) Symmetry.}
The time-ordered products are symmetric under a permutation of
the tensor factors.

\medskip
\noindent
{\bf T7) Unitarity.}
Let $\bar \T_n(\otimes_i \O_i(x_i)) =
[\T_n(\otimes_i \O_i(x_i)^*)]^\dagger$ be
the \lq anti-time-ordered' product.
\footnote{It can be shown that the anti-time-ordered product in T7) satisfies the causal factorisation property T8) with the reversed time-orientation.}
Then we require
\begin{equation}
\bar \T_n \big( \otimes_{i=1}^n \O_i(x_i) \big) =
\sum_{I_1 \, \sqcup \, ...\, \sqcup \, I_j \, =\, \underline{n}} (-1)^{n + j} \; \T_{| I_1 |} \big( \otimes_{i \in I_1} \O_i(x_i) \big) \; \dots \; \T_{|I_j|}\big(\otimes_{j \in I_j} \O_j(x_j) \big)~,
\label{atoprod}
\end{equation}
where the sum runs over all partitions of the set $\underline{n} = \{1, \dots, n\}$ into pairwise disjoint subsets $I_1, ..., I_j$. This is a version of the optical theorem and formally implies ${\Sm}^\dagger={\Sm}^{-1}$ for the scattering matrix.

\medskip
\noindent
{\bf T8) Causal Factorisation.}
Let $\{x_1, \dots, x_i\}
\cap \eJ^-(\{x_{i+1},\dots,x_n\}) = \emptyset$. Then we require
\begin{equation}
\T_n(\O_1(x_1) \otimes \dots \otimes \O_n(x_n)) = \T_i(\O_1(x_1) \otimes \dots \otimes \O_i(x_i)) \; \T_{n-i}(\O_{i+1}(x_{i+1}) \otimes \dots \otimes \O_n(x_n))~.
\end{equation}
For $n=2$, this means
\ben\label{cases}
\T_2(\O_1(x_1) \otimes \O_2(x_2)) =
\begin{cases}
\T_1(\O_1(x_1)) \T_1(\O_2(x_2)) & \text{if $x_1 \notin \eJ^-(x_2)$}\\
\T_1(\O_2(x_2)) \T_1(\O_1(x_1)) & \text{if $x_2 \notin \eJ^-(x_1)$}~.
\end{cases}
\een

\medskip
\noindent
{\bf T9) Commutator.}
We require
\ben
\left[\T_n \big( \otimes_i^n \O_i(x_i) \big), \phi(x) \right] =
i\, \hbar\sum_{k=1}^n \T_n\bigg(\O_1(x_1) \otimes \dots \int_\M E(x,y)
\frac{\delta \O_k(x_k)}{\delta \phi(y)} \otimes \dots \O_n(x_n) \bigg)~,
\label{T9}
\een
where $E$ is the causal propagator (defined by the advanced minus retarded Green's function for the operator $\square - \tfrac{1}{6} R$).

\medskip
\noindent
{\bf T10) Schwinger-Dyson equation.}
The free field equation $\delta S_0/\delta \phi =0$ holds, in the sense
that
\ben
\T_{n+1}\bigg(
\frac{\delta S_0}{\delta \phi(x)} \otimes \bigotimes_{i=1}^n \O_i(x_i)
\bigg)
= \sum_{i=1}^n
\T_n
\bigg(
\O_1(x_1) \otimes \cdots \frac{\delta \O_i(x_i)}{\delta \phi(x)} \otimes \cdots \O_n(x_n) \bigg) \ ,
\een
for module algebra elements containing $\delta S_0/\delta \phi$ as a factor\footnote{
This defines an ideal in $\W(\M,g)$. Representations $\pi$ of $\W(\M,g)$ on
a Hilbert space will map this ideal to zero, which corresponds to the fact that
free particles are \lq on-shell'.}, i.e. \lq on-shell'.

\medskip
\noindent
{\bf T11) Action Ward identity.}
$\T_n$ must commute with derivatives.

Further explanations of these conditions may be found in~\cite{bf,hw1,hw2,hw3,h1}.
The first fundamental fact is that time-ordered
products $\T_n$ with $n\ge 1$ actually exist:

\begin{theorem} (Existence)
There exist maps $\T_n$ satisfying T0)-T11).
\end{theorem}

The constructive proof of this theorem was given in~\cite{hw2}, key parts of which were based on the foundational paper~\cite{bf}. While the details are rather complicated, the basic idea -- going back to~\cite{eg,bog} -- is as follows. For $n=1$, the time-ordered products are just the Wick-products and are hence known. For $n=2$, the causal factorisation condition
T8) gives $\T_2(\O_1(x_1) \otimes \O_2(x_2))$ except when $x_1 = x_2$, because in all other cases $x_1 \notin \eJ^-(x_2)$ or $x_2 \notin \eJ^-(x_1)$ holds true, and
we can apply~\eqref{cases}. Thus, in a sense, we only need an \lq extension' to the
\lq diagonal' locus $x_1=x_2$ in $\M {\times} \M$. This non-trivial step corresponds to renormalization,
and it must be performed so as to be consistent with the other requirements. The
extension of the argument to $n>2$ is similar.

The next question is whether $\T_{n\ge 1}$ is unique. Thus, suppose
we have another renormalization scheme $\hat \T_{n\ge 1}$ which also satisfies T0)-T11).
To characterise the difference, we introduce a hierarchy $\D_n$ of linear functionals with the following properties. Each $\D_n$ is a linear map
\ben\label{Dndef}
\D_n: \P^{k_1}(\M) \otimes \dots \otimes \P^{k_n}(\M) \to
\P^{k_1/.../k_n}(\M^n) \rf{\hbar}~,
\een
where we denote by $\P^{k_1/\dots/k_n}(\M^n)$ the space of all
distributional local, covariant functionals of $\phi$ (and its covariant
derivatives), of $g$, and of the Riemann tensor (and its covariant derivatives),
which are supported on the total diagonal (i.e. of delta-function type), and which are a $k_i$-form in the $i$th argument $x_i$, for all $i=1,...,n$. The difference between $\T_n$ and $\hat \T_n$ for time-ordered products satisfying T0)-T11) may now be expressed in terms of a hierarchy $\D_n$ as follows. Let $F= \int f \wedge \O$ be an integrated local
functional $\O \in \P(\M)$, and formally combine the
time-ordered functionals into a generating functional
written
\ben
\T(\exp_\otimes \left( \tfrac{1}{\hbar} F \right) ) = \sum_{n=0}^\infty \frac{\hbar^n}{n!} \T_n(F^{\otimes n}) \in \W \rf{\hbar}~,
\een
where $\exp_\otimes$ is the
standard map from the vector space  of local actions to the tensor algebra
over the space of local action functionals.
We similarly write $\D(\exp_\otimes (F))$ for the corresponding generating functional
obtained from $\D_n$. The difference between the time-ordered
products $\T$ and $\hat \T$ may now be expressed in the following way~\cite{hw1}:
\ben
\label{unique1}
\hat \T ( \exp_\otimes \left( \tfrac{i}{\hbar} F \right) ) = \T (
\exp_\otimes \left( \tfrac{i}{\hbar} \left( F + \D({\rm exp}_\otimes ( F ) ) \right) \right) )  \, .
\een
Each $\D_n$ is a formal power series
in $\hbar$, and if each $\O_i = O(\hbar^0)$, then it can be shown
that $\D_n(\otimes \O_i) = O(\hbar)$, essentially because there are no
ambiguities of any kind in the underlying classical theory.
The expression $\D(\exp_\otimes (F))$ may be viewed as being equal to
the finite counterterms that characterise the difference between the
two prescriptions for the time-ordered products.

The counterterms, i.e. the maps $\D_n$, satisfy a number of
properties corresponding to T0)-T11) for the time-ordered products~\cite{hw1,hw2,hw3}. As we have already said, $\D_n$
are supported on the total diagonal, and this corresponds to the
causal factorisation property T8). The $\D_n$ are local and covariant
functionals of the field $\phi$ and metric $g$ in the following sense. Let $\psi : \M \to \M'$ be any causality and orientation preserving isometric embedding, so $\psi^* g' = g$. If $\D_n$ and $\D_n'$ denote the
functionals on $\M$ respectively $\M'$, then we have
that $\psi^* \circ \D_n' =
\D_n \circ (\psi^* \otimes \dots \otimes \psi^*)$. This follows from T1). It follows from the
smoothness and analyticity properties T4), T5) and the scaling property T2)
that $\D_n$ have polynomial dependence on the Riemann curvature tensor and on the field $\phi$.
Since there is no ambiguity in defining Wick products (i.e. $\T_1$), then $\D_1=0$.
As a consequence of the symmetry of the time-ordered products T6), the maps
$\D_n$ are symmetric (respectively graded symmetric when
Grassmann valued fields would be present), and as a
consequence of T9), they must satisfy
\ben\label{fieldindep}
\frac{\delta}{\delta \phi(y)}
\D_n ( \O_1(x_1) \otimes \cdots \otimes \O_n(x_n) ) =
\sum_{i=1}^n \D_n \! \left( \O_1(x_1) \otimes \cdots \frac{\delta
\O_i(x_i)}{\delta \phi(y)}
  \otimes \cdots \O_n(x_n) \right) \, .
\een
In particular, $\D_n$
depend polynomially on $\phi$.
As a consequence of the scaling
property T2) of time-ordered products, the
engineering dimension of each term appearing in $\D_n$
must satisfy the following constraint
\ben\label{dimension}
({\tt N}_{\rm dim} + \Delta_s) \, \D_n \left( \O_1(x_1) \otimes \cdots \otimes \O_n(x_n) \right)
= \sum_{i=1}^n
D_n \left( \O_1(x_1) \otimes \cdots {\tt N}_{\rm dim} \O_i(x_i) \otimes \dots
\O_n(x_n) \right)~.
\een
where $\Delta_s$ is the {\bf scaling degree} of a distribution.
The unitarity requirement T7) on the time-ordered products
yields the constraint
\ben\label{unitary}
\D_n \left( \O_1(x_1) \otimes \cdots \otimes \O_n(x_n) \right)^*
=
-\D_n \left( \O_1(x_1)^* \otimes \cdots \otimes \O_n(x_n)^* \right)~.
\een
The action Ward identity T11) implies that one can freely
pull a derivative into $\D_n$,
\ben\label{pullin}
\nabla_{x_i}\D_n \left( \O_1(x_1) \otimes \cdots \otimes \O_n(x_n) \right)
=
\D_n \left( \O_1(x_1) \otimes \cdots \nabla_{x_i} \O_i(x_i) \dots \otimes \O_n(x_n) \right)~.
\een

\begin{example}
The meaning of the above considerations about $\D_n$ are best illustrated in a simple example. For example, if $F=\int f \phi^2 \dv$, to lowest order the formula~\eqref{unique1} gives
\ben
\hat \T_2(\phi^2(x) \otimes \phi^2(y)) = \T_2(\phi^2(x) \otimes \phi^2(y)) +
\tfrac{\hbar}{i} \,
\T_1(\D_2(\phi^2(x) \otimes \phi^2(y)))~,
\een
By the constraints on $\D_2$, this must be given by
\ben
\D_2(\phi^2(x) \otimes \phi^2(y)) = c_0 \, \delta(x,y)~,
\een
for some real constant $c_0$ of order $\hbar$, because the
scaling degree of the delta function is $4$ and the dimension of $\phi$ is 1, and because
the right side must be generally covariant and analytic in the metric. Similarly
\ben
\D_2(\phi^3(x) \otimes \phi^3(y)) = c_1 \delta(x,y) \phi^2(y) + (c_2 R + c_3 \square) \delta(x,y) \ ,
\een
because the scaling degree of $\square \delta(x,y)$ is 6, and the dimension of $R$ is 2.
An example with 3 factors is
\ben
\D_3(\phi^2(x) \otimes \phi^3(y) \otimes \phi^3(z)) = c_4 \delta(x,y,z) \ ,
\een
because the scaling degree of the delta function with three arguments is 8.
The general pattern should be clear.
\end{example}

We summarise the renormalization ambiguities in the {\bf main theorem of renormalization theory}:

\begin{theorem}\label{Tuniqueness}\cite{hw1,hw2,hw3}
(Uniqueness) If
$\T_n$ and $\hat \T_n$ are two different renormalization schemes, both satisfying conditions T0)-T11), then their difference is given by \eqref{unique1}, for any $F = \int \O \wedge f$, $\O \in \P^p(\M)$ and $f \in \bigwedge_0^{4-p}(\M)$. The functionals $\D_n$ are maps as specified in~\eqref{Dndef} and satisfy:
\begin{enumerate}
\item[(i)] $\D(\e^F_\otimes)=O(\hbar)$.
\item[(ii)] Each $\D_n$ is locally and covariantly constructed from $g$.
\item[(iii)] Each $\D_n$ is an analytic functional of $g$.
\item[(iv)]  Each $\D_n(\O_1(x_1) \otimes \dots \otimes \O_n(x_n))$ is a distribution that is supported on the total diagonal ($=$ \lq contact term' $=$ \lq delta-function type').
\item[(v)] The maps $\D_n$ are real in the sense of~\eqref{unitary}.
\item[(vi)] Each $\D_n$ is symmetric.
\item[(vii)] Each $\D_n$ satisfies the dimension constraint in \eqref{dimension}.
\item[(viii)] Derivatives can be pulled into $\D_n$, as in \eqref{pullin}.
\item[(ix)] $\hbar \tfrac{\rm d}{{\rm d} \hbar} \D(\e^F_\otimes) = -\D[(S_0+F) \otimes \e_\otimes^F]$, where $S_0$ is the action of the free theory.
\end{enumerate}
Conversely, if $\D_n$ has these properties, then any $\hat \T$ given by~\eqref{unique1} defines a new renormalization scheme with the properties T0)-T11).
\end{theorem}

\noindent
{\em Proof:} The complete proof is given in \cite{hw1,hw2,hw3}, except for (ix).
This is related to the principle of perturbative agreement, which is yet another
renormalization condition that may be imposed~\cite{hw3}. We omit the proof.


\subsection{Time-ordered products in supersymmetric gauge theory}

So far, we have only considered the toy model of a scalar field $\phi$ with $\eL_0 = - \half g^{\mu\nu} \partial_\mu \phi \partial_\nu \phi - \tfrac{1}{12} R\phi^2$. We now move on to consider supersymmetric gauge theory in curved spacetime, in the particular context of the $\N =2$ theory described in section~\ref{sec:BRSTNequals2}. To render this theory amenable to perturbative techniques, we incorporate the familiar ghosts $( X,\alpha^{AB},\epsilon^A_+)$ for the conformal symmetry superalgebra $\cS$, $c$ for the gauge algebra $\cG$ and the $(B, {\overline c} )$-system, together with all the associated anti-fields, just as in section~\ref{sec:QSYM}.

The starting point is the extended action
\ben
\hat S = \int_\M \hat \eL \dv - \int_\M \Q \Phi \cdot \hat \Phi \, \dv + \Q \eG~,
\een
where $\hat \eL$ is defined in \eqref{Lhat}, and the gauge-fermion
\ben
\eG = \int_\M \left( ( \nabla^\mu A_\mu ) \bar c - \frac{t}{2} B \bar c \right) \dv~,
\een
 for some $t \in (0,1]$. Taking $t=1$ corresponds to Feynman gauge while the $t \to 0$ limit corresponds to Laudau gauge. All the dynamical fields in Tables~\ref{table2} and~\ref{table3} are written collectively as $\Phi$, with associated anti-fields ${\hat \Phi}$. The gauge-fixing term $\Q \eG$ yields the standard kinetic terms for $\Phi$.

Now split
\ben
\hat S=\hat S_0 + \hat S_1 \ ,
\een
where $\hat S_0$ contains all terms of up to quadratic order in $( \Phi , {\hat \Phi})$, but with arbitrary dependence on the non-dynamical ghosts $(X,\alpha^{AB},\epsilon^A_+ )$. All non-dynamical fields and anti-fields are considered as classical sources. Their products are taken to be the classical products in the definition of the algebra $\W(\M,g)$ for $\hat S_0$. We omit the lengthy details of the precise definitions. Let $\hQ_0$ denote the part of the differential $\hQ$ (obtained from \eqref{qhatdef} , \eqref{b1}, \eqref{b2}, \eqref{b3}, \eqref{b4}) which is linear in all fields except the non-dynamical ghosts $( X,\alpha^{AB},\epsilon^A_+ )$. It follows that $\hQ_0^2=0$, with the following fundamental \lq zero curvature' equation
\ben
\label{zcurvature}
\hQ_0 \hat S_1 - \half (\hat S_1, \hat S_1) = 0~,
\een
which expresses that
\ben
\label{hq1}
\hQ = \e^{(\eG,-)} \circ \Q \circ \e^{-(\eG,-)} = \hQ_0 - (\hat S_1, - )~,
\een
is a nilpotent differential. The last relation follows from the formalism in section~\ref{sec:brstextension} (c.f. \eqref{qhatdef}). Whence, $H^n(\Q | \d, \M)$ and $H^n(\hQ | \d, \M)$ are just related by $\eG$-equivariance.

The following key theorem describes the relationship between a choice of renormalization scheme and symmetries of the corresponding quantum field theory:

\begin{theorem}\label{Athm} ({\bf Anomalous Ward Identity}) \cite{h1} Let $( \M ,g )$ be a globally hyperbolic Lorentzian four-manifold which admits a twistor spinor. $\hQ_0$ can be extended to a graded derivation of $\W( \M ,g)$. For any renormalization scheme satisfying T0)-T11), it follows that
\ben\label{WA}
\boxed{
\\
\hQ_0 \T \left( \e^{iF/\hbar}_\otimes \right) = \tfrac{i}{\hbar}
\T\left( (\hQ_0 F - \half (F,F)) \otimes \e^{iF/\hbar}_\otimes \right) +
\tfrac{i}{\hbar} \T \left( \A (\e^{F}_\otimes ) \otimes \e^{iF/\hbar}_\otimes \right)~,
\\
}
\een
where $F = \int f \O$ for any smeared local field with $\O \in \P^4(\M)$, and any smooth $f$ with compact support. The {\bf anomaly}
\ben
\A(\e^F_\otimes) = \sum_{n \ge 0} \frac{1}{n!} \A_n(F^{\otimes n})~,
\een
where $\A_n: \P^{k_1}(\M) \otimes \dots \otimes \P^{k_n}(\M) \rightarrow
\P^{k_1/.../k_n}(\M^n)\rf{\hbar}$, which obeys (c.f. Theorem~\ref{Tuniqueness}):
\begin{enumerate}
\item[(i)] $\A(\e^F_\otimes)$ is $O(\hbar)$.
\item[(ii)] Each $\A_n$ is locally and covariantly constructed from $g$, and is an analytic functional of $g$.
\item[(iii)] Each $\A_n$ increases the ghost number by one unit.
\item[(iv)]  Each $\A_n(\O_1(x_1) \otimes \dots \otimes \O_n(x_n))$ is supported on the total diagonal.
\item[(v)] The maps $A_n$ are real, in the sense that $\A(\e^F)^* = \A(\e^{F^*})$.
\item[(vi)] Each $\A_n$ is graded symmetric.
\item[(vii)] Each $\A_n$ satisfies the dimension constraint analogous to~\eqref{dimension}.
\item[(viii)] Derivatives can be pulled into $\A_n$.
\item[(ix)] $\hbar \tfrac{\rm d}{{\rm d} \hbar} \A(\e_\otimes^F) = -
\A((\hat S_0 + F) \otimes \e^F_\otimes)$.
\item[(x)] $\A_n(\O_1(x_1) \otimes \dots \otimes \O_n(x_n))=0$ if one entry contains no dynamical field in Tables~\ref{table2} and~\ref{table3}.
\end{enumerate}
\end{theorem}

The proof proceeds by expanding the equation out to arbitrary powers $n$ in $F$. It is inductive in nature, showing that the anomalous Ward identity holds at order $n$ if it holds up to order $n-1$, modulo a contribution supported on the total
diagonal. That contribution is defined to be $\A_n$. The details are as in the proof of Proposition 3 in \cite{h1} (see also \cite{reijzner}), which was based partly on \cite{df}.

The connection with the criterion for the preservation of symmetries in quantum field theory described at the beginning of this section may now be explained.
Let us fix a representation $\pi: \W(\M,g) \to {\rm End}(\H)$ on a Hilbert space such that all anti-fields are represented trivially by 0. Now one can show that there exists a Hermitian operator ${\mathscr Q}_0 \in {\rm End}(\H)$
implementing the graded derivation $\hat \Q_0$, i.e. such that $\frac{i}{\hbar} [{\mathscr Q}_0, \pi(-)] = \pi(\hat \Q_0 -)$. In an asymptotically flat $(\M,g)$, one might hope that a scattering matrix can be defined by the \lq adiabatic limit'
\ben\label{adiabatic}
{\Sm} = \pi(\T(\exp_\otimes \left( \tfrac{i}{\hbar} \hat S_1 \right) )) = \lim_{f \to 1}
\sum_n \tfrac{1}{n!} \left( \tfrac{i}{\hbar} \right)^n
\int_{\M^n} \pi(\T_n(\O_1(x_1) \otimes \cdots \otimes \O_1(x_n)))f(x_1) \cdots f(x_n) \ ,
\een
where $\int \O_1 = \hat S_1$, and $f$ is a smooth cutoff function which tends to $1$ in the limit. We will not address the rather difficult question of whether (and in what precise sense) this limit really exists. Indeed, even in Minkowski space, this is a difficult question to answer because it concerns infra-red properties of the S-matrix.
\footnote{For a thorough mathematical existence proof in the context of massive scalar fields in Minkowski space, see \cite{eg2}.}
The expression \eqref{adiabatic} for the S-matrix corresponds exactly to \eqref{smatrix} in our framework. Suppose now that $\A(\e_\otimes^{\hat S_1} )=0$. Then taking formally $F=\hat S_1$ in the Ward identity~\eqref{WA} and using the zero curvature condition~\eqref{zcurvature}, we get $[{\mathscr Q}_0, {\Sm} ]=0$. Since ${\mathscr Q}_0$ incorporates all the symmetries of the theory (i.e. the BRST structure of $\cS$ and $\cG$) then, if the adiabatic limit exists, it follows that $\Sm$ is both gauge-invariant and invariant under the conformal symmetry superalgebra.

Thus, the key question is whether there exists a renormalization scheme in which $\A(\e_\otimes^{\hat S_1} )=0$? Note that this question is mathematically well-posed, irrespective of whether $\Sm$ is well-defined. Indeed, while the
adiabatic limit~\eqref{adiabatic} may not exist due to infra-red divergences, these may never occur in the corresponding series expansion for $\A(\e_\otimes^{\hat S_1} )$, which is always well-defined -- and can in fact be interpreted as being determined by the ultra-violet behaviour. The point is that $\A_n$ is supported on the diagonal, as opposed to $\T_n$. The answer to the question depends on the nature of $( \M,g)$ -- or rather the nature of the associated conformal symmetry superalgebra $\cS$ -- and on the representation-theoretic data $(\fg,U)$ for the vector and hyper multiplets. For this, we must learn more about the possible structure of the anomaly. The following consistency condition is essential:

\medskip
\noindent
\begin{proposition}
({\bf Consistency condition})~\cite{h1} The anomaly satisfies the equation
\begin{equation}\label{anomaly}
\boxed{
\\
\hQ_0 \A(\e^F_\otimes) -\left( F, \A(\e^F_\otimes) \right) -
\A\left( (\hQ_0 F - \half (F,F)) \otimes \e^F_\otimes \right)
=  \A \left( \A(\e^F_\otimes)
\otimes \e^F_\otimes \right)~,
\\
}
\end{equation}
for any $F=\int f\O$, with $\O \in \P^4(\M)$ and $f$ having compact support.
\end{proposition}

\medskip
\noindent
{\em Proof:}
In order to give a flavour of the arguments, we repeat the proof of this proposition. First act with $\hQ_0$ on the anomalous Ward identity \eqref{WA} and use that
$\hQ_0^2=0$. This gives
\ben
0= \hQ_0 \T\bigg( \A(\e^F_\otimes) \otimes \e^{iF/\hbar}_\otimes \bigg) +
\hQ_0 \T \bigg(
(\hQ_0 F - \half (F,F)) \otimes \e_\otimes^{iF/\hbar}
\bigg) = {\rm (I)} + {\rm (II)}
\een
The trick is now to apply the anomalous Ward identity one more
time to each of the terms on the right side. For simplicity,
we assume that $F$ has Grassmann parity 0. We can then write the
first term as (derivative at $\tau=0$)
\bena
{\rm (I)} &=& \frac{\hbar}{i}
\frac{\d}{\d \tau} \hQ_0 \T \bigg( \e^{i(F + \tau \A(\e^F))/\hbar}_\otimes \bigg)  \non\\
&=&
\frac{\d}{\d \tau}
\T \bigg( \Big(\hQ_0(F + \tau \A(\e^F_\otimes))
- \half (F + \tau \A(\e^F_\otimes),F + \tau \A(\e^F_\otimes)\Big) \otimes
\e^{i(F + \tau A(\e^F_\otimes))/\hbar}_\otimes \bigg) \non\\
&&
+\frac{\d}{\d \tau}\T \bigg( \A(\e^{\tau \A(\e^F_\otimes)}) \otimes \e^{i(F + \tau \A(\e^F_\otimes))/\hbar}_\otimes\bigg)
\non\\
&=&
\T\bigg( (\hQ_0\A(\e^F_\otimes)-(F,\A(\e_\otimes^F))) \otimes \e^{iF/\hbar}_\otimes \bigg)
+ \frac{i}{\hbar} \T\bigg( \A(\e^F_\otimes) \otimes (\hQ_0 F - \half (F,F)) \otimes
\e^{iF/\hbar} \bigg) \non\\
&&
-\T \bigg( \A(\A(\e^F_\otimes) \otimes \e^F_\otimes) \otimes \e^{iF/\hbar}_\otimes \bigg)
+ \T \bigg( \A(\e^F_\otimes) \otimes \A(\e^F_\otimes ) \otimes \e^{iF/\hbar}_\otimes \bigg)~.
\eena
Since $F$ has Grassmann parity 0, $\A(\e^F_\otimes)$ has Grassmann parity 1,
so by the anti-symmetry of the time-ordered products for such
elements, the last term vanishes. Next, we apply
the anomalous Ward identity to term (II). We now obtain
\bena
{\rm (II)} &=& \frac{\hbar}{i}
\frac{\d}{\d \tau} \hQ_0 \T \bigg( \e^{i(F + \tau (\hQ_0 F - \half (F,F))/\hbar}_\otimes\bigg) \\
&=&
\T \bigg( (\hQ_0(\hQ_0 F - \half (F,F))  \half (F,\hQ_0 F - \half (F,F))) \otimes
\e^{i(F + \tau (\hat S_0+F,\hat S_0+F))/\hbar}_\otimes \bigg) \non\\
&&+
\frac{\d}{\d \tau}
\T \bigg( \A(\e^{\tau (\hQ_0 F - \half (F,F))}_\otimes) \otimes \e^{i(F + \tau
  (\hQ_0 F - \half (F,F)))_\otimes/\hbar}
\bigg)
\non\\
&=&
\frac{i}{\hbar} \T\bigg( (\hQ_0 F - \half (F,F)) \otimes (\hQ_0 F - \half (F,F)) \otimes
\e^{iF/\hbar}_\otimes \bigg) \non\\
&&
- \T \bigg( \A((\hQ_0 F - \half (F,F)) \otimes \e^F_\otimes) \otimes \e^{iF/\hbar}_\otimes \bigg)
- \frac{i}{\hbar}
\T \bigg( \A(\e^F_\otimes) \otimes (\hQ_0 F - \half (F,F)) \otimes \e^{iF/\hbar}_\otimes \bigg)~. \non
\eena
Now, the first term on the right side vanishes using the \lq Bianchi identity'
for the \lq curvature' $\hQ_0 F - \half (F,F)$. The second term vanishes due
to the anti-symmetry of time-ordered products, since $\hQ_0 F - \half (F,F)$ has Grassmann parity 1. Adding terms (I) and (II) then gives
\ben
\T \bigg( \bigg(
\hQ_0 \A(\e^F_\otimes) - (F,\A(\e^F_\otimes))
-\A((\hQ_0 F - \half (F,F)) \otimes \e^{F}_\otimes)
-\A(\A(\e^{F}_\otimes) \otimes \e^F_\otimes)
\bigg) \otimes \e^{iF/\hbar}_\otimes \bigg) = 0 \ . \non
\een
The desired consistency condition~\eqref{anomaly} follows. \qed
\medskip
\noindent

An immediate consequence of the consistency condition is

\begin{proposition}
({\bf $\hbar$-expanded consistency condition})
Let
\ben
\A(\e^F_\otimes) = \sum_{n \ge m} \hbar^n \A^{(n)}(\e^F_\otimes)~,
\een
be the $\hbar$-expansion of the anomaly of the Ward identity. For any $F=\int f\O$, the first term in this $\hbar$-expansion satisfies
\ben\label{con}
\boxed{
\\
\hQ_0 \A^{(m)} \left(\e^{F}_\otimes \right) - \left(F, \A^{(m)} \left(\e^{F}_\otimes \right) \right)-\A^{(m)} \left( \left(\hQ_0 F - \half (F,F) \right) \otimes \e^{F}_\otimes
\right) = 0~.
\\
}
\een
\end{proposition}
This proposition is a direct consequence of the previous proposition, noting that the term on the right side of~\eqref{anomaly} must be $O(\hbar^{2m})$.

In particular, for $F=\hat S_1$, using the zero curvature condition \eqref{zcurvature}, it follows that
\ben
\hQ \A^{(m)}(\e^{\hat S_1}_\otimes) = 0~,
\een
where $\hat \Q$ is given by \eqref{hq1}. Now, from (ii), (iii), (iv) and (vii) in Theorem~\ref{Athm}, it follows that $\A^{(m)}(\e_\otimes^{\hat S_1})$ is an integral of a four-form in $\P^4_1(\M)$ of dimension $4$ and ghost number $1$. This four-form must be $\hat \Q$-closed modulo $\d$-exact terms, whence in $H^5(\hat \Q|\d, \M)$. Elements of $H^5(\hat \Q|\d, \M)$ can be obtained
simply by applying $\e^{(\eG,-)}$ to the elements in $H^5(\Q|\d, \M)$ given in part 2) of Theorem~\ref{cohothm}. Actually, due to the simple form of $\eG$,
$\e^{(\eG,-)}$ does not change them at all.

If $(\M,g)$ is conformally flat, these generate {\em all} the elements in $H^5(\hat \Q|\d, \M)$, whence
\ben\label{Acondition}
\A^{(m)}(\e_\otimes^{\hat S_1}) = -\half \beta^{(m)} \int_\M \eA \qquad \text{mod $\hat \Q$-exact}~.
\een
The real constant $\beta^{(m)}$ will soon be identified with the leading order (in $\hbar$) contribution to the $\beta$-function. Furthermore, if $(\M,g)$ is not conformally flat, it follows from the covariant nature of the anomaly that the form of \eqref{Acondition} may only be violated by terms with explicit dependence on the Weyl tensor. The left hand side must still be $\hat \Q$-closed, and \eqref{consistency} implies that it is impossible to construct $\hat \Q$-closed expressions at dimension 4 and ghost number 1, containing the Weyl-tensor if the spacetime admits a twistor spinor. Whence, the form of \eqref{Acondition} is valid on any of the spacetimes of interest. The $\hat \Q$-exact term may always be removed by changing to a new renormalization scheme, via \eqref{Dndef} and \eqref{unique1}, for a suitable choice of $\D$ (see~\cite{h1} for details of this in the present formalism in curved spacetime).

Now suppose that either:
\vspace*{-.1cm}
\begin{itemize}
\item $\eA=0$, modulo $\d$-exact terms. This occurs if $\cS$ contains only Killing vectors and parallel spinors, e.g. if $(\M, g)$ is a pp-wave. (We have already assumed that the principle $G$-bundle is trivial, so that $\Tr (F \wedge F)$ is globally exact.) It also occurs by taking $\hat \Q$ with respect to a restricted Lie superalgebra $\cS^\prime <\cS$, defined such that the only conformal Killing vectors in its even part are Killing vectors in $\cS$ and the only twistor spinors in its odd part are parallel spinors in $\cS$. For example, in Minkowski space, $\cS^\prime$ corresponds to the $\eN =2$ Poincar\'{e} superalgebra contained in $\fsu(2,2|2)$.
\item The $\beta$-function $\sum_n \hbar^n \beta^{(n)}$ vanishes at all orders in $\hbar$ (see below for the definition in our framework in curved spacetime). As is well-known, and as we will argue in our framework below, this occurs only if $h^\vee (\fg) = c(U)$. That is, if the dual Coxeter number $h^\vee (\fg)$ of the simple Lie algebra $\fg$ for the vector multiplet equals the Dynkin index $c(U)$ of the complex representation $U$ for the hyper multiplet. For completeness, the classification of all solutions to this condition is presented in detail in Appendix~\ref{app:A}.
\end{itemize}
In either case, it follows that $\A^{(n)}(\e^{\hat S_1}_\otimes )=0$ for all orders $n \le m+1$. Iterating the argument at each order yields a renormalization scheme with vanishing anomaly at all orders.

In summary, on a globally hyperbolic Lorentzian four-manifold $(\M,g)$ which admits a twistor spinor,  the conformal symmetry superalgebra $\cS$ is realised at the quantum level if and only if either:

--- $(\eM,g)$ is a pp-wave that is not conformally flat and has no \lq proper' conformal Killing vectors.

--- The data $(\fg,U)$ for the vector and hyper multiplet obeys $h^\vee(\fg) = c(U)$.

Or else only a Lie superalgebra $\cS^\prime <\cS$, with Killing vectors from the even part of $\cS$ and parallel spinors from the odd part of $\cS$ can be realised at the quantum level.

Accepting for the moment our interpretation of $\beta^{(m)}$ in \eqref{Acondition} as the leading order term in the $\beta$-function allows us to provide a simple argument why the $\beta$-function must be one-loop exact and given by~\eqref{beta}. It suffices to consider the special case of~\eqref{Acondition} where the ghosts $X$ and $\epsilon_+^A$ are set to zero. Then $\A(\e_\otimes^{\hat S_1})$ is nothing but the global anomaly for the ${\mathrm{U}}(1)$ part of the R-symmetry. This is known to be one-loop exact, i.e. linear in $\hbar$, by the Adler-Bardeen theorem (see~\cite{kopper1} for a rigorous argument). Using the R-charges in Table~\ref{table2} and the representation-theoretic data $(\fg,U)$ for the vector and hyper multiplets, one recovers the standard expression for the one-loop contribution to the anomaly
\ben\label{Acondition1}
\A^{(1)}(\e_\otimes^{\hat S_1}) = \tfrac{i}{4\pi^2} \, \alpha \left( h^\vee(\fg )- c(U) \right) \int_\M \tr(F \wedge F)~,
\een
whereas $\A^{(m>1)}(\e_\otimes^{\hat S_1})=0$, having set $X$ and $\epsilon_+^A$ to zero. Comparing this with \eqref{Acondition}, and using the form of
$\eA$ from Theorem~\ref{cohothm}, leads to the result that $\beta = \hbar \, \beta^{(1)}$ is one-loop exact, and given by~\eqref{beta}.

We may also contemplate what happens if $(\M,g)$ does not admit a twistor spinor. In that case, the odd part of $\cS$ is absent and symmetries are described by the Lie algebra $\cB = \fX^c(\M) \oplus \fu(2)$. The right hand side of \eqref{Acondition} can now have additional terms which are $\hQ$-closed and vanish if $(\M,g)$ admits a twistor spinor. These are precisely the curvature terms in 2b) of Theorem~\ref{cohothm}.
Thus, even if $\beta=0$, \eqref{Acondition} becomes
\ben\label{Acondition2}
\A^{(m)}(\e_\otimes^{\hat S_1}) = \re \Big( c^{(m)} \alpha \int_\M \tr({}^* \bm{C} \wedge \bm{C}
+ i \; \bm{C} \wedge \bm{C}) \Big)
 \qquad \text{mod $\hat \Q$-exact}~,
\een
for some $c^{(m)} \in \CC$, at leading order $\hbar^m$. The part of the right hand side which is not $\hat \Q$-exact cannot be removed by changing the renormalization scheme, but fortunately it does not contribute. Indeed, by a version of the Adler-Bardeen theorem one finds that the only non-trivial term is $c^{(1)}$, which is imaginary. Hence, the corresponding anomaly term is proportional to the Pontryagin class, which vanishes identically on any globally hyperbolic spacetime.


\subsection{Renormalization group}

Here we relate the $\beta$-function, as defined via the renormalization group, to the numerical coefficient on the right hand side of the anomaly in \eqref{Acondition}. In Minkowski space, the renormalization group is often defined by the behaviour of operators in quantum field theory under dilatations,  which map coordinates $x \mapsto \mu x$, for some positive real number $\mu$. On a generic curved spacetime, this procedure would not be covariant since there is no conformal Killing vector analogous to $x^\alpha \partial_\alpha$. However, in Minkowski space, the dilatation $x \mapsto \mu x$ is of course equivalent to a rescaling of the metric by the constant conformal factor $\mu^2$. On a generic curved spacetime, it is this notion which leads to a meaningful definition of the renormalization group \cite{hw4}, without reference to conformal Killing vectors. We now recall how this is done, using the notion of renormalization scheme that was defined above.

Given a renormalization scheme $\T_n$ on a curved spacetime $(\M,g)$, then changing $g \mapsto \mu^2 g$ implies that $\T_n \mapsto \T^{(\mu)}_n$, as in \eqref{tmudef}. We would like to understand more clearly how $\T^{(\mu)}_n$ differs from $\T_n$. Since $\T^{(\mu)}_n$ is just another renormalization scheme satisfying T0) to T11) (but which happens to be parameterised by $\mu$), we can apply the main theorem of renormalization theory, given by Theorem~\ref{Tuniqueness}. According to \eqref{unique1}, this gives
\ben
\label{unique2}
\T^{(\mu)}\left(\e^{iF/\hbar}_\otimes  \right) = \T \left(
\exp_\otimes \left( \tfrac{i}{\hbar}(F + \D^{(\mu)}(\e_\otimes^F)) \right) \right)~,
\een
where $\D^{(\mu)}_n$ characterises the difference between the two renormalization schemes. The scattering matrix follows by taking $F=S_1$ for the interaction terms in the action (e.g. $S_1 = \frac{\lambda}{4!} \int_\M \phi^4 \dv$ for a scalar field $\phi$ with quartic self-interaction). We refer to
\ben
S^{(\mu)} = S + \D^{(\mu)}(\e_\otimes^{S_1}) = S +
\sum_{n=2}^\infty \frac{1}{n!} \, \D_n^{(\mu)}(S_1^{\otimes n})~,
\een\label{smu}
as the {\bf running action}.

\begin{example}
For $S_1 = \frac{\lambda}{4!} \int_\M \phi^4 \dv$, let us present the
running action to lowest order $\lambda^2$ in $\lambda$. We need $\D_2^{(\mu)}(S_1 \otimes S_1)$, so we must compute $\D_2^{(\mu)}(\phi^4 \otimes \phi^4)$. By Theorem~\ref{Tuniqueness}, this must be linear in $\log \mu$ and the sum of its dimension and scaling degree must be eight. Using the methods of \cite{hw4}, one finds that
\ben
\D_2^{(\mu)}(\phi^4(x) \otimes \phi^4(y)) = \frac{36}{\pi^2} \log \mu \ \delta(x,y) \phi^2(x) \phi^2(y) -\log \mu \left( \frac{3}{2 \pi^4} R
-\frac{9}{8\pi^4}  \square \right) \delta(x,y) \phi(x)\phi(y) ...
\een
Dots represent similar terms quadratic
in the curvature, or terms containing derivatives of $\phi$. The
running action at order $\lambda^2$ is obtained by multiplying by
$(\tfrac{1}{24} \lambda)^2$ and integrating over $x$ and $y$.
\end{example}
Generalising the above construction, we may also let a general operator $\O \in \P(\M)$ run, with $\O^{(\mu)} = \O + \D^{(\mu)}(\O \otimes \e_\otimes^{S_1})$. The infinitesimal
version of this is
\ben\label{gammadef}
\mu \frac{\d}{\d \mu} \O^{(\mu)} \ \Big|_{\mu=1} = \gamma \O \equiv
\sum_{n=1}^\infty \frac{1}{n!} \, \D_n^{(\mu)}( \O \otimes S_1^{\otimes n}) \ .
\een
The operation $\gamma: \P(\M) \to \P(\M)$ is a linear map which preserves dimension. Since there are only a finite number of operators at a given dimension $\Delta$, we may expand
$\gamma$ in a basis, leading to a matrix, sometimes referred to as the {\bf mixing matrix}. At operator dimension 1, there is only one field $\phi$, so
the mixing matrix is just a number.

In principle, for a supersymmetric gauge theory, the definition of the renormalization group is just the same. However, there are constraints on the running action due to the presence of further symmetries. To derive these, we need the following proposition:

\begin{proposition}
Let $F = \int_\M f \wedge \O$, with $f$ having compact support and $\O \in \P^4(\M)$.
Then
\ben\label{anomalymu}
\begin{split}
&\hQ_0 \D^{(\mu)}(\e_\otimes^F) - \left( F , \D^{(\mu)}(\e_\otimes^F) \right)
+ \frac{1}{2}\, \left(\D^{(\mu)}(\e_\otimes^F), \D^{(\mu)}(\e_\otimes^F)\right) -  \D^{(\mu)} \left(
(\hQ_0 - \half (F,F)) \otimes \e^F_\otimes
\right) \\
&= \D^{(\mu)}\left( \A^{(\mu)}(\e_\otimes^F) \otimes \e^F_\otimes \right) -
\A \left(
\exp_\otimes(F + \D^{(\mu)}(\e_\otimes^F)) \right) + \A^{(\mu)}\left( \e^F_\otimes \right)~,
\end{split}
\een
in the sense of formal power series in $F$.
\end{proposition}

\noindent
{\em Proof:}
We evaluate the term on the left hand side of the anomalous Ward identity $\hQ_0 \T^{(\mu)} \left(\e^{iF/\hbar}_\otimes \right)$ in two different ways: (i) Apply the anomalous Ward identity for the time-ordered product $\T_n^{(\mu)}$, and then apply identity \eqref{unique2}. (ii) Apply the same steps in the opposite order.

Method (i) gives
\bena
\hQ_0 \T^{(\mu)}\left(\e^{iF/\hbar}_\otimes \right) &=& \frac{i}{\hbar}
\T^{(\mu)}\left( (\hQ_0 F - \half (F,F)) \otimes \e^{iF/\hbar}_\otimes \right) +
\frac{i}{\hbar}
\T^{(\mu)} \left( \A^{(\mu)}(\e^{F}_\otimes) \otimes \e^{iF/\hbar}_\otimes \right) \non\\
&=&
\frac{i}{\hbar}
\T\left( (\hQ_0 F - \half (F,F)) \otimes \exp_\otimes \left( \frac{i}{\hbar} (F+
\D^{(\mu)}(\e_\otimes^F)) \right) \right) \nonumber \\
&&+
\frac{i}{\hbar}
\T\left( \D^{(\mu)}\left( (\hQ_0 F - \half (F,F)) \otimes \e^F_\otimes \right) \otimes \exp_\otimes \left( \frac{i}{\hbar} (F+
\D^{(\mu)}(\e_\otimes^F)) \right) \right) \nonumber \\
&&+
\frac{i}{\hbar}
\T\left( \A^{(\mu)}(\e^{F}_\otimes) \otimes  \exp_\otimes \left( \frac{i}{\hbar} (F+
\D^{(\mu)}(\e_\otimes^F)) \right) \right) \nonumber \\
&&+
\frac{i}{\hbar}
\T\left( \D^{(\mu)}\left( \A^{(\mu)}(\e^{F}_\otimes) \otimes \e^F_\otimes \right) \otimes  \exp_\otimes \left( \frac{i}{\hbar} (F+
\D^{(\mu)}(\e_\otimes^F)) \right) \right)~.
\eena
Method (ii) gives
\bena
\hQ_0 \T^{(\mu)}\left(\e^{iF/\hbar}_\otimes \right) &=&\hQ_0 \T \left(
\exp_\otimes \left( \frac{i}{\hbar}(F + \D^{(\mu)}(\e_\otimes^F)) \right) \right)\nonumber\\
&=& \frac{i}{\hbar}
\T\left( \left( \hQ_0(F+ \D^{(\mu)}(\e_\otimes^F)) - \half (F+\D^{(\mu)}(\e_\otimes^F) ,F+ \D^{(\mu)}(\e_\otimes^F)) \right) \right. \nonumber\\
&&\hspace*{8cm} \left. \otimes
\exp_\otimes \left( \frac{i}{\hbar}(F + \D^{(\mu)}(\e_\otimes^F)) \right) \right)  \nonumber\\
&&+\frac{i}{2\hbar}
\T\left( \A \left( \exp_\otimes (F+\D^{(\mu)}(\e_\otimes^F) ) \right) \otimes
\exp_\otimes \left( \frac{i}{\hbar}(F + \D^{(\mu)}(\e_\otimes^F)) \right) \right)~.
\eena
Comparing both results yields the desired expression. \qed

An application of the functional identity is as follows.
\begin{lemma}\label{commutatorlemma}
Let $F=\hat S_1$ be the interaction and suppose that $\A(\e_\otimes^F)=0$, in addition to T0)-T11). Then there holds the following functional equation
\ben\label{funcid}
[\hat \Q_\hbar, \gamma] \O = \A'(\O \otimes
\e^F_\otimes) + (\hat S', \O) + \A(\O \otimes \hat S' \otimes
\e^F_\otimes)~,
\een
where $\hat S'= \mu \tfrac{\rm d}{{\rm d} \mu} \D^{(\mu)}(\e^{F}_\otimes)$, $\A'(\e^F_\otimes) = \mu \tfrac{\rm d}{{\rm d} \mu} \A^{(\mu)}(\e^F_\otimes)$, and  $\hat \Q_\hbar = \hat \Q + \A( - \otimes \e^F_\otimes) $.
Moreover, $\hat \Q_\hbar^2 = 0$ and $\hat \Q_\hbar {\rm d} +
{\rm d} \hat \Q_\hbar =0$.
\end{lemma}

\begin{remark}
The map $\hat \Q_\hbar: \P^p_q \to \P^p_{q+1} \otimes \mc \rf{\hbar}$ is called the {\bf quantum differential}. Its cohomology can be seen to be isomorphic to that of $\hat \Q$, in the sense that each cohomology class of $\hQ$ can be perturbed (in the sense of a formal power series in $\hbar$) to a cohomology class of $\hQ_\hbar$.
\end{remark}

\noindent
{\em Proof:} In~\eqref{funcid}, any commutator with $\d$ is automatically zero as a consequence of (viii) in Theorem~\ref{Tuniqueness}. In the consistency condition~\eqref{anomalymu}, take $F=\hat S_1 + \int f\O$. Then act with $\mu \tfrac{\d}{\d \mu}$ at $\mu=1$, functionally differentiate with respect to $f$, and use~\eqref{zcurvature}. The functional equation $\hat \Q_\hbar^2=0$ follows from \eqref{anomaly}. \qed

Another lemma is as follows.

\begin{lemma}\label{stabilityRG}
In Minkowski space, if $\hat \Q$ is defined with respect to the $\N=2$ Poincar\'{e} superalgebra in $\fsu(2,2|2)$, there is a renormalization scheme such that
\ben
\hat S' = \beta \int_\M \hat \eL \dv + \hat \Q \Psi \ ,
\een
where $\hat \eL$ is as in Theorem~\ref{cohothm} and $\hat S'=\mu \tfrac{\rm d}{{\rm d} \mu} \hat S^{(\mu)}$, in terms of the running action \eqref{smu}. The prefactor $\beta = \sum_n \hbar^n \beta^{(n)}$ is a formal power series in $\hbar$ and corresponds to the unique $\beta$-function of the theory. On a curved spacetime $(\M,g)$, the same is true if all the conformal Killing vectors in $\cS$ are Killing vectors and all the twistor spinors in $\cS$ are parallel spinors.
\end{lemma}

\begin{remark}
$\;$ \\ [.1cm]
1) The term $\hat \Q \Psi$ may always be removed by passing to a new
renormalization scheme via the same type of argument in the proof based on \eqref{anomaly}. \\ [.1cm]
2) In the literature, it is not uncommon to set $\hbar={\tt g}^2$ (in terms of the gauge coupling ${\tt g}$) and to rescale all the fields $\Phi \leadsto  {\tt g} \Phi$. This eliminates all dependence on ${\tt g}$ from quadratic terms in the action. Furthermore, $\beta \leadsto {\tt g}^2 \beta$ yields a formal power series in ${\tt g}^2$ (rather than $\hbar$) which can be identified with the $\beta$-function for the gauge coupling.
\end{remark}

\noindent
{\em Proof:} Take $F=\hat S_1$ in \eqref{anomalymu} and differentiate with respect to $\log \mu$ at $\mu=1$. In Minkowski space, if
$\Q$ is defined with respect to the $\N =2$ Poincar\'{e} superalgebra in $\fsu(2,2|2)$, the anomaly candidate $\eA=0$ (c.f. \eqref{Adef}). This is because, in the $\N =2$ Poincar\'{e} superalgebra, all conformal transformations are translations and Lorentz rotations, and there are only Poincar\'{e} supersymmetry transformations generated by constant spinors. That is, all the conformal Killing vectors are Killing vectors and all the twistor spinors are parallel spinors. Consequently, $\A(\e^F_\otimes)=0$. The same is true for a pp-wave with no \lq proper' conformal isometries.

With this, we get
\ben
0=\hQ_\hbar \hat S' = \hQ \hat S' + \A(\hat S' \otimes \e^{F})~,
\een
where $\hat S'$ has dimension $4$ and ghost number $0$. Since the cohomologies of $\hQ$ and $\hQ_\hbar$ are isomorphic, we conclude from Theorem~\ref{cohothm} that $\hat S'$ must be a multiple of a suitable $\hbar$-deformation of $\Omega = \int \hat \eL \dv$ in the kernel of $\hQ_\hbar$, modulo $\hQ_\hbar$-exact terms. We claim that $\Omega$, which is a priori in the kernel of $\hQ$, is also in the kernel of $\hQ_\hbar$. Using (ix) of Theorem~\ref{Athm}, it also follows that $\A(\hat S \otimes \e^F_\otimes)=\A((\hat S_0+F)\otimes \e^F_\otimes) = -\hbar \tfrac{\rm d}{{\rm d} \hbar} \A(\e^F_\otimes) = 0$. Whence, $\A(\Omega \otimes \e^F_\otimes) - \A(( \Q \Phi \cdot \hat \Phi - \Q \eG) \otimes \e^F_\otimes) = 0$, and since these terms have different anti-field dependence, they must be zero individually. Therefore $\A(\Omega \otimes \e^F_\otimes )=0$, and since $\hQ \Omega=0$ from Theorem~\eqref{cohothm}, it follows that $\hQ_\hbar \Omega=0$.

In other words, $\hat S' = C \Omega + \hQ_\hbar \Psi$ for some formal power series $C = \sum_n
C^{(n)} \hbar^n$, or equivalently $\hat S' = C \Omega + \hat\Q\Psi+\A(\Psi \otimes \e_\otimes^F)$. We would like to remove the unwanted last term by changing the renormalization scheme. As before, we expand $\A(\Psi \otimes \e^F_\otimes) = \sum \hbar^n A^{(n)}(\Psi \otimes \e^F_\otimes)$ in a formal power series in $\hbar$. We may assume
\footnote{
We may always achieve that $\A(\hQ \O \otimes \e_\otimes^F) = 0$ by a change of renormalization scheme. This may be proved by noting
that, to leading order in $\hbar$, $\A^{(m)}(\hat \Q \O \otimes \e^F_\otimes) = \hat \Q \A(\O \otimes \e^F_\otimes)$. We then set
$\D_m(\hat \Q \O \otimes F \dots \otimes F) = -\A^{(m)}(\O \otimes \e^F_\otimes)/m!$, which achieves that for the new renormalization scheme
defined by \eqref{unique1}, $\hat \A(\hat \Q \O \otimes \e^F_\otimes)$ starts at order $O(\hbar^{m+1})$. We may continue this process iteratively
and remove the anomaly to all orders in $\hbar$.
}
that $\A(\hQ \Psi \otimes \e^F_\otimes) = 0$. Then the first non-trivial $\hbar$-contribution must satisfy
$\hat \Q \A^{(m)}(\Psi \otimes \e^F_\otimes) = 0$, from our $\hbar$-expanded consistency condition. It follows that $\A^{(m)}(\Psi \otimes \e^F_\otimes) = b^{(m)} \hbar^m \Omega + \hat \Q\Phi^{(m)}$. We define a new renormalization scheme $\hat \T$ for time-ordered products by setting $\D_{m}(\Psi \otimes F \dots \otimes F) = -m! \Phi^{(m)}$ in \eqref{unique1}. By construction, the new renormalization scheme (we omit the hat) has
$\A^{(m)}(\Psi \otimes \e^F_\otimes) = \beta^{(m)} \hbar^m \Omega$. We proceed  inductively with this process, such that
\ben\label{induct}
\A(\Psi \otimes \e^F_\otimes) = (b^{(m)} \hbar^m + \cdots + b^{(m+j)} \hbar^{m+j}) \Omega + O(\hbar^{m+j+1}) \ .
\een
Taking $\hQ$ of this identity and using the $\hbar$-expanded consistency condition gives
\ben
\hat \Q \A^{(m+j+1)}(\Psi \otimes \e^F_\otimes) = \A^{(m+j+1)}(\hat \Q \Psi \otimes \e^F_\otimes) +
\sum_{i=0}^j \A^{(i+1)}(\A^{(m+j-i)}(\Psi \otimes \e^F_\otimes) \otimes \e^F_\otimes) =0~,
\een
using the inductive assumption, and using $\A(\Omega \otimes \e^F_\otimes) = 0 = \A(\hat \Q\Psi \otimes \e^F_\otimes)$. It follows that $\A^{(m+j+1)}(\Psi \otimes \e^F_\otimes) = b^{(m+j+1)} \hbar^{m+j+1} \Omega + \hQ\Phi^{(m+j+1)}$. We define a new renormalization scheme $\hat \T$ for time-ordered products by setting $\D_{m+j+1}(\Psi \otimes F \dots \otimes F) = -(m+j+1)! \Phi^{(m+j+1)}$ in \eqref{unique1}. By construction, the new renormalization scheme (we omit the hat) satisfies \eqref{induct}, with $j$ increased by one unit. Therefore, we may inductively assume that \eqref{induct} is
satisfied for all $j$, and we conclude that $\A(\Psi \otimes \e^F_\otimes) = \left( \sum_j b^{(j)} \hbar^j \right) \Omega$ for this new renormalization scheme. Consequently, the result stated in the lemma holds, with $\beta = C + \sum_j b^{(j)} \hbar^j$. \qed

As we have already argued, to leading order in $\hbar$, the anomaly $\A(\e^F_\otimes)$ is proportional to the anomaly candidate $\int \eA$ plus a $\hat \Q$-exact contribution (see \eqref{Acondition}) which we may remove by passing to a new renormalization scheme. It only remains to show that the coefficient in front of $\int \eA$ is indeed the one-loop $\beta$-function. If this is true in Minkowski space then, since the expression for $\A(\e^F_\otimes)$ is generally covariant and the coefficient is constant, it must also be true on any curved spacetime which admits a twistor spinor.

On $\M =\RR^{3,1}$, consider the conformal Killing vector $X = x^\alpha \partial_\alpha$. Its one-parameter flow $\psi_\mu$ describes dilatations, and clearly $\psi_\mu^* g = \mu^2 g$ for the Minkowski metric $g=\eta$. Recall that the running time-ordered products $\T^{(\mu)}$ on $(\M,g)$ were defined from those on $(\M,\mu^2 g)$ in T2). Furthermore, there exists a relationship between the time-ordered products associated with $g$ and $\mu^2 g$ due to T1) via $\psi_\mu$. Combining these relations, one gets a relationship between $\T^{(\mu)}$ and $\T$ on the same spacetime $(\M,g)$, in this case Minkowski space. Differentiating this relationship with respect to $\log \mu$, and using \eqref{unique1}, it follows that
\ben
\label{eq:Nequals2anomaly}
\T\left(\sigma_X \mu \tfrac{\d}{\d\mu} \D^{(\mu)}(\e^F_\otimes) \otimes \e^{iF/\hbar}_\otimes\right) =
\delta_X \T\left(\e_\otimes^{iF/\hbar}\right) - \frac{i}{\hbar} \T\left(\delta_X F \otimes \e^{iF/\hbar}_\otimes \right)~,
\een
for $X \propto x^\alpha \partial_\alpha$, where $F$ corresponds to those terms in $\hat S_1$ which do not depend on $(X,\alpha^{AB},\epsilon^A_+)$, and $\delta_X = \int (\cL_X + w_\Phi \sigma_X) \Phi \cdot \frac{\delta}{\delta \Phi}$. The right hand side of this equation resembles the defining relation for the anomaly, but a few terms are missing. Let $s_0$ be the free BRST operator, defined by setting $(X,\alpha^{AB},\epsilon^A_+)$ to zero in $\hat \Q_0$. The defining relation \eqref{anomaly} for the anomaly reduces to
\ben
s_0 \T\left(\e^{iF/\hbar} \right) - \frac{i}{\hbar}\T\left(s_0 F \otimes \e^{iF/\hbar}_\otimes \right)
- \frac{i}{2\hbar} \T\left( (F,F) \otimes \e^{iF/\hbar}_\otimes \right)  =0~.
\een
There is no gauge anomaly. Adding this expression to the right hand side of \eqref{eq:Nequals2anomaly} gives
\ben
\begin{split}
\T\left(\sigma_X \, \mu\tfrac{\d}{\d \mu} \D^{(\mu)}(\e^F_\otimes) \otimes \e^{iF/\hbar}_\otimes \right) &=
\hQ_0 \T\left( \e_\otimes^{iF/\hbar} \right) - \frac{i}{\hbar} \T
\left( \left( \hat \Q_0 F - \frac{1}{2} (F,F) \right) \otimes \e^{iF/\hbar}_\otimes \right) \\
&= \T
\left(\A(\e^F_\otimes) \otimes \e^{iF/\hbar}_\otimes \right)~,
\end{split}
\een
which is valid when all ghosts except $c$ and $X \propto x^\alpha \partial_\alpha$ are zero. Using \eqref{smu}, we may infer
\ben
\sigma_X \, \mu\tfrac{\d}{\d \mu} \hat S^{(\mu)}  = \A(\e^F_\otimes)~,
\een
under the same conditions. From Lemma~\ref{stabilityRG}, we see that the left hand side involves $\beta \int \sigma_X \hat \eL \dv$ while the leading $\hbar$ contribution to the right hand side is $C \int \eA$, keeping only those terms in $\eA$ which contain either $c$ or $X$, and where $C=\sum_n C^{(n)} \hbar^n$ is a numerical coefficient. From the concrete expression for $\eA$ given in Theorem~\ref{cohothm}, it follows that, to leading order in $\hbar$, the right hand side is  $-2C^{(1)} \hbar \int \sigma_X \eL \dv$. Comparing both sides therefore gives $C^{(1)}=-\half \beta^{(1)}$. This is the claim demonstrating \eqref{Acondition} for $m=1$, with no $m>1$ contributions.


\section{Non-perturbative effects, chiral rings and localisation}

Even in Minkowski space, precise non-perturbative results in quantum field theory are a rare commodity. Their acquisition for theories on the curved spacetimes considered here should be no less difficult. A non-perturbative effect that is frequently encountered for gauge theories in Minkowski space comes from instanton configurations. Instantons exist as real solutions only in Euclidean signature, so their definition in spacetime requires an analytic continuation. Certain spacetimes which admit a twistor spinor -- and hence support field theories with rigid conformal supersymmetry -- admit an analytic continuation with suitable Riemannian counterparts. For example, de Sitter space $\dS_4$ and the Einstein static universe $\RR \times S^3$ (both conformally flat) are of this type. Unfortunately, for generic non-conformally flat spacetimes with a twistor spinor (i.e. cases {\bf (2a)} and {\bf (2b)} in section~\ref{sec:spacetimes}), there is simply no Riemannian counterpart (i.e. a real Riemannian section of a suitable analytic continuation). Indeed, this is the case even for generic pp-waves~\eqref{ppwave}. It is therefore unclear to us what r\^{o}le instantons might play in field theories on these more general curved spacetimes.

A special class of operators in a supersymmetric gauge theory in Minkowski space is defined by the {\bf chiral ring} \cite{Cachazo:2002ry,Witten:2003ye}. This well-known construction proceeds by writing a Poincar\'{e} supersymmetry transformation $\delta_\epsilon = \delta_{\epsilon_+} + \delta_{\epsilon_-}$, in terms of the complex chiral projections $\epsilon_\pm$ of a constant Majorana spinor $\epsilon$. It follows that $\delta_{\epsilon_\pm}^2 =0$ while $\delta_{\epsilon_+} \delta_{\epsilon_-} + \delta_{\epsilon_-} \delta_{\epsilon_+}$ generates a translation along a vector $\xi_\epsilon$ in Minkowski space. The {\bf classical chiral ring} $\eR$ is then defined by the space of all complex gauge-invariant operators in the kernel of $\delta_{\epsilon_-}$, with any pair of such operators considered equivalent if they differ by a $\delta_{\epsilon_-}$-exact operator. (The classical anti-chiral ring is defined in the same way, but with respect to $\delta_{\epsilon_+}$.) In Minkowski space, the translations generated by $\delta_{\epsilon_+} \delta_{\epsilon_-} + \delta_{\epsilon_-} \delta_{\epsilon_+}$, for all constant Majorana spinors $\epsilon$, span all of $\RR^{3,1}$. Whence, the derivative $\partial_\mu \O$ of any $\O \in \R$ is trivial in $\R$. For an $\eN =1$  supersymmetric gauge theory, $\eR$ is generated by monomials built from the bosonic fields in the matter supermultiplet and the (positive) chiral projection $\lambda_+$ of the gaugino in the gauge supermultiplet. These monomials are subject to certain relations defined by the representation-theoretic data and the superpotential for the theory. For $\eN =2$ supersymmetric Yang-Mills theory, $\eR$ is generated by monomials in the complex scalar $\varphi$ (see \eqref{eq:4dsusyvectorcurved} evaluated in Minkowski space) subject to certain algebraic relations defined by the gauge Lie algebra $\fg$ (e.g. for $\fg = \fsu(N)$, $\Tr ( \varphi^{N+1} ) =0$ in $\eR$). In a theory with $\eN = 2$ supersymmetry, one may also define a chiral ring with respect to an $\eN = 1$ Poincar\'{e} superalgebra that is specified by fixing a vector $l \in \CC^2$ in the $\eN = 2$ Poincar\'{e} superalgebra. The odd part of this $\eN = 1$ Poincar\'{e} superalgebra is spanned by elements of the form $\epsilon^A_+ = l^A \epsilon_+$, for all constant positive-chirality spinors $\epsilon_+$ on $\RR^{3,1}$. For $\eN =2$ supersymmetric Yang-Mills theory, the associated $\eN =1$ chiral ring is generated by gauge-invariant monomials in $\varphi$ and $\lambda_+ = l_A \lambda_+^A$ (where $l_A = (l^A)^*$).

Analogs of the $\eN = 1$ and $\eN = 2$ chiral rings can be defined on any curved spacetime admitting twistor spinors. To embed this notion into the formalism 
employed in the body of the paper, we may also formulate its definition in terms of suitable BRST differential (for definiteness, we restrict attention to the
$\eN=1$ chiral ring). Let $\q$ denote the BRST operator defined by setting $\epsilon_+^A = l^A \epsilon_+$ in \eqref{b1}, \eqref{b2}, \eqref{b3} and \eqref{b4}. It admits a decomposition $\q = \q_- + \q_0 + \q_+$, defined such that $\q_\pm$ increases/decreases by one unit the number of $\epsilon_-$ minus the number of $\epsilon_+$ factors, while $\q_0$ leaves this difference invariant. It follows that $\q_\pm^2 =0$ while $\q_+ \q_- + \q_- \q_+ = \delta_{\xi}$, where $\xi^\mu = 2 \overline \epsilon_- \Gamma^\mu \epsilon_+$. The classical chiral ring is identified with the cohomology $H(\q_+ , \M)=\{ \ker \ \q_+ \}/\{ \im \ \q_+ \}$. The nature of this ring depends critically on the nature of the conformal symmetry superalgebra $\cS$ for the spacetime in question. For example, let $(\M, g)$ be either a pp-wave or a Fefferman space (i.e. cases {\bf (2a)} and {\bf (2b)} in section~\ref{sec:spacetimes}). In that case, $\xi$ is the unique null Killing vector field on $(\M, g)$, and the local operators $\O \in H(\q_+ , \M)$ for the Yang-Mills sector are generated by gauge-invariant monomials in $\varphi$ and $\lambda_+ = l_A \lambda_+^A$. An example of a non-local operator in the chiral ring is given by
\ben\label{nonlocalchiralringops}
\O[C] = \Tr \left( {\overline \lambda}_+ \, {\rm P} \exp \left( \int_C A \right) \lambda_+ \right)~,
\een
for any curve $C$ in $\M$ that runs parallel to the null conformal Killing vector $\xi^\mu = 2 \overline \epsilon_- \Gamma^\mu \epsilon_+$. The descent equations for $H(\q_+ | \d , \M)$ may also be defined, yielding appropriate ladders for the chiral ring. One such ladder is provided by the Lagrangian $\hat \eL$ in \eqref{Lhat}. 

The {\bf quantum chiral ring} can be defined after incorporating the relevant anomalies into $\q_+$, in a similar manner to Lemma~\ref{commutatorlemma}. These anomalies typically modify the form of the relations in the classical chiral ring, though we omit the details. Since $\q_+ \q_- + \q_- \q_+ = \delta_{\xi}$, it follows that any correlation function of chiral ring operators containing a factor $\delta_{\xi} \O$, for any chiral ring operator $\O$, must vanish identically. Thus,
any correlation function of a product of local operators in the chiral ring is invariant under a shift of any individual insertion points in the $\xi$-direction. The precise implications of this property depend on the spacetime being considered. In Minkowski space, if $\q_+$ is defined with respect to either the $\eN=1$ or $\eN=2$ Poincar\'{e} superalgebras in $\cS$, then $\xi$ can generate an arbitrary translation on $\RR^{3,1}$ (with constant Grassmann-valued coefficients). Whence, in Minkowski space, we recover the well-known fact that any correlation function of local operators in the chiral ring is independent of the insertion points. On the other hand, for a generic pp-wave spacetime, $\xi = \tfrac{\partial}{\partial v}$ is the unique null Killing vector and correlation functions of local operators in the chiral ring do not depend on the coordinate $v$. Similar remarks apply to the correlation function $\langle \O[C_1] \dots \O[C_n] \rangle$ of non-local operators in \eqref{nonlocalchiralringops}, which is invariant under shifting individually any of the curves $C_1$,...,$C_n$ in the $v$-direction.

One method which has led to a number of important non-perturbative results in the context of supersymmetric quantum field theory is {\bf localisation} \cite{Witten:1988ze,Nekrasov:2002qd,Nekrasov:2003rj,Pestun:2007rz}. In this setup, one argues that the only contributions to the path integral come from the (typically finite-dimensional) space of field configurations which describe fixed points of the supersymmetry transformations for a particular choice of rigid supercharge. Expectation values of various local and non-local operators (including those in the chiral ring) may sometimes be computed exactly using this technique. For the partition function, this is often the case when the underlying action is exact with respect to the relevant supercharge (at least up to topological terms like $\int_\M \Tr ( F \wedge F )$ for the instanton number) \cite{Witten:1988ze}. However, in order to obtain meaningful results via localisation, it is typically essential for the background space to be either compact or flat, and Riemannian. Thus, for the class of Lorentzian manifolds we have been considering that are not conformally flat, being non-compact and with no natural analytic continuation, there appears to be no obvious way to make use of localisation to obtain exact results.


\section*{Acknowledgments}

The financial support provided by ERC Starting Grant QC \& C 259562 is gratefully acknowledged. We would like to thank F. Brandt for some helpful comments. 

{\bf Note added in proof.} After this paper was published, it was pointed out to us by the authors of \cite{Anous:2014lia} that a remark on the non-existence of minimal conformal supersymmetry in four-dimensional de Sitter space we had made in the final paragraph of section~\ref{sec:spacetimes} in a previous version of our paper was erroneous. Indeed, the offending remark was inconsistent with our general results in section 2 on the classification of conformal symmetry superalgebras. We thank the authors of \cite{Anous:2014lia} for bringing this mistake to our attention.


\appendix


\section{Data for $\eN = 2$ superconformal quantum field theories}
\label{app:A}

Let $U$ be a complex representation of a (real) simple Lie algebra $\fg$. Let $c(U)$ denote the Dynkin index of $U$. For $\fg$ simple, $c(\fg ) = h^\vee (\fg)$ is the dual Coxeter number of $\fg$. In this appendix, we shall describe the classification of pairs $(U,\fg)$ for which $c(U) = h^\vee (\fg)$. This condition is necessary and sufficient for the vanishing of the $\beta$-function in a quantum field theory with rigid $\eN \geq 2$ supersymmetry.

Table~\ref{tab:Dynkin} shows the Dynkin index $c$ of the adjoint and fundamental representation $\ff$ of each simple Lie algebra $\fg$ (where $N>1$ for $\fsu(N)$ and $\fsp(N)$ and $N>6$ for $\fso(N)$). Since $U$ is completely reducible, we can write $U = \bigoplus_i m_i U_i$, where each $U_i$ is irreducible and occurs with multiplicity $m_i$. Solutions of $c(U) = \sum_i m_i c(U_i) = h^\vee (\fg)$ can therefore involve only irreducible representations $U_i$ with $c( U_i ) \leq h^\vee (\fg)$. We omit the generic solution with $U \oplus U^* = \fg_\CC$, which yields the data for $\eN =4$ supersymmetric Yang-Mills theory. Notice that the adjoint representation cannot occur in $U$, except for the $\eN =4$ solution which we have just omitted. Table~\ref{tab:Dynkin} shows that $c(\ff) < h^\vee (\fg)$, with $\frac{h^\vee (\fg)}{c(\ff)}$ a positive integer for all simple $\fg$. A generic $\eN =2$ solution is therefore given by
\begin{equation}\label{eq:Fsol}
U = \tfrac{h^\vee (\fg)}{c(\ff)} \; \ff~,
\end{equation}
except for $\fg = E_8$ where $\ff \cong \fg$ and this just recovers the $\eN =4$ solution.

\begin{table}
\begin{center}
\begin{tabular}{|c|c|c||c|c|}
  \hline
  $\fg$ & $\mbox{dim} (\fg )$ &  $c( \fg ) = h^\vee (\fg)$ & $\ff$  & $c(\ff)$  \\
  \hline \hline
  $\fsu(N)$ & $N^2 -1$ & $N$ & ${\bf N}$ & $\half$  \\
  $\fso(N)$ & $\half N (N-1)$ & $N-2$ & ${\bf N}$ & $1$  \\ [.05cm]
  $\fsp(N)$ & $N (2N+1)$ & $N+1$ & ${\bf 2N}$ & $\half$  \\
  $E_6$ & 78 & 12 & {\bf 27} & 3  \\
  $E_7$ & 133 & 18  & {\bf 56} & 6  \\
  $E_8$ & 248 & 30 & {\bf 248} & 30  \\
  $F_4$ & 52 & 9 & {\bf 26} & 3  \\
  $G_2$ & 14 & 4 & {\bf 7} & $1$  \\
  \hline
\end{tabular} \vspace*{.2cm}
\caption{Dynkin index of adjoint and fundamental representations.}
\label{tab:Dynkin}
\end{center}
\end{table}

The remaining irreducible representations $U^{\mathrm{irr}} \neq \ff$ with $c(U^{\mathrm{irr}}) < h^\vee (\fg)$ are shown in Table~\ref{tab:Nequals2irreps}. They were obtained using the {\tt LieART} Mathematica package~\cite{LieART}. There are no such irreducible representations of any of the exceptional Lie algebras. The notation $\bigwedge^k \ff$ and ${\sf S}^k \ff$ indicates the rank $k$ totally (skew)symmetric tensor representations for $\ff$. For $\fso(N)$, $\$$ denotes the spinor representation for $N$ odd and $\$_\pm$ denotes the $\pm$ chirality spinor representation for $N$ even (with $\$_+ \cong \$_- \cong \ff$ for $N=8$). For $\fsp(N)$, $\bigwedge^k_0 \ff$ denotes the $\fsp(N)$-invariant subspace of $\bigwedge^k \ff$ consisting of rank $k$ skewsymmetric tensors which are traceless with respect to contraction of any pair of indices with the inverse symplectic form. If $U^{\mathrm{irr}}$ is such that $c(U^{\mathrm{irr}}) < h^\vee (\fg)$ only for some particular values of $N$ within the assumed ranges, the values are indicated in the column headed \lq $N$' in Table~\ref{tab:Nequals2irreps}.

\begin{table}
\begin{center}
\begin{tabular}{|c|c|c|c|}
  \hline
  $\fg$ & $U^{\mathrm{irr}}$ & $N$ & $c( U^{\mathrm{irr}} )$  \\
  \hline \hline
  $\fsu(N)$ & $\bigwedge^2 \ff$ &  & $\tfrac{N}{2}-1$ \\ [.1cm]
  & ${\sf S}^2 \ff$ &  & $\tfrac{N}{2}+1$ \\ [.1cm]
  & $\bigwedge^3 \ff$ & $6,7,8$ & $3,5,\tfrac{15}{2}$ \\ [.1cm]
  \hline
  $\fso(N)$ & $\$$ & $7,9,11,13$ & $1,2,4,8$ \\ [.1cm]
  & $\$_\pm$ & $8,10,12,14$ & $1,2,4,8$  \\ [.1cm]
  \hline
  $\fsp(N)$ & $\bigwedge^2_0 \ff$ & & $N-1$   \\ [.1cm]
  & $\bigwedge^3_0 \ff$ & 3 & $\tfrac{5}{2}$ \\ [.1cm]
  \hline
\end{tabular} \vspace*{.2cm}
\caption{Irreducible representations $U^{\mathrm{irr}} \neq \ff$ with $c(U^{\mathrm{irr}}) < h^\vee (\fg)$.}
\label{tab:Nequals2irreps}
\end{center}
\end{table}

In addition to the $\eN=2$ solution in \eqref{eq:Fsol}, the data in Table~\ref{tab:Nequals2irreps} yields five more generic solutions: \\ [.2cm]
$\bullet$ $\fg =\fsu(N)$ with $U$ either $\ff \otimes \ff$, $(N+2)\, \ff \oplus \bigwedge^2 \ff$, $4\, \ff \oplus 2\, \bigwedge^2 \ff$ or $(N-2)\, \ff \oplus {\sf S}^2 \ff$. \\ [.1cm]
$\bullet$ $\fg = \fsp(N)$ with $U = 4\, \ff \oplus \bigwedge^2_0 \ff$.\\ [.2cm]
Plus several more isolated solutions:
\\ [.2cm]
$\bullet$ $\fg =\fsu(4)$ with $U$ either $2 \, \ff \oplus 3 \bigwedge^2 \ff$ or $4 \bigwedge^2 \ff$. \\ [.1cm]
$\bullet$ $\fg =\fsu(5)$ with $U = \ff \oplus 3 \bigwedge^2 \ff$. \\ [.1cm]
$\bullet$ $\fg =\fsu(6)$ with $U$ either $3 \bigwedge^2 \ff$, $6\, \ff \oplus \bigwedge^3 \ff$, $2\, \ff \oplus \bigwedge^2 \ff \oplus \bigwedge^3 \ff$ or $2 \bigwedge^3 \ff$. \\ [.1cm]
$\bullet$ $\fg =\fsu(7)$ with $U = 4\, \ff \oplus \bigwedge^3 \ff$. \\ [.1cm]
$\bullet$ $\fg =\fsu(8)$ with $U = \ff \oplus \bigwedge^3 \ff$. \\ [.1cm]
$\bullet$ $\fg =\fso(7)$ with $U = (5-n) \, \ff \oplus n \, \$$, for any $n=1,2,3,4,5$. \\ [.1cm]
$\bullet$ $\fg =\fso(9)$ with $U = (7-2n) \, \ff \oplus n \, \$$, for any $n=1,2,3$. \\ [.1cm]
$\bullet$ $\fg =\fso(10)$ with $U = 2(4- n_+ - n_- ) \, \ff \oplus n_+ \, \$_+ \oplus n_- \, \$_-$, for any $n_\pm$ such that $n_+ + n_- =1,2,3,4$. \\ [.1cm]
$\bullet$ $\fg =\fso(11)$ with $U$ either $5 \, \ff \oplus \$$ or $\ff \oplus 2\, \$$. \\ [.1cm]
$\bullet$ $\fg =\fso(12)$ with $U = 2(5- 2n_+ - 2n_- ) \, \ff \oplus n_+ \, \$_+ \oplus n_- \, \$_-$, for any $n_\pm$ such that $n_+ + n_- =1,2$. \\ [.1cm]
$\bullet$ $\fg =\fso(13)$ with $U = 3 \, \ff \oplus \$$. \\ [.1cm]
$\bullet$ $\fg =\fso(14)$ with $U = 4 \, \ff \oplus \, \$_\pm$. \\ [.1cm]
$\bullet$ $\fg = \fsp(2)$ with $U$ either $2\, \ff \oplus 2 \bigwedge^2_0 \ff$ or $3 \bigwedge^2_0 \ff$.\\ [.1cm]
$\bullet$ $\fg = \fsp(3)$, with $U$ either $3\, \ff \oplus \bigwedge^3_0 \ff$ or $2 \bigwedge^2_0 \ff$.\\ [.2cm]


\section{Descent equations for the vector multiplet in curved spacetime}
\label{app:B}

Here we give explicitly all elements of the ladder \eqref{ladder}, starting with $\O^4_0=\hat \eL \dv$ (see Theorem~\ref{cohothm}). For simplicity, we restrict attention to the vector multiplet. The elements in the ladder can be written
\ben\label{opq}
\O^3_1 = \eL^3_1 + i_X \eL^4_0 \ , \quad
\O^2_2 = \eL^2_2 + i_X \eL^3_1 + i_X i_X \eL^4_0 \ , \quad
\text{etc.},
\een
where $\eL^4_0=\O^4_0$ and
\begin{align}\label{eq:4dvectordescent}
\eL^{3}_1 &= \tfrac{1}{6} {\rm Re} \left( {\overline \epsilon}_{-\; A} \Gamma_\mu \left( -{\slashed F} \lambda_+^A - 2\, [ \varphi , \varphi^* ] \lambda_+^A + 2\, {\slashed D} \varphi \lambda_-^A + Y^{AB} \lambda_{+\; B} \right) \right.   \nonumber \\
&\hspace*{1.5cm} \left.  + {\overline{{\slashed \nabla} \epsilon}}_{-\; A} \Gamma_\mu \lambda_-^A \varphi \right) \varepsilon^\mu{}_{\nu\rho\sigma} {\rm d} x^\nu \wedge {\rm d} x^\rho \wedge {\rm d} x^\sigma \nonumber \\
\eL^{2}_2 &= \tfrac{1}{4} {\rm Re} \left( 2i \varepsilon_{\mu\nu\alpha\beta} \, \varphi^*  ( \xi^\alpha D^\beta \varphi + ( {\overline \epsilon}_+^A \epsilon_{+\; A} )\, F^{\alpha\beta} )    + \half ( {\overline \epsilon}_+^A \epsilon_{+\; A} ) ( {\overline \lambda}_-^B \Gamma_{\mu\nu} \lambda_{-\; B} ) \right. \\
&\hspace*{1.5cm} \left. -( {\overline \epsilon}_+^A  \Gamma_{\mu\nu} \epsilon_+^B ) ( {\overline \lambda}_{-\; A} \lambda_{-\; B} ) - i \varepsilon_{\mu\nu\alpha\beta} \, ( {\overline \epsilon}_{-\; A} \Gamma^\alpha \epsilon_{+\; B} ) ( {\overline \lambda}_-^A \Gamma^{\beta} \lambda_+^B ) \right) \varepsilon^{\mu\nu}{}_{\rho\sigma} {\rm d} x^\rho \wedge {\rm d} x^\sigma \nonumber \\
\eL^{1}_3 &= 4 {\rm Re} \left( 4i\varphi \, \left( 2\, ( {\overline \epsilon}_-^{(A} \Gamma_\mu \epsilon_+^{B)} )  ( {\overline \epsilon}_{-\; A} \lambda_{-\; B} )  - ( {\overline \epsilon}_-^{A} \epsilon_{-\; A} )  ( {\overline \epsilon}_{-\; B} \Gamma_\mu \lambda_+^B ) \right) \right) {\rm d} x^\mu \nonumber \\
  \eL^0_4 &= 4  {\rm Re} \left( \! -2i \, ( {\overline \epsilon}_-^{A} \epsilon_{-\; A} )^2 \varphi^2 \right)~. \nonumber
\end{align}
We have omitted the obvious $(-,-)$ inner product symbols in \eqref{eq:4dvectordescent} in order to minimise clutter. All elements in the ladder are Weyl-invariant.

In Minkowski space, there have been several claims in the literature relating one-loop exactness of the $\beta$-function to properties of this ladder, see e.g.~\cite{Baulieu:2007dk}
and references therein. In essence, these arguments are based on the idea of relating renormalization properties of the bottom element in the ladder $\O^0_4$ to renormalization properties of
the top element $\O^4_0=\hat \eL \dv$, i.e. the Lagrangian. Concretely, their strategy is to relate $\gamma \O^4_0$ to $\gamma \O^0_4$ -- the interest being in the former object due to a relation with the $\hbar$-derivative of the $\beta$-function (which may be derived using the results of our Section~\ref{smu}, \eqref{gammadef}, (ix) of Theorem~\ref{Tuniqueness} and Lemma~\ref{stabilityRG}). It is claimed that $\gamma \O^4_0$, which contains the operator $\varphi^2$ in the chiral ring, has special renormalization properties. Now, in order to relate $\gamma \O^4_0$ to $\gamma \O^0_4$, one must clearly try to show that they belong to the same ladder. One may attempt to prove this by applying $\gamma$ to the relations $\Q \O^p_{4-p} = \d \O^{p-1}_{5-p}$, and using~\eqref{funcid} to commute $\gamma$ through $\Q$. Unfortunately, this strategy runs into a number of difficulties. One reason (which can be dealt with) is that \eqref{funcid} involves $\hat \Q_\hbar$, rather than $\Q$. A more fundamental difficulty is that the right hand side of \eqref{funcid} also involves $\hat S'$ which, from Lemma~\ref{stabilityRG}, may involve a rather arbitrary term $\Psi$. For this reason, it is our opinion that the aforementioned strategy is not rigorous unless one is able make more detailed statements about $\Psi$. It is unclear how this could be accomplished in a framework based solely on consistency conditions like ours.

\bibliographystyle{utphys}
\bibliography{CurvedSCFT}
\end{document}